\documentclass{LMCS}

\def\doi{8 (1:26) 2012}
\lmcsheading%
{\doi}
{1--63}
{}
{}
{Sep.~\phantom07, 2011}
{Mar.~23, 2012}
{}

\usepackage[utf8]{inputenc}
\usepackage{mathbbol}
\usepackage{stmaryrd,amsfonts,amsmath,amssymb,amsbsy}
\usepackage[svgnames]{xcolor}
\usepackage{subfigure}
\usepackage{graphicx}
\usepackage[matrix,arrow,curve,frame]{xy}
\usepackage{enumerate}

\ifx\hascolor\undefined
\usepackage{hyperref}
\def\mycolor#1{}
\else
\usepackage{hyperref}
\def\mycolor#1{\color{#1}}
\fi

\DeclareSymbolFontAlphabet{\mathbbx}{bbold}

\def\mC#1{\mathcal{#1}}
\def\mB#1{\mathbb{#1}}
\def\mBx#1{\mathbbx{#1}}
\def\mb#1{\mathbf{#1}}

\def\mi#1{\mathit{#1}}

\def\P{\mC{P}}
\def\N{\mB{N}}

\def\R{\mB{R}}

\def\myfun#1{\operatorname{{\it\mycolor{DarkGreen}#1}}}
\def\sy#1{{\mycolor{Maroon}\mb{#1}}}
\def\V{\mC{V}}      
\def\D{\mC{D}}      
\def\E{\mC{E}}      
\def\L{\mC{L}}      
\def\I{\mC{I}}      
\def\s{\sharp}
\def\RR{\mC{N}}     
\def\H{\mC{H}}      
\def\C{\mC{C}}      
\def\T{\mC{T}}      
\def\M{\mC{M}}      
\def\race{{\C\mC{R}}}
\def\lfp{\myfun{lfp}}
\def\lim{\myfun{lim}}
\def\lval{\myfun{lval}}
\def\nonblock{\myfun{nonblock}}
\def\noerror{\myfun{noerror}}
\def\var{\myfun{var}}
\def\proj{\myfun{proj}}
\def\enabled{\myfun{enabled}}
\def\sched{\myfun{sched}}
\def\widen{\mathbin{\triangledown}} 
\def\body{\myfun{body}}
\def\get{\myfun{get}}
\def\asexpr{\myfun{as-expr}}
\def\apply{\myfun{apply}}
\def\excl{\myfun{intf}}
\def\ready{\myfun{ready}}
\def\yield{\myfun{yield}}
\def\wait{\myfun{wait}}
\def\funin{\myfun{in}}
\def\funout{\myfun{out}}
\def\weak{\myfun{weak}}
\def\sync{\myfun{sync}}

\def\dsymb#1{\;\,{\mycolor{blue}\stackrel{\mbox{\rm\tiny def}}{#1}}\;\,}
\def\deq{\dsymb{=}}
\def\diff{\dsymb{\iff}}
\def\letin#1#2{\mbox{\mycolor{blue}let }#1{\mycolor{blue}\;=\;}#2\mbox{ \mycolor{blue}in }}

\def\becomes{\;\rightsquigarrow\;}

\def\br{\linebreak[4]}

\def\set#1{\{\,#1\,\}}
\def\sset#1{\{#1\}}
\def\setst#1#2{\{\,#1\;|\;#2\,\}}

\def\lbd#1{\lambda #1:}
\def\lbb#1#2#3{\mathbin{{\mycolor{FireBrick}#1}#2{\mycolor{FireBrick}#3}}}
\def\lb#1#2{\lbb{\mB{#1}\llbracket}{\,#2\,}{\rrbracket}}
\def\lbp#1{\lbb{\mB{#1}}{}{}}
\def\lbx#1#2{\lbb{\mBx{#1}\llbracket}{\,#2\,}{\rrbracket}}
\def\lbpx#1{\lbb{\mBx{#1}}{}{}}
\def\ex#1{\exists #1:}
\def\fa#1{\forall #1:}

\def\com#1{\mycolor{darkgray}{\text{\textsl{\small(#1)}}}}

\begin{document}

\title[Static Analysis of Run-Time Errors in Parallel C Programs]{Static Analysis of Run-Time Errors in Embedded Real-Time Parallel C Programs\rsuper*}
\author{Antoine Miné}
\address{CNRS \& École Normale Supérieure, 45 rue d'Ulm, 75005 Paris, France}
\email{mine@di.ens.fr}
\thanks{This work was partially supported by the INRIA project ``Abstraction" common to CNRS and ENS in France, and by the project ANR-11-INSE-014 from the French {\em Agence nationale de la recherche}.}

\begin{abstract}
We present a static analysis by Abstract Interpretation to check for
run-time errors in parallel and multi-threaded C programs.
Following our work on Astrée, we focus on embedded critical programs without
recursion nor dynamic memory allocation, but extend the analysis to
a static set of threads communicating implicitly through a shared memory
and explicitly using a finite set of mutual exclusion locks, and
scheduled according to a real-time scheduling policy and fixed priorities.
Our method is thread-modular.
It is based on a slightly modified non-parallel analysis that, when
analyzing a thread, applies and enriches an abstract set of thread 
interferences.
An iterator then re-analyzes each thread in turn until interferences stabilize.
We prove the soundness of our method with respect to the sequential
consistency semantics, but also with respect to a reasonable weakly consistent 
memory semantics.
We also show how to take into account mutual exclusion and thread priorities
through a partitioning over an abstraction of the scheduler state.
We present preliminary experimental results analyzing an industrial program with
our prototype, Thésée, and demonstrate the scalability of our approach.
\end{abstract}


\keywords{Abstract interpretation, parallel programs, run-time errors, static analysis}
\subjclass{D.2.4, F.3.1, F.3.2}
\titlecomment{{\lsuper*}This article is an extended version of our article 
\cite{mine:esop11} published in the Proceedings of the 20th European Symposium 
on Programming (ESOP'11).}

\maketitle


\section{Introduction}
\label{sec:intro}

Ensuring the safety of critical embedded software is important as a single
``bug'' can have catastrophic consequences.
Previous work on the Astrée analyzer \cite{blanchet-al-PLDI03} demonstrated
that static analysis by Abstract Interpretation could help, when specializing
an analyzer to a class of properties and programs --- namely in that case,
the absence of run-time errors (such as arithmetic and memory errors)
on synchronous control~/ command embedded avionic C software.
In this article, we describe ongoing work to achieve similar results for
{\em multi-threaded\/} and {\em parallel\/} embedded C software.
Such an extension is demanded by the current trend in critical embedded 
systems to switch from large numbers of single-program processors 
communicating through a 
common bus to single-processor multi-threaded applications communicating 
through a shared memory --- for instance, in the context of Integrated Modular 
Avionics \cite{ima}.
Analyzing each thread independently with a tool such as Astrée would not be 
sound and could miss bugs that only appear when threads interact.
In this article, we focus on detecting the same kinds of run-time errors as 
Astrée does, while taking thread communications into account in a sound way,
including accesses to the shared memory and synchronization primitives.
In particular, we correctly handle the effect of concurrent threads
accessing a common variable without enforcing mutual exclusion by
synchronization primitives, and we report such accesses --- these will be called
{\em data-races\/} in the rest of the article.
However, we ignore other concurrency hazards such as dead-locks, 
live-locks, and priority inversions, which are considered to be
orthogonal issues.

\medskip

Our method is based on Abstract Interpretation \cite{cc-POPL77}, a general 
theory of the approximation of semantics which allows designing static
analyzers that are fully automatic and sound by construction --- i.e., 
consider a superset of all program behaviors.
Such analyzers cannot miss any bug in the class of errors they analyze.
However, they can cause spurious alarms due to 
over-approximations, an unfortunate effect we wish to minimize while keeping
the analysis efficient.

To achieve scalability, our method is thread-modular and performs
a rely-guarantee reasoning, where rely and guarantee conditions are
inferred automatically.
At its core, it performs a sequential analysis of each thread considering 
an abstraction of the effects of the other threads, called interferences.
Each sequential analysis also collects a new set of interferences generated by 
the analyzed thread. It then serves as input when analyzing the other threads.
Starting from an empty set of interferences, threads are re-analyzed
in sequence until a fixpoint of interferences is reached for all threads.
Using this scheme, few modifications are required to a sequential analyzer 
in order to analyze multi-threaded programs.
Practical experiments suggest that few thread re-analyses are required in 
practice, resulting in a scalable analysis.
The interferences are considered in a flow-insensitive and non-relational
way: they store, for each variable, an abstraction of the set of all
values it can hold at any program point of a given thread.
Our method is however quite generic in the way individual threads are analyzed.
They can be analyzed in a fully or partially flow-sensitive, context-sensitive,
path-sensitive, and relational way (as is the case in our prototype).

As we target embedded software, we can safely assume that there is no recursion,
dynamic allocation of memory, nor dynamic creation of threads nor locks, 
which makes the analysis easier.
In return, we handle two subtle points.
Firstly, we consider a weakly consistent memory model: memory accesses
not protected by mutual exclusion (i.e., data-races)
may cause behaviors that are not the result of any thread interleaving to 
appear.
The reason is that arbitrary observation by concurrent threads can expose 
compiler and processor optimizations (such as instruction reordering) that are 
designed to be transparent on non-parallel programs only.
We prove that our semantics is invariant by large classes
of widespread program transformations, so that an analysis of the original
 program is also sound with respect to reasonably 
compiled and optimized versions.
Secondly, we show how to take into account the effect of a real-time scheduler 
that schedules the threads on a single processor following strict, fixed
priorities.
According to this scheduling algorithm, which is quite common in the realm of
embedded real-time software --- e.g., in the real-time thread extension of the
POSIX standard \cite{posix-threads}, or in the ARINC 653 avionic operating 
system standard \cite{ARINC} ---
only the unblocked thread of highest priority may run.
This ensures some lock-less mutual exclusion properties that are actually 
exploited in real-time embedded programs and relied on for their correctness
(this includes the industrial application our prototype currently targets).
We show how our analysis can take these properties into account, but we
also present an analysis that assumes less properties on the scheduler
and is thus sound for true multi-processors and non-real-time schedulers.
We handle synchronization properties (enforced by either locks or priorities)
through a partitioning with respect to an abstraction of
the global scheduling state.
The partitioning recovers some kind of inter-thread flow-sensitivity
that would otherwise be completely abstracted away by the interference
abstraction.

The approach presented in this article has been implemented and used at the 
core of a prototype analyzer named Thésée.
It leverages the static analysis techniques developed in Astrée 
\cite{blanchet-al-PLDI03} for single-threaded programs, 
and adds the support for multiple threads.
We used Thésée to analyze in 27~h
a large (1.7~M lines) multi-threaded industrial 
embedded C avionic application, which illustrates the scalability of our 
approach.

\subsection*{Organisation}
Our article is organized as follows.
First, Sec.~\ref{sec:proc} presents a classic non-parallel semantics and
its static analysis.
Then, Sec.~\ref{sec:shared} extends them to several threads in a shared memory
and discusses weakly consistent memory issues.
A model of the scheduler and support for locks and priorities are
introduced in Sec.~\ref{sec:sched}.
Our prototype analyzer, Thésée, is presented in Sec.~\ref{sec:result},
as well as some experimental results.
Finally, Sec.~\ref{sec:relwork} discusses related work, 
and Sec.~\ref{sec:conclusion} concludes and envisions future work.

This article defines many semantics.
They are summarized in Fig.~\ref{fig:sumsem}, using $\subseteq$ to denote
the ``is less abstract than'' relation.
We alternate between two kinds of concrete semantics: semantics 
based on {\em control paths\/} ($\lbp{P_\pi}$, $\lbp{P_*}$, $\lbp{P_\H}$),
that can model precisely thread interleavings and are also useful to
characterize weakly consistent memory models
($\lbp{P'_*}$, $\lbp{P'_\H}$), and semantics by
{\em structural induction\/} on the syntax 
($\lbp{P}$, $\lbp{P_\I}$, $\lbp{P_\C}$),
that give rise to effective abstract interpreters
($\lbp{P^\s}$, $\lbp{P_\I^\s}$, $\lbp{P_\C^\s}$).
Each semantics is presented in its subsection and adds
some features to the previous ones, so that the final abstract analysis
$\lbp{P_\C^\s}$ presented in Sec.~\ref{sec:schedabs} should hopefully not 
appear as too complex nor artificial, but rather as the logical conclusion of a 
step-by-step construction.

Our analysis has been mentioned first, briefly and informally, in 
\cite[\S~VI]{bertrane-al-aiaa10}.
We offer here a formal, rigorous treatment by presenting all the
semantics fully formally, albeit on an idealised language, and 
by studying their relationship.
The present article is an extended version of \cite{mine:esop11}
and includes a more comprehensive description of the semantics as well
as the proof of all theorems, that were omitted in the 
conference proceedings due to lack of space.

\begin{figure}[t]
  \centering
  \begin{tabular}{c@{\qquad}c@{\qquad}c}
    non-parallel semantics & parallel semantics \\[5pt]
    \xy
    \xymatrix"*"@R-1pc{
      \txt{$\lbp{P}$\\(\S\ref{sec:structsem})} \ar[r]^\subseteq &
      \txt{$\lbp{P^\s}$\\(\S\ref{sec:procabs})}
      \\
      \txt{$\lbp{P_\pi}$\\(\S\ref{sec:pathsem})} \ar@{-}[u]^=
    }
    \POS*\frm{^)}
    \endxy
    &
    \xy
    \xymatrix"*"@R-1pc{
      &
      \txt{$\lbp{P_\I}$\\(\S\ref{sec:interfersem})} \ar[r]^\subseteq &
      \txt{$\lbp{P_\I^\s}$\\(\S\ref{sec:sharedabs})}
      \\
      \txt{$\lbp{P_*}$\\(\S\ref{sec:interleavesem})} \ar[ur]^\subseteq &
      \txt{$\lbp{P'_*}$\\(\S\ref{sec:weaksem})} \ar[u]^\subseteq
      &
      \\\\
      &
      \txt{$\lbp{P_\C}$\\(\S\ref{sec:schedinterfersem})} \ar[r]^\subseteq \ar@{.>}@/_2pc/[uuu]_\subseteq &
      \txt{$\lbp{P_\C^\s}$\\(\S\ref{sec:schedabs})} 
      \\
      \txt{$\lbp{P_\H}$\\(\S\ref{sec:schedinterleavesem})} \ar[ru]^\subseteq \ar@{.>}[uuu]^\subseteq &
      \txt{$\lbp{P'_\H}$\\(\S\ref{sec:schedweaksem})} \ar[u]^\subseteq \ar@{.>}@/^2pc/[uuu]^\subseteq
      &
    }
    \POS*\frm{^)}
    \endxy
    &
    \xymatrix"*"@R-1.25pc{
      \txt{non-scheduled\\structured\\semantics}\\
      \txt{non-scheduled\\path-based\\semantics\\{\small~}}\\
      \txt{scheduled\\structured\\semantics}\\
      \txt{scheduled\\path-based\\semantics}
    }
 \end{tabular}
  \caption{Semantics defined in the article.}
  \label{fig:sumsem}
\end{figure}

\subsection*{Notations}

In this article, we use the theory of {\em complete lattices}, denoting their
partial order, join, and least element respectively as $\sqsubseteq$, $\sqcup$,
and $\bot$, possibly with some subscript to indicate which lattice is
considered.
All the lattices we use are actually constructed by taking the Cartesian
product of one or several {\em powerset\/} lattices --- i.e., $\P(S)$ for some
set $S$ --- $\sqsubseteq$, $\sqcup$, and $\bot$ are then respectively the 
set inclusion $\subseteq$, the set union $\cup$, and the empty set $\emptyset$,
applied independently to each component.
Given a monotonic operator $F$ in a complete lattice, we denote by 
$\lfp F$ its least fixpoint --- i.e., $F(\lfp F)=\lfp F$ and
$\fa{X}F(X)=X\Longrightarrow \lfp F \sqsubseteq X$ --- which exists 
according to Tarski \cite{tarski-PJM55,cc-PJM79}.
We denote by $A\rightarrow B$ the set of functions from a set $A$ to a set $B$,
and by $A\stackrel{\sqcup}{\longrightarrow}B$ the set of complete 
$\sqcup-$morphisms from a complete lattice $A$ to a complete lattice $B$, i.e.,
such that
$F(\sqcup_A\,X)=\bigsqcup_B\;\setst{F(x)}{x\in X}$ for any finite 
or infinite set $X\subseteq A$. Additionally, such a function is 
monotonic.
We use the theory of Abstract Interpretation by Cousot and Cousot and, 
more precisely,
its concretization-based ($\gamma$) formalization \cite{cc-JLC92}. We
use widenings ($\widen$) to ensure termination \cite{cc-PLILP92}.
The abstract version of a domain, operator, or function is denoted with
a $\sharp$ superscript.
We use the lambda notation $\lbd{x}f(x)$ to denote functions.
If $f$ is a function, then $f[x\mapsto v]$ is the function with the
same domain as $f$ that maps $x$ to $v$, and all other elements $y\neq x$ to 
$f(y)$.
Likewise, $f[\fa{x\in X}x\mapsto g(x)]$ denotes the function that maps
any $x\in X$ to $g(x)$, and other elements $y\notin X$ to $f(y)$.
Boldface fonts are used for syntactic elements, such as ``$\sy{while}$'' in 
Fig.~\ref{fig:syntax}.
Pairs and tuples are bracketed by parentheses, as in $X=(A,B,C)$, and
can be deconstructed (matched) with the notation ``$\letin{(A,-,C)}{X}\cdots$''
where the ``$-$'' symbol denotes irrelevant tuple elements.
The notation ``$\letin{\fa{x\in X}y_x}{\cdots}\cdots$'' is used to bind a 
collection of variables $(y_x)_{x\in X}$ at once.
Semantic functions are denoted with double brackets,
as in $\lb{X}{y}$, where $y$ is an (optional) syntactic object, and
$\lbp{X}$ denotes the kind of objects ($\lbp{S}$ for statements,
$\lbp{E}$ for expressions, $\lbp{P}$ for programs, $\lbpx{\Pi}$ for
control paths).
The kind of semantics considered (parallel, non-parallel, abstract, etc.)
is denoted by subscripts and superscripts over $\lbp{X}$, as exemplified in 
Fig.~\ref{fig:sumsem}.
Finally, we use finite words over arbitrary sets, using
$\epsilon$ and $\cdot$ to denote, respectively, the empty word and word
concatenation. The concatenation $\cdot$ is naturally extended to 
sets of words: $A\cdot B\deq \setst{a\cdot b}{a\in A,\,b\in B}$.

\section{Non-parallel Programs}
\label{sec:proc}

This section recalls a classic static analysis by Abstract Interpretation 
of the run-time errors of \emph{non-parallel\/} programs, as performed for 
instance by Astrée \cite{blanchet-al-PLDI03}.
The formalization introduced here will be extended later to parallel 
programs, and it will be apparent that an analyzer for parallel 
programs can be constructed by extending an analyzer for
non-parallel programs with few changes.

\subsection{Syntax}
\label{sec:syn}

For the sake of exposition, we reason on a vastly simplified programming 
language.
However, the results extend naturally to a realistic language, such
as the subset of C excluding recursion and dynamic memory allocation
considered in our practical experiments (Sec.~\ref{sec:result}).
We assume a fixed, finite set of variable names $\V$.
A program is a single structured statement, denoted $\body\in\mi{stat}$.
The syntax of statements $\mi{stat}$ and of expressions $\mi{expr}$ is
depicted in Fig.~\ref{fig:syntax}.
Constants are actually constant intervals $[c_1,c_2]$, which return
a new arbitrary value between $c_1$ and $c_2$ every time the expression 
is evaluated.
This allows modeling non-deterministic expressions, such as inputs from
the environment, or stubs for expressions that need not be handled
precisely, e.g., $\sin(x)$ could be replaced with $[-1,1]$.
Each unary and binary operator $\diamond_\ell$ is tagged with a syntactic 
location $\ell\in\L$ and we denote by $\L$ the finite set of all syntactic 
locations.
The output of an analyzer will be the set of locations $\ell$ with errors
--- or rather, a superset of them, due to approximations.

\begin{figure}
  \centering
  $\begin{array}{l}
    \begin{array}{lcl@{\qquad}l}
      \mi{stat} & ::= & X\leftarrow\mi{expr} & \com{assignment into $X\in\V$}\\
      &|& \sy{if}\;\mi{expr}\bowtie 0\;\sy{then}\;\mi{stat} & \com{conditional}\\
      &|& \sy{while}\;\mi{expr}\bowtie 0\;\sy{do}\;\mi{stat} & \com{loop}\\
      &|& \mi{stat};\,\mi{stat} & \com{sequence}\\
      \\
      \mi{expr} & ::= & X & \com{variable $X\in\V$}\\
      &|&[c_1,c_2] & \com{constant interval, $c_1,c_2\in\R\cup\sset{\pm\infty}$}\\
      &|&-_\ell\,\mi{expr} & \com{unary operation, $\ell\in\L$}\\
      &|&\mi{expr}\diamond_\ell\mi{expr} & \com{binary operation, $\ell\in\L$}\\
      \\
      \bowtie & ::= & = | \neq | < | > | \leq | \geq\\
      \diamond & ::= & + | - | \times |\; /\\
    \end{array}
  \end{array}$
  \caption{Syntax of programs.}
  \label{fig:syntax}
\end{figure}

For the sake of simplicity, we do not handle procedures.
These are handled by inlining in our prototype.
We also focus on a single data-type (real numbers in $\R$)
and numeric expressions, which are sufficient to provide interesting properties to
express, e.g., variable bounds, although in the following we will only
discuss proving the absence of division by zero.
Handling of realistic data-types (machine integers, floats
arrays, structures, pointers, etc.) and more complex properties
(such as the absence of numeric and pointer overflow) as done in our prototype 
is orthogonal,
and existing methods apply directly --- for instance~\cite{bertrane-al-aiaa10}.

\subsection{Concrete Structured Semantics \texorpdfstring{$\lbp{P}$}{P}}
\label{sec:structsem}

As usual in Abstract Interpretation, we start by providing a concrete
semantics, that is, the most precise mathematical
expression of program semantics we consider.
It should be able to express the properties of interest to us, i.e., which
run-time errors can occur --- only divisions by zero for the simplified 
language of Fig.~\ref{fig:syntax}.
For this, it is sufficient that our concrete semantics tracks numerical 
invariants.
As this problem is undecidable, it will be abstracted in the next
section to obtain a sound static analysis.

\medskip

A program environment $\rho\in\E$ maps each variable to a value, i.e., 
$\E\deq\V\rightarrow \R$.
The semantics $\lb{E}{e}$ of an expression $e\in\mi{expr}$ takes as input
a single environment $\rho$, and outputs a set of values, in $\P(\R)$, and a
set of locations of run-time errors, in $\P(\L)$.
It is defined by structural induction in Fig.~\ref{fig:exprstructsem}.
Note that an expression can evaluate to one value, several values
(due to non-determinism in $[c_1,c_2]$) or no value at all 
(in the case of a division by zero).

\begin{figure}
  \centering
  $\begin{array}{l}
    \underline{\lb{E}{e}:\E\rightarrow(\P(\R)\times\P(\L))}
    \\[4pt]
    \lb{E}{X}\rho\deq(\set{\rho(X)},\,\emptyset)
    \\[3pt]
    \lb{E}{[c_1,c_2]}\rho\deq(\setst{c\in\R}{c_1\leq c\leq c_2},\,\emptyset)
    \\[3pt]
    \lb{E}{-_\ell\,e}\rho\deq
    \letin{(V,\Omega)}{\lb{E}{e}\rho}
    (\setst{-x}{x\in V},\,\Omega)
    \\[3pt]
    \lb{E}{e_1\diamond_\ell e_2}\rho\deq\\
    \qquad\letin{(V_1,\Omega_1)}{\lb{E}{e_1}\rho}\\
    \qquad\letin{(V_2,\Omega_2)}{\lb{E}{e_2}\rho}\\
    \qquad(\setst{x_1\diamond x_2}{x_1\in V_1,\,x_2\in V_2,\,\diamond\neq /\vee x_2\neq 0},\\
    \qquad\;\Omega_1\cup \Omega_2\cup\setst{\ell}{\diamond=/\wedge 0\in V_2})\\
    \text{where }\diamond\in\set{+,-,\times,/}
  \end{array}$
  \caption{Concrete semantics of expressions.}
  \label{fig:exprstructsem}
\end{figure}

To define the semantics of statements, we consider as semantic domain
the complete lattice:
\begin{equation}
  \label{eq:d}
  \D\deq\P(\E)\times\P(\L)
\end{equation}
with partial order $\sqsubseteq$ defined as the pairwise set inclusion:
$(A,B)\sqsubseteq(A',B')\diff A\subseteq A'\wedge B\subseteq B'$.
We denote by $\sqcup$ the associated join, i.e., pairwise set union.
The {\em structured\/} semantics $\lb{S}{s}$ of a statement $s$ is a morphism
in $\D$ that, given a set of environments $R$ and errors $\Omega$ before a 
statement $s$, returns the reachable environments after $s$, as well 
as $\Omega$ enriched with the errors encountered during the execution of $s$.
It is defined by structural induction in Fig.~\ref{fig:statstructsem}.
We introduce the new statements $e\bowtie 0?$
(where $\bowtie\,\in\set{=,\neq,<,>,\leq,\geq}$ is a comparison operator)
which we call ``guards.''
These  statements do not appear stand-alone in programs, but are
useful to  factor the semantic definition of conditionals and loops
(they are similar to the guards used in Dijkstra's Guarded Commands
\cite{dijkstra-EWD472}).
Guards will also prove useful to define control paths in Sec.~\ref{sec:pathsem}.
Guards filter their argument and keep only those environments where the
expression $e$ evaluates to a set containing a value $v$ satisfying 
$v\bowtie 0$.
The symbol $\not\bowtie$ denotes the negation of $\bowtie$, i.e.,
the negation of $=$, $\neq$, $<$, $>$, $\leq$, $\geq$ is, respectively,
$\neq$, $=$, $\geq$, $\leq$, $>$, $<$.
Finally, the semantics of loops computes a loop invariant using the 
least fixpoint operator $\lfp$.
The fact that such fixpoints exist, and the related fact that the semantic
functions are complete $\sqcup-$morphisms, i.e., 
$\lb{S}{s}(\sqcup_{i\in I} X_i)=\sqcup_{i\in I}\;{\lb{S}{s}X_i}$, is stated
in the following theorem:
\begin{thm}
  \label{thm:morphism}
  $\fa{s\in\mi{stat}}\lb{S}{s}$ is well defined and a complete $\sqcup-$morphism.
\end{thm}
\proof In Appendix~\ref{proof:morphism}.\qed

\begin{figure}
  \centering
  $\begin{array}{l}
    \underline{\lb{S}{s}:\D\stackrel{\sqcup}{\longrightarrow}\D}
    \\[4pt]
    \lb{S}{X\leftarrow e}(R,\Omega)\deq
    (\emptyset,\Omega)\;\sqcup\;
    \bigsqcup_{\rho\in R}\;
    \letin{(V,\Omega')}{\lb{E}{e}\rho}
    (\setst{\rho[X\mapsto v]}{v\in V},\,\Omega')
    \\[3pt]
    \lb{S}{e\bowtie 0?}(R,\Omega)\deq
    (\emptyset,\Omega)\;\sqcup\;
    \bigsqcup_{\rho\in R}\;
    \letin{(V,\Omega')}{\lb{E}{e}\rho}
    (\setst{\rho}{\ex{v\in V}v\bowtie 0},\,\Omega')
    \\[3pt]
    \lb{S}{s_1;\,s_2}(R,\Omega)\deq (\lb{S}{s_2}\circ\lb{S}{s_1})(R,\Omega)
    \\[3pt]
    \lb{S}{\sy{if}\;e\bowtie 0\;\sy{then}\;s}(R,\Omega)\deq
    (\lb{S}{s}\circ\lb{S}{e\bowtie 0?})(R,\Omega)\sqcup\lb{S}{e\not\bowtie 0?}(R,\Omega)
    \\[3pt]
    \lb{S}{\sy{while}\;e\bowtie 0\;\sy{do}\;s}(R,\Omega)\deq
    \lb{S}{e\not\bowtie 0?}(\lfp\lbd{X}(R,\Omega)\sqcup(\lb{S}{s}\circ\lb{S}{e\bowtie 0?})X)
    \\[5pt]
    \text{where }\bowtie\,\in\sset{=,\neq,<,>,\leq,\geq}\\
  \end{array}$
  \caption{Structured concrete semantics of statements.}
  \label{fig:statstructsem}
\end{figure}

\noindent
We can now define the concrete structured semantics of the program as follows:
\begin{equation}
  \label{eq:progstructsem}
  \lbp{P}\deq\Omega,\text{ where }(-,\Omega)=\lb{S}{\body}(\E_0,\emptyset)
\end{equation}
where $\E_0\subseteq \E$ is a set of initial environments.
We can choose, for instance, $\E_0=\E$ or 
$\E_0\deq\set{\lbd{X\in\V}{0}}$.
Note that all run-time errors are collected while traversing
the program structure; they are never discarded and all of them eventually 
reach the end of $\body$, and so, appear in $\lbp{P}$, even if
$\lb{S}{\body}(\E_0,\emptyset)$ outputs an empty set of environments.
Our program semantics thus observes the set of run-time errors that can appear 
in any execution starting at the beginning of $\body$ in an initial 
environment.
This includes errors occurring in executions that
loop forever (such as infinite reactive loops in control~/ command 
software) or that halt before the end of $\body$.

\subsection{Abstract Structured Semantics \texorpdfstring{$\lbp{P^\s}$}{P\#}}
\label{sec:procabs}

The semantics $\lbp{P}$ is not computable as it involves least fixpoints
in an infinite-height domain $\D$, and not all elements in $\D$ are 
representable in a computer as $\D$ is uncountable.
Even if we restricted variable values to a more realistic, large but
finite, subset --- such as machine integers or floats --- naive computation 
in $\D$ would be unpractical.
An effective analysis will instead compute an abstract semantics 
over-approximating the concrete one.

\medskip

The abstract semantics is parametrized by the choice of an abstract domain of 
environments obeying the signature presented in Fig.~\ref{fig:absdomain}.
It comprises a set $\E^\s$ of computer-representable abstract environments, 
with a 
partial order $\subseteq^\s_\E$ (denoting abstract entailment)
and an abstract environment $\E^\s_0\in\E^\s$ representing initial environments.
Each abstract environment represents a set of concrete environments 
through a monotonic concretization function $\gamma_\E:\E^\s\rightarrow\P(\E)$.
We also require an effective abstract version $\cup^\s_\E$ of the set
union $\cup$, as well as effective abstract versions $\lb{S^\s}{s}$ of the
semantic operators $\lb{S}{s}$ for assignment and guard statements.
Only environment sets are abstracted, while error sets are represented 
explicitly, so that
the actual abstract semantic domain for $\lb{S^\s}{s}$ is 
$\D^\s\deq\E^\s\times\P(\L)$, with
concretization $\gamma$ defined in Fig.~\ref{fig:absdomain}.
Figure~\ref{fig:absdomain} also presents the soundness conditions that
state that an abstract operator outputs a superset of the environments 
and error locations returned by its concrete version.
Finally, when $\E^\s$ has infinite strictly increasing chains, we require
a widening operator $\widen_\E$, i.e., a sound abstraction of the join
$\cup$ with a termination guarantee to ensure the convergence of abstract 
fixpoint computations in finite time.
There exist many abstract domains $\E^\s$, for instance the interval domain
\cite{cc-POPL77}, where an abstract environment in $\E^\s$ associates an interval to each
variable, the octagon domain \cite{mine-HOSC06}, where an abstract environment
in $\E^\s$ is a conjunction of constraints of the form
$\pm X\pm Y\leq c$ with $X,Y\in\V$, $c\in\R$, or the polyhedra domain
\cite{ch:popl78}, where an abstract environment in $\E^\s$ is a convex, closed
(possibly unbounded) polyhedron.

\begin{figure}
  \centering
  $\begin{array}{ll}
    \E^\s & \com{set of abstract environments}
    \\[3pt]
    \gamma_\E:\E^\s\rightarrow\P(\E) & \com{concretization}
    \\[3pt]
    \bot^\s_\E\in\E^\s & \com{empty abstract environment}\\
    \qquad\text{ s.t. }\gamma_\E(\bot^\s_\E)=\emptyset 
    \\[3pt]
    \E^\s_0\in\E^\s & \com{initial abstract environment}\\
    \qquad\text{ s.t. }\gamma_\E(\E^\s_0)\supseteq\E_0 
    \\[3pt]
    \subseteq^\s_\E:(\E^\s\times\E^\s)\rightarrow\set{\myfun{true},\myfun{false}}
    & \com{abstract entailment}\\
    \qquad \text{s.t. }X^\s\subseteq^\s_\E Y^\s\Longrightarrow\gamma_\E(X^\s)\subseteq\gamma_\E(Y^\s)
    \\[3pt]
    \cup^\s_\E:(\E^\s\times\E^\s)\rightarrow\E^\s
    & \com{abstract join}\\
    \qquad \text{s.t. }\gamma_\E(X^\s\cup^\s_\E Y^\s)\supseteq\gamma_\E(X^\s)\cup\gamma_\E(Y^\s)
    \\[3pt]
    \widen_\E:(\E^\s\times\E^\s)\rightarrow\E^\s
    & \com{widening}\\
    \qquad \text{s.t. }\gamma_\E(X^\s\widen_\E Y^\s)\supseteq\gamma_\E(X^\s)\cup\gamma_\E(Y^\s)\\
    \qquad \text{and }\fa{(Y^\s_i)_{i\in\N}}\text{ the sequence }X^\s_0=Y^\s_0,\,
    X^\s_{i+1}=X^\s_i\widen_\E Y^\s_{i+1}\hspace*{-1.5cm}\\
    \qquad\text{reaches a fixpoint }X^\s_k=X^\s_{k+1}\text{ for some }k\in\N
    \\[5pt]
    \D^\s\deq\E^\s\times\P(\L) & \com{abstraction of $\D$}
    \\[3pt]
    \gamma:\D^\s\rightarrow\D & \com{concretization for $\D^\s$} \\[-2pt]
    \qquad\text{s.t. }\gamma(R^\s,\Omega)\deq(\gamma_\E(R^\s),\Omega)
    \\[3pt]
    \lb{S^\s}{s}:\D^\s\rightarrow\D^\s\\
    \multicolumn{2}{l}{
    \qquad\text{s.t. }\fa{s\in\set{X\leftarrow e,\,e\bowtie 0?}}
    (\lb{S}{s}\circ\gamma)(R^\s,\Omega)\sqsubseteq (\gamma\circ\lb{S^\s}{s})(R^\s,\Omega)}\\
  \end{array}$
  \caption{Abstract domain signature, and soundness and termination conditions.}
  \label{fig:absdomain}
\end{figure}

In the following, we will refer to assignments and guards collectively as
{\em primitive statements}.
Their abstract semantics $\lb{S^\s}{s}$ in $\D^\s$ depends on the choice of 
abstract domain;
we assume it is provided as part of the abstract domain definition and do
not discuss it.
By contrast, the semantics of non-primitive statements can be derived in
a generic way, as presented in Fig.~\ref{fig:absstatstructsem}.
Note the similarity between these definitions and the concrete semantics 
of Fig.~\ref{fig:statstructsem}, except for the semantics of loops that
uses additionally a widening operator $\widen$ derived from $\widen_\E$.
The termination guarantee of the widening ensures that, given any
(not necessarily monotonic) function $F^\s:\D^\s\rightarrow\D^\s$, the sequence
$X^\s_0\deq(\bot^\s_\E,\emptyset)$, $X^\s_{i+1}\deq X^\s_i\widen F^\s(X^\s_i)$
reaches a fixpoint $X^\s_k=X^\s_{k+1}$ in finite time $k\in\N$.
We denote this limit by $\lim \lbd{X^\s}X^\s\widen F^\s(X^\s)$.
Note that, due to widening, the semantics of a loop is generally not a join
morphism, and even not monotonic \cite{cc-PLILP92}, even if the semantics 
of the loop body is.
Hence, there would be little benefit in imposing that the semantics of 
primitive statements provided with $\D^\s$ is monotonic, and we do not impose 
it in Fig.~\ref{fig:absdomain}.
Note also that $\lim F^\s$ may not be the least fixpoint of $F^\s$
(in fact, such a least fixpoint may not even exist).

\begin{figure}
  \centering
  $\begin{array}{l}
    \underline{\lb{S^\s}{s}:\D^\s\rightarrow\D^\s}
    \\[4pt]
    \lb{S^\s}{s_1;\,s_2} (R^\s,\Omega)\deq(\lb{S^\s}{s_2}\circ\lb{S^\s}{s_1})(R^\s,\Omega)
    \\[3pt]
    \lb{S^\s}{\sy{if}\;e\bowtie 0\;\sy{then}\;s} (R^\s,\Omega)\deq\\
    \qquad
    (\lb{S^\s}{s}\circ\lb{S^\s}{e\bowtie0?})(R^\s,\Omega)\;\cup^\s\;\lb{S^\s}{e\not\bowtie0?}(R^\s,\Omega)
    \\[3pt]
    \lb{S^\s}{\sy{while}\;e\bowtie0\;\sy{do}\;s} (R^\s,\Omega)\deq\\
    \qquad
    \lb{S^\s}{e\not\bowtie0?}(\lim\lbd{X^\s}X^\s\widen((R^\s,\Omega)\;\cup^\s\;(\lb{S^\s}{s}\circ\lb{S^\s}{e\bowtie0?})X^\s))
    \\[3pt]
    \text{where:}\\
    \quad(R^\s_1,\Omega_1)\,\cup^\s\,(R^\s_2,\Omega_2)\deq(R^\s_1\cup^\s_\E R^\s_2,\;\Omega_1\cup\Omega_2)\\
    \quad(R^\s_1,\Omega_1)\widen(R^\s_2,\Omega_2)\deq(R^\s_1\widen_\E R^\s_2,\;\Omega_1\cup\Omega_2)
  \end{array}$
  \caption{Derived abstract functions for non-primitive statements.}
  \label{fig:absstatstructsem}
\end{figure}

The abstract semantics of a program can then be defined, similarly to 
(\ref{eq:progstructsem}), as:
\begin{equation*}
  \lbp{P^\s} \deq \Omega,\text{ where }(-,\Omega)=\lb{S^\s}{\body}(\E^\s_0,\emptyset)\enspace.
\end{equation*}
The following theorem states the soundness of the abstract semantics:
\begin{thm}
  \label{thm:sound}
  $\lbp{P}\subseteq\lbp{P^\s}.$
\end{thm}
\proof In Appendix~\ref{proof:sound}.\qed

The resulting analysis is flow-sensitive.
It is relational whenever $\E^\s$ is --- e.g., with octagons \cite{mine-HOSC06}.
The iterator follows, in the terminology of \cite{bourdoncle-FMPA93}, a
recursive iteration strategy.
The advantage of this strategy is its efficient use of memory:
few abstract elements need to be kept in memory during the analysis.
Indeed, apart from the current abstract environment, a clever implementation
of Fig.~\ref{fig:absstatstructsem} exploiting tail recursion would only need
to keep one extra environment per $\sy{if}\;e\bowtie 0\;\sy{then}\;s$
statement --- to remember the $(R^\s,\Omega)$ argument while evaluating $s$ ---
and two environments per $\sy{while}\;e\bowtie0\;\sy{do}\;s$ statement
--- one for $(R^\s,\Omega)$ and one for the accumulator $X^\s$ ---
in the call stack of the abstract interpreter function $\lbp{S^\s}$.
Thus, the maximum memory consumption is a function of the maximum 
nesting of conditionals and loops in the analyzed program, which is generally
low.
This efficiency is key to analyze large programs, as demonstrated by Astrée 
\cite{blanchet-al-PLDI03}.

\subsection{Concrete Path-Based Semantics \texorpdfstring{$\lbp{P_\pi}$}{P-Pi}}
\label{sec:pathsem}

The structured semantics of Sec.~\ref{sec:structsem} is defined
as an interpretation of the program by induction on its
syntactic structure, which can be conveniently transformed into
a static analyzer, as shown in Sec.~\ref{sec:procabs}.
Unfortunately, the execution of a parallel program does not follow 
such a simple syntactic structure; it is rather defined as an interleaving of 
control paths from distinct threads (Sec.~\ref{sec:interleavesem}).
Before considering parallel programs, we start by proposing in this section
an alternate concrete semantics of non-parallel programs
based on control paths.
While its definition is different from the structured semantics of
Sec.~\ref{sec:structsem}, its output is equivalent.

\medskip

A {\em control path} $p$ is any finite sequence of primitive statements, among
$X\leftarrow e$, $e\bowtie0?$.
We denote by $\Pi$ the set of all control paths.
Given a statement $s$, the set of control paths it spawns $\pi(s)\subseteq\Pi$
is defined by structural induction as follows:
\begin{equation}
  \label{eq:path}
  \begin{array}{l}
    \pi(X\leftarrow e)\deq\set{X\leftarrow e}
    \\
    \pi(s_1;\,s_2)\deq \pi(s_1)\cdot\pi(s_2)
    \\
    \pi(\sy{if}\;e\bowtie0\;\sy{then}\;s)\deq
    (\set{e\bowtie0?}\cdot\pi(s))\cup\set{e\not\bowtie0?}
    \\
    \pi(\sy{while}\;e\bowtie0\;\sy{do}\;s)\deq
    (\lfp\lbd{X}\sset{\epsilon}\cup(X\cdot\set{e\bowtie0?}\cdot\pi(s)))\cdot\set{e\not\bowtie0?}
  \end{array}
\end{equation}
where $\epsilon$ denotes then empty path, and $\cdot$ denotes path 
concatenation, naturally extended to sets of paths.
A primitive statement spawns a singleton path of length one, while a
conditional spawns two sets of paths --- one set where the 
$\sy{then}$ branch is taken, and one where it is not taken --- and loops
spawn an infinite number of paths --- corresponding to all possible
unrollings.
Although $\pi(s)$ is infinite whenever $s$ contains a loop,
it is possible that many control paths in $\pi(s)$ are actually infeasible, 
i.e., have no corresponding execution. In particular, even if a loop $s$
is always bounded, $\pi(s)$ contains unrollings of arbitrary length.

We can now define the semantics 
$\lbx{\Pi}{P}\in\D\stackrel{\sqcup}{\longrightarrow}\D$ of a set of paths 
$P\subseteq\Pi$ as follows, reusing the semantics of primitive statements
from Fig.~\ref{fig:statstructsem} and the pairwise join $\sqcup$ on 
sets of environments and errors:
\begin{equation}
  \label{eq:pathsem}
  \lbx{\Pi}{P}(R,\Omega)\deq\bigsqcup\;\setst{(\lb{S}{s_n}\circ\cdots\circ\lb{S}{s_1})(R,\Omega)}{s_1\cdot\ldots\cdot s_n\in P}\enspace.
\end{equation}
The path-based semantics of a program is then:
\begin{equation}
  \label{eq:progpathsem}
  \lbp{P_\pi}\deq\Omega,\text{ where }
  (-,\Omega)=\lbx{\Pi}{\pi(\body)}(\E_0,\emptyset)\enspace.
\end{equation}
Note that this semantics is similar to the standard \emph{meet over all paths\/} solution\footnote{The lattices used in data-flow analysis and in abstract interpretation are dual: the former use a meet to join paths --- hence the expression ``meet over all paths'' --- while we employ a join $\sqcup$. Likewise, the greatest fixpoint solution of a data-flow analysis corresponds to our least fixpoint.} of
data-flow problems --- see, e.g., \cite[\S~2]{nielson-al} --- but 
for concrete executions 
in the infinite-height lattice $\D$.
The meet over all paths and maximum fixpoint solutions of data-flow
problems are equal for distributive frameworks; 
similarly, our structured and path-based concrete semantics 
(based on complete $\sqcup-$morphisms) are equal:
\begin{thm}
  \label{thm:structpath}
  $\fa{s\in\mi{stat}}\lbx{\Pi}{\pi(s)}=\lb{S}{s}.$
\end{thm}
\proof In Appendix~\ref{proof:structpath}.\qed
\noindent
An immediate consequence of this theorem is that $\lbp{P}=\lbp{P_\pi}$,
hence the two semantics compute, in different ways, the same set of errors.

\section{Parallel Programs in a Shared Memory}
\label{sec:shared}

In this section, we consider several threads that communicate through
a shared memory, without any synchronization primitive yet --- they will
be introduced in Sec.~\ref{sec:sched}.
We also discuss here the memory consistency model, and its effect
on the semantics and the static analysis.

A program has now a fixed, finite set $\T$ of threads.
To each thread $t\in\T$ is associated a statement body $\body_t\in\mi{stat}$.
All the variables in $\V$ are shared and can be accessed by all threads.

\subsection{Concrete Interleaving Semantics \texorpdfstring{$\lbp{P_*}$}{P*}}
\label{sec:interleavesem}

The simplest and most natural model of parallel program execution
considers all possible interleavings of control paths from all threads.
These correspond to \emph{sequentially consistent executions\/}, as coined
by Lamport \cite{lamport-TC79}.

\medskip

A {\em parallel control path} $p$ is a finite sequence of pairs $(s,t)$, where
$s$ is a primitive statement (assignment or guard)
and $t\in\T$ is a thread that executes it.
We denote by $\Pi_*$ the set of all parallel control paths.
The  semantics $\lbx{\Pi_*}{P}\in\D\stackrel{\sqcup}{\longrightarrow}\D$ 
of a set of parallel control paths $P\subseteq\Pi_*$ is
defined as in the case of regular control paths (\ref{eq:pathsem}), ignoring
thread identifiers:
\begin{equation}
  \label{eq:interleavepathsem}
  \lbx{\Pi_*}{P}(R,\Omega)\deq
  \bigsqcup\;\setst{(\lb{S}{s_n}\circ\cdots\circ\lb{S}{s_1})(R,\Omega)}{(s_1,-)\cdot\ldots\cdot(s_n,-)\in P}\enspace.
\end{equation}

We now denote by $\pi_*\subseteq\Pi_*$ the set of parallel control paths
spawned by the whole program.
It is defined as:
\begin{equation}
  \label{eq:interleave}
  \pi_*\deq\setst{p\in\Pi_*}{\fa{t\in\T}\proj_t(p)\in\pi(\body_t)}
\end{equation}
where the set $\pi(\body_t)$ of regular control paths of
a thread is defined in (\ref{eq:path}), and $\proj_t(p)$ extracts
the maximal sub-path of $p$ on thread $t$ as follows:
\begin{equation*}
  \begin{array}{l}
    \proj_t((s_1,t_1)\cdot\ldots\cdot(s_n,t_n))\deq
    s_{i_1}\cdot\ldots\cdot s_{i_m}\\
    \text{such that }\fa{j}\begin{array}[t]{@{}l}
      1\leq i_j<i_{j+1}\leq n\;\wedge\;t_{i_j}=t\;\wedge\\
      \fa{k}(k<i_1\vee k>i_m\vee i_j<k<i_{j+1})\Longrightarrow t_k\neq t
      \enspace.
    \end{array}
  \end{array}
\end{equation*}

The semantics $\lbp{P_*}$ of a parallel program becomes,
similarly to (\ref{eq:progpathsem}):
\begin{equation}
  \label{eq:interleavesem}
  \lbp{P_*}\deq\Omega,\text{ where }
  (-,\Omega)=\lbx{\Pi_*}{\pi_*}(\E_0,\emptyset)
\end{equation}
i.e., we collect the errors that can appear in any interleaved execution
of all the threads, starting in an initial environment.

Because we interleave primitive statements, a thread can only interrupt
another one between two primitive statements, and not in the middle of 
a primitive statement.
For instance, in a statement such as $X\leftarrow Y+Y$, no thread can
interrupt the current one and change the value of $Y$ between the evaluation
of the first and the second $Y$ sub-expression, while it can if the
assignment is split into $X\leftarrow Y; X\leftarrow X+Y$.
Primitive statements are thus {\em atomic} in $\lbp{P_*}$.
By contrast, we will present a semantics where primitive statements are not
atomic in Sec.~\ref{sec:weaksem}.

\subsection{Concrete Interference Semantics \texorpdfstring{$\lbp{P_\I}$}{P-I}}
\label{sec:interfersem}

Because it reasons on infinite sets of paths, the concrete interleaving
semantics from the previous section is not easily amenable to abstraction.
In particular, replacing the concrete domain $\D$ in $\lbp{P_*}$
with an abstract one $\D^\s$ (as defined in Sec.~\ref{sec:procabs})
is not sufficient to obtain an effective and efficient static analyzer as 
we still have a large or infinite number of paths to analyze separately and 
join.
By contrast, we propose here a (more abstract) concrete semantics that can be 
expressed by induction on the syntax. It will lead naturally, after
further abstraction in Sec.~\ref{sec:sharedabs}, to an effective static
analysis.

\subsubsection{Thread semantics}
We start by enriching the non-parallel structured semantics of 
Sec.~\ref{sec:structsem} with a notion of interference.
We call \emph{interference\/} a triple $(t,X,v)\in\I$, where
$\I\deq\T\times\V\times\R$, indicating that the thread $t$ can set the variable 
$X$ to the value $v$. However, it does not say at which program point the
assignment is performed, so, it is a flow-insensitive information.

The new semantics of an expression $e$, denoted $\lb{E_\I}{e}$, takes
as argument the current thread $t\in\T$ and an interference set 
$I\subseteq\I$ in addition to an environment $\rho\in\E$.
It is defined in Fig.~\ref{fig:interfexprsem}.
The main change with respect to the interference-free semantics 
$\lb{E}{e}$ of
Fig.~\ref{fig:exprstructsem} is that, when fetching a variable 
$X\in\V$, each interference $(t',v,X)\in I$ on $X$ from any other thread
$t'\neq t$ is applied.
The semantics of constants and operators is not changed, apart from 
propagating $t$ and $I$ recursively.
Note that the choice of evaluating $\lb{E_\I}{X}(t,\rho,I)$ to $\rho(X)$ or
to some interference in $I$, as well as the choice of the interference in 
$I$, is non-deterministic.
Thus, distinct occurrences of the same variable in an expression may 
evaluate, in the same environment, to different values.

\begin{figure}
  \centering
  $\begin{array}{l}
    \underline{\lb{E_\I}{e}:(\T\times\E\times\P(\I))\rightarrow(\P(\R)\times\P(\L))}
    \\[4pt]
    \lb{E_\I}{X}(t,\rho,I)\deq(\set{\rho(X)}\cup\setst{v}{\ex{t'\neq t}(t',X,v)\in I},\,\emptyset)
    \\[3pt]
    \lb{E_\I}{[c_1,c_2]}(t,\rho,I)\deq(\setst{c\in\R}{c_1\leq c\leq c_2},\,\emptyset)
    \\[3pt]
    \lb{E_\I}{-_\ell\,e}(t,\rho,I)\deq
    \letin{(V,\Omega)}{\lb{E_\I}{e}(t,\rho,I)}
    (\setst{-x}{x\in V},\,\Omega)
    \\[3pt]
    \lb{E_\I}{e_1\diamond_\ell e_2}(t,\rho,I)\deq\\
    \qquad\letin{(V_1,\Omega_1)}{\lb{E_\I}{e_1}(t,\rho,I)}\\
    \qquad\letin{(V_2,\Omega_2)}{\lb{E_\I}{e_2}(t,\rho,I)}\\
    \qquad(\setst{x_1\diamond x_2}{x_1\in V_1,\,x_2\in V_2,\,\diamond\neq /\vee x_2\neq 0},\\
    \qquad\;\Omega_1\cup \Omega_2\cup\setst{\ell}{\diamond=/\wedge 0\in V_2})\\
    \text{where }\diamond\in\set{+,-,\times,/}
  \end{array}$
  \caption{Concrete semantics of expressions with interference.}
  \label{fig:interfexprsem}
\end{figure}

The semantics of a statement $s$ executed by a thread $t\in\T$ 
is denoted $\lb{S_\I}{s,t}$. It is presented in Fig.~\ref{fig:interfstatsem}.
This semantics is enriched with interferences and is thus
a complete $\sqcup_\I-$morphism in the complete lattice:
\begin{equation*}
  \D_\I\deq\P(\E)\times\P(\L)\times\P(\I)
\end{equation*}
where the join $\sqcup_\I$ is the pairwise set union.
The main point of note is the semantics of assignments $X\leftarrow e$.
It both uses its interference set argument, passing it to 
$\lb{E_\I}{e}$, and enriches it with the interferences generated on the 
assigned variable $X$.
The semantics of guards simply uses the interference set, while the
semantics of conditionals, loops, and sequences is
identical to the non-interference one from Fig.~\ref{fig:statstructsem}.
The structured semantics of a thread $t$ with interferences $I$ is then
$\lb{S_\I}{\body_t,\,t}(\E_0,\emptyset,I)$.

\begin{figure}
  \centering
  $\begin{array}{l}
    \underline{\lb{S_\I}{s,t}:\D_\I\stackrel{\sqcup_\I}{\longrightarrow}\D_\I}
    \\[4pt]
    \lb{S_\I}{X\leftarrow e,\,t}(R,\Omega,I)\deq\\
    \qquad(\emptyset,\Omega,I)\;\sqcup_\I\;\underset{\rho\in R}{\bigsqcup_\I}\;
    \begin{array}[t]{l}
      \letin{(V,\Omega')}{\lb{E_\I}{e}(t,\rho,I)}\\
      (\setst{\rho[X\mapsto v]}{v\in V},\,\Omega',\,
      \setst{(t,X,v)}{v\in V})\\[3pt]
    \end{array}
    \\
    \lb{S_\I}{e\bowtie 0?,\,t}(R,\Omega,I)\deq\\
    \qquad(\emptyset,\Omega,I)\;\sqcup_\I\;\underset{\rho\in R}{\bigsqcup_\I}\;
    \begin{array}[t]{l}
      \letin{(V,\Omega')}{\lb{E_\I}{e}(t,\rho,I)}\\
      (\setst{\rho}{\ex{v\in V}v\bowtie 0},\,\Omega',\emptyset)\\[3pt]
    \end{array}
    \\
    \lb{S_\I}{\sy{if}\;e\bowtie0\;\sy{then}\;s,\,t}(R,\Omega,I)\deq\\
    \qquad
    (\lb{S_\I}{s,t}\circ\lb{S_\I}{e\bowtie 0?,\,t})(R,\Omega,I)\sqcup_\I\lb{S_\I}{e\not\bowtie0?,\,t}(R,\Omega,I)
    \\[3pt]
    \lb{S_\I}{\sy{while}\;e\bowtie0\;\sy{do}\;s,\,t}(R,\Omega,I)\deq\\
    \qquad
    \lb{S_\I}{e\not\bowtie0?,\,t}(\lfp\lbd{X}(R,\Omega,I)\sqcup_\I(\lb{S_\I}{s,t}\circ\lb{S_\I}{e\bowtie0?,\,t})X)
    \\[3pt]
    \lb{S_\I}{s_1;\,s_2,\,t}(R,\Omega,I)\deq (\lb{S_\I}{s_2,t}\circ\lb{S_\I}{s_1,t})(R,\Omega,I)
  \end{array}$
  \caption{Concrete semantics of statements with interference.}
  \label{fig:interfstatsem}
\end{figure}

\subsubsection{Program semantics}
The semantics $\lb{S_\I}{\body_t,\,t}$ still only analyzes the 
effect of a single thread $t$.
It assumes {\it a priori\/} knowledge of the other threads, through 
$I$, and contributes to this knowledge, by enriching $I$.
To solve this dependency and take into account multiple threads, we iterate 
the analysis of all threads until interferences stabilize.
Thus, the semantics of a multi-threaded program is:
\begin{equation}
  \label{eq:intersem}
  \begin{array}{l}
    \lbp{P_\I}\deq\Omega,\text{ where }
    (\Omega,-)\deq\\
    \quad
    \lfp\lbd{(\Omega,I)}
    \bigsqcup_{t\in\T}\,
      \letin{(-,\Omega',I')}{\lb{S_\I}{\body_t,\,t}(\E_0,\Omega,I)}
      (\Omega',I')
  \end{array}
\end{equation}
where the join $\sqcup$ is the componentwise set union in the complete 
lattice $\P(\L)\times\P(\I)$.

\subsubsection{Soundness and completeness}
Before linking our interference semantics $\lbp{P_\I}$ (by structural induction)
to the interleaving semantics $\lbp{P_*}$ of Sec.~\ref{sec:interleavesem}
(which is path-based), we remark that
we can restate the structured interference semantics $\lb{S_\I}{\body_t,\,t}$
of a single thread $t$ in a path-based form, as we did
in Sec.~\ref{sec:pathsem} for non-parallel programs.
Indeed, we can replace $\lbp{S}$ with $\lbp{S_\I}$ in
(\ref{eq:pathsem}) and derive a path-based semantics with interference
$\lbx{\Pi_\I}{P,\,t}\in\D_\I\stackrel{\sqcup_\I}{\longrightarrow}\D_\I$ 
of a set of (non-parallel) control paths $P\subseteq\Pi$
in a thread $t$ as follows:
\begin{equation}
  \label{eq:interpathsem}
  \textstyle
  \lbx{\Pi_\I}{P,t}(R,\Omega,I)\deq\bigsqcup_\I\;\setst{(\lb{S_\I}{s_n,t}\circ\cdots\circ\lb{S_\I}{s_1,t})(R,\Omega,I)}{s_1\cdot\ldots\cdot s_n\in P}\enspace.
\end{equation}
These two forms are equivalent, and Thm.~\ref{thm:structpath} naturally becomes:
\begin{thm}
  \label{thm:structpath2}
  $\fa{t\in\T,\,s\in\mi{stat}}\lbx{\Pi_\I}{\pi(s),\,t}=\lb{S_\I}{s,\,t}.$
\end{thm}
\proof In Appendix~\ref{proof:structpath2}.\qed

The following theorem then states that the semantics $\lbp{P_\I}$ computed
with an interference fixpoint is indeed sound with respect to 
the interleaving semantics
$\lbp{P_*}$ that interleaves paths from all threads:
\begin{thm}
  \label{thm:interf}
  $\lbp{P_*}\subseteq\lbp{P_\I}$.
\end{thm}
\proof In Appendix~\ref{proof:interf}.\qed
The equality does not hold in general.
Consider, for instance, the program fragment in Fig.~\ref{fig:weakex1} inspired
from Dekker's mutual exclusion algorithm \cite{dijkstra-EWD123}.
According to the interleaving semantics, both threads can never be in their 
critical section simultaneously.
The interference semantics, however, does not ensure mutual exclusion.
Indeed, it computes the following set of
interferences: $\set{(t_1,\mathrm{flag1},1),(t_2,\mathrm{flag2},1)}$.
Thus, in thread $t_1$, $\mathrm{flag2}$ evaluates to $\sset{0,1}$.
The value 0 comes from the initial state 
$\E_0$ and the value 1 comes from the interference
$(t_2,\mathrm{flag2},1)$.
Likewise, $\mathrm{flag1}$ evaluates to $\sset{0,1}$ in thread $t_2$.
Thus, both conditions $\mathrm{flag1}=0$ and $\mathrm{flag2}=0$ can be 
simultaneously true.
This imprecision is due to the flow-insensitive treatment of interferences.
We now present a second example of incompleteness where the loss of
precision is amplified by the interference fixpoint.
Consider the program in Fig.~\ref{fig:weakex2} where two threads increment 
the same zero-initialized variable $x$.
According to the interleaving semantics, either the value 1 or 2 is
stored into $y$.
However, in the interference semantics, the interference fixpoint builds a 
growing set of interferences, up to 
$\setst{(t,x,i)}{t\in\{t_1,t_2\},\,i\geq 1}$, 
as each thread increments the possible values written by the other thread.
Note that the program features no loop and $x$ can thus be incremented 
only finitely many times (twice), but the interference abstraction is 
flow-insensitive and forgets how many times an action can be performed.
As a consequence, any positive value can be stored into $y$, instead of
only $1$ or $2$.

\begin{figure}[t]
  \centering
  \subfigure[Mutual exclusion algorithm.]{
    \begin{tabular}{l@{\quad}|@{\quad}l}
      \multicolumn{2}{c}{$\E_0:\mathrm{flag1}=\mathrm{flag2}=0$}\\[4pt]
      \quad \underline{\rm thread $t_1$} & 
      \quad \underline{\rm thread $t_2$}\\[2pt]
      $\mathrm{flag1} \leftarrow 1;$ & $\mathrm{flag2} \leftarrow 1;$\\
      $\sy{if}\;\mathrm{flag2}=0\;\sy{then}$ & $\sy{if}\;\mathrm{flag1}=0\;\sy{then}$\\
      \quad{\rm\em critical section} & \quad{\rm\em critical section}\\[5pt]
    \end{tabular}
    \label{fig:weakex1}
  }
  \subfigure[Parallel incrementation.]{
    \begin{tabular}{l@{\quad}|@{\quad}l}
      \multicolumn{2}{c}{$\E_0:x=y=0$}\\[4pt]
      \quad \underline{\rm thread $t_1$} & 
      \quad \underline{\rm thread $t_2$}\\[2pt]
      $x \leftarrow x + 1;$ & $x \leftarrow x + 1$\\
      $y \leftarrow x$&\\
      \multicolumn{2}{c}{}\\[5pt]
    \end{tabular}
    \label{fig:weakex2}
  }
  \caption{Incompleteness examples for the concrete interference semantics.}
\end{figure}

Our interference semantics is based on a decomposition of the
invariant properties of parallel programs into a local invariant at each 
thread program point and a global interference invariant.
This idea is not new, and complete methods to do so have already
been proposed.
Such methods date back to the works of Owicki, Gries, and Lamport
\cite{owicki-gries-AI76,lamport-TSE77,lamport-AI80} and have been formalized
in the framework of Abstract Interpretation by Cousot and Cousot
\cite{cc-APCT84}.
We would say informally that our interference semantics is an incomplete 
abstraction of such complete methods, where interferences are abstracted in
a flow-insensitive and non-relational way.
Our choice to abstract away these information is a deliberate move 
that eases considerably the construction of an
effective and efficient static analyzer, as shown in Sec.~\ref{sec:sharedabs}.
Another strong incentive is that the interference semantics is compatible
with the use of weakly consistent memory models, as shown in
Sec.~\ref{sec:weaksem}.
Note finally that Sec.~\ref{sec:schedinterfersem} will present a method to
recover a 
weak form of flow-sensitivity (i.e., mutual exclusion)
on interferences, without loosing the
efficiency nor the correctness with respect to weak memory models.

\subsection{Abstract Interference Semantics \texorpdfstring{$\lbp{P_\I^\s}$}{P\#-I}}
\label{sec:sharedabs}

The concrete interference semantics $\lbp{P_\I}$  introduced in the previous
section is defined by structural induction.
It can thus be easily abstracted to provide an effective, always-terminating,
and sound static analysis.

\medskip

We assume, as in Sec.~\ref{sec:procabs}, the existence of an abstract
domain $\E^\s$ abstracting sets of environments --- see
Fig.~\ref{fig:absdomain}.
Additionally, we assume the existence of an abstract domain
$\RR^\s$ that abstracts sets of reals, which will be useful to abstract
interferences.
Its signature is presented in Fig.~\ref{fig:absrealdomain}.
It is equipped with a concretization $\gamma_\RR:\RR^\s\rightarrow\P(\R)$, 
a least element $\bot^\s_\RR$, an abstract join $\cup^\s_\RR$ and,
if it has strictly increasing infinite chains, a widening
$\widen_\RR$.
We also require two additional functions that will be necessary to
communicate information between $\E^\s$ and $\RR^\s$.
Firstly, a function $\get(X,R^\s)$ that extracts from
an abstract environment $R^\s\in\E^\s$ the set of values a variable
$X\in\V$ can take, and abstracts this set in $\RR^\s$.
Secondly, a function $\asexpr(V^\s)$ able to synthesize a 
(constant) expression approximating any non-empty abstract value 
$V^\s\in\RR^\s\setminus\sset{\bot^\s_\RR}$.
This provides a simple way to use an abstract value from
$\RR^\s$ in functions on abstract environments in $\E^\s$.
For instance, $\lb{S^\s}{X\leftarrow\asexpr(V^\s)}(R^\s,\Omega)$ 
non-deterministically sets the variable $X$ in the environments 
$\gamma_\E(R^\s)$ to any value in $\gamma_\RR(V^\s)$.

Any non-relational domain on a single variable can be used as $\RR^\s$.
One useful example is the interval domain \cite{cc-POPL77}.
In this case, an element in $\RR^\s$ is either $\bot^\s_\RR$, or
a pair consisting of a lower and an upper bound.
The function $\asexpr$ is then straightforward because intervals can be 
directly and exactly represented in the syntax of expressions.
Moreover, the function $\get$ consists in extracting the range
of a variable from an abstract environment $R^\s\in\E^\s$, an
operation which is generally available in the implementations of
numerical abstract domains, e.g., in the Apron library
\cite{mine:cav09}.

\begin{figure}
  \centering
  $\begin{array}{ll}
    \RR^\s & \com{abstract sets of reals}
    \\[3pt]
    \gamma_\RR:\RR^\s\rightarrow\P(\R) & \com{concretization function}
    \\[3pt]
    \bot^\s_\RR\in\RR^\s & \com{abstract empty set}\\
    \qquad\text{ s.t. }\gamma_\RR(\bot^\s_\RR)=\emptyset
    \\[3pt]
    \cup^\s_\RR:(\RR^\s\times\RR^\s)\rightarrow\RR^\s
    & \com{abstract join}\\
    \qquad \text{s.t. }\gamma_\RR(V^\s\cup^\s_\RR W^\s)\supseteq\gamma_\RR(V^\s)\cup\gamma_\RR(W^\s)
    \\[3pt]
    \widen_\RR:(\RR^\s\times\RR^\s)\rightarrow\RR^\s
    & \com{widening}\\
    \qquad \text{s.t. }\gamma_\RR(V^\s\widen_\RR W^\s)\supseteq\gamma_\RR(V^\s)\cup\gamma_\RR(W^\s)\\
    \qquad \text{and }\fa{(W^\s_i)_{i\in\N}}\text{ the sequence }V^\s_0=W^\s_0,\,
    V^\s_{i+1}=V^\s_i\widen_\RR W^\s_{i+1}\hspace*{-2cm}\\
    \qquad\text{reaches a fixpoint }V^\s_k=V^\s_{k+1}\text{ for some }k\in\N
    \\[3pt]
    \get:(\V\times\E^\s)\rightarrow\RR^\s
    & \com{variable extraction}\\
    \qquad \text{s.t. }\gamma_\RR(\get(X,R^\s))\supseteq
    \setst{\rho(X)}{\rho\in\gamma_\E(R^\s)}
    \\[3pt]
    \asexpr:(\RR^\s\setminus\sset{\bot^\s_\RR})\rightarrow\mi{expr}
    & \com{conversion to expression}\\
    \qquad \text{s.t. }\fa{\rho}
    \letin{(V,-)}{\lb{E}{\asexpr(V^\s)}\rho}
    V\supseteq\gamma_\RR(V^\s)\hspace*{-0.5cm}
  \end{array}$
  \caption{Signature, soundness and termination conditions for a domain $\RR^\s$ abstracting sets of reals.}
  \label{fig:absrealdomain}
\end{figure}

\medskip

We now show how, given these domains, we can construct an abstraction
$\lbp{P_\I^\s}$ of $\lbp{P_\I}$.
We first construct, using $\RR^\s$, an abstraction
$\I^\s$ of interference sets from $\P(\I)$, as presented in 
Fig.~\ref{fig:absinterfdomain}.
It is simply a partitioning of abstract sets of real values with
respect to threads and variables:
$\I^\s\deq(\T\times \V)\rightarrow\RR^\s$,
together with pointwise concretization $\gamma_\I$, join $\cup^\s_\I$, and
widening $\widen_\I$.
Note that $\I^\s$ is not isomorphic to a non-relational
domain on a set $\T\times\V$ of variables.
Indeed, the former abstracts $(\T\times\V)\rightarrow\P(\R)\simeq\P(\T\times\V\times\R)=\P(\I)$, 
while the latter would abstract $\P((\T\times\V)\rightarrow\R)$.
In particular, the former can express abstract states where the value set of
some but not all variables is empty, 
while $\bot^\s_\RR$ elements in the later coalesce to a
single element representing $\emptyset$.
We then construct an abstraction $\D^\s_\I$ of the semantic domain $\D_\I$, 
as presented in Fig.~\ref{fig:absinterfdomain2}.
An element of $\D^\s_\I$  is a triple $(R^\s,\Omega,I^\s)$ composed of an
abstraction $R^\s\in\E^\s$ of environments, a set $\Omega\subseteq\L$ of errors, and an 
abstraction $I^\s\in\I^\s$ of interferences. 
The concretization $\gamma$, join $\cup^\s$, and widening
$\widen$ are defined pointwise.

\begin{figure}
  \centering
  $\begin{array}{l}
    \I^\s\deq(\T\times \V)\rightarrow\RR^\s
    \\[3pt]
    \gamma_\I:\I^\s\rightarrow\P(\I)\\
    \qquad \text{s.t. }\gamma_\I(I^\s)\deq\setst{(t,X,v)}{t\in\T,\,X\in\V,\,v\in\gamma_\RR(I^\s(t,X))}
    \\[3pt]
    \bot^\s_\I\deq\lbd{(t,X)}\bot^\s_\RR
    \\[3pt]
    I_1^\s\cup^\s_\I I^\s_2\deq\lbd{(t,X)}I_1^\s(t,X)\cup^\s_\RR I_2^\s(t,X)
    \\[3pt]
    I_1^\s\widen_\I I^\s_2\deq\lbd{(t,X)}I_1^\s(t,X)\widen_\RR I_2^\s(t,X)
  \end{array}$
  \caption{Abstract domain $\I^\s$ of interferences, derived from $\RR^\s$.}
  \label{fig:absinterfdomain}
\end{figure}

\begin{figure}
  \centering
  $\begin{array}{l}
    \D^\s_\I\deq \E^\s\times\P(\L)\times\I^\s
    \\[3pt]
    \gamma:\D^\s_\I\rightarrow\D_\I\\
    \qquad\text{s.t. }\gamma(R^\s,\Omega,I^\s)\deq(\gamma_\E(R^\s),\Omega,\gamma_\I(I^\s))
    \\[3pt]
    (R^\s_1,\Omega_1,I^\s_1)\cup^\s (R^\s_2,\Omega_2,I^\s_2)\deq
    (R^\s_1\cup^\s_\E R^\s_2,\Omega_1\cup\Omega_2,I^\s_1 \cup^\s_\I I^\s_2)
    \\[3pt]
    (R^\s_1,\Omega_1,I^\s_1)\widen (R^\s_2,\Omega_2,I^\s_2)\deq
    (R^\s_1\widen_\E R^\s_2,\Omega_1\cup\Omega_2,I^\s_1 \widen_\I I^\s_2)
  \end{array}$
  \caption{Abstract semantic domain $\D^\s_\I$, derived from $\E^\s$ and $\I^\s$.}
  \label{fig:absinterfdomain2}
\end{figure}

The abstract semantics $\lb{S_\I^\s}{s,t}$ of a statement $s$ executed
in a thread $t\in\T$ should be a function from $\D^\s_\I$ to $\D^\s_\I$
obeying the soundness condition:
\begin{equation*}
  (\lb{S_\I}{s,t}\circ\gamma)(R^\s,\Omega,I^\s)\sqsubseteq_\I
  (\gamma\circ\lb{S_\I^\s}{s,t})(R^\s,\Omega,I^\s)
\end{equation*}
i.e., the abstract function over-approximates the sets of environments, errors, 
and interferences.
Such a function is defined in a generic way in Fig.~\ref{fig:absinterfstat}.
The semantics of assignments and guards with interference is defined based on
their non-interference semantics $\lb{S^\s}{s}$, provided as part of the 
abstract domain $\E^\s$.
In both cases, the expression $e$ to assign or test is first modified
to take interferences into account, using the $\apply$ function.
This function takes as arguments a thread $t\in\T$, an abstract environment
$R^\s\in\E^\s$, an abstract interference $I^\s\in\I^\s$, and an expression $e$.
It first collects, for each variable $Y\in\V$, the relevant interferences 
$V_Y^\s\in\RR^\s$ from $I^\s$, i.e., concerning the variable 
$Y$ and threads $t'\neq t$.
If the interference for $Y$ is empty, $\bot^\s_\RR$, the occurrences of $Y$ in $e$
are kept unmodified.
If it is not empty, then the occurrences of $Y$ are replaced with a constant
expression encompassing all the possible values that can be read from $Y$,
from either the interferences or the environments $\gamma_\E(R^\s)$.
Additionally, the semantics of an assignment $X\leftarrow e$ enriches 
$I^\s$ with new interferences corresponding to the values of $X$ after the
assignment.
The semantics of non-primitive statements is identical to the interference-free
case of Fig.~\ref{fig:absstatstructsem}.

\begin{figure}
  \centering
  $\begin{array}{l}
    \underline{\lb{S_\I^\s}{s,t}:\D^\s_\I\rightarrow\D^\s_\I}
    \\[4pt]
    \lb{S_\I^\s}{X\leftarrow e,\,t}(R^\s,\Omega,I^\s)\deq\\
    \qquad
    \letin{(R^\s{}',\Omega{}')}{\lb{S^\s}{X\leftarrow\apply(t,R^\s,I^\s,e)}(R^\s,\Omega)}\\
    \qquad
    (R^\s{}',\,\Omega{}',\,I^\s[(t,X)\mapsto I^\s(t,X)\cup^\s_\RR\get(X,R^\s{}')])
    \\[3pt]
    \lb{S_\I^\s}{e\bowtie 0?,\,t}(R^\s,\Omega,I^\s)\deq\\
    \qquad
    \letin{(R^\s{}',\Omega{}')}{\lb{S^\s}{\apply(t,R^\s,I^\s,e)\bowtie 0?}(R^\s,\Omega)}
    (R^\s{}',\Omega{}',I^\s)
    \\[3pt]
    \lb{S_\I^\s}{\sy{if}\;e\bowtie0\;\sy{then}\;s,\,t} (R^\s,\Omega,I^\s)\deq\\
    \qquad
    (\lb{S_\I^\s}{s,t}\circ\lb{S_\I^\s}{e\bowtie0?,\,t})(R^\s,\Omega,I^\s)\;\cup^\s\;\lb{S_\I^\s}{e\not\bowtie0?,\,t} (R^\s,\Omega,I^\s)
    \\[3pt]
    \lb{S_\I^\s}{\sy{while}\;e\bowtie0\;\sy{do}\;s,\,t} (R^\s,\Omega,I^\s)\deq\\
    \qquad
    \lb{S_\I^\s}{e\not\bowtie0?,\,t}(\lim\lbd{X^\s}X^\s\widen((R^\s,\Omega,I^\s)\;\cup^\s\;(\lb{S_\I^\s}{s,\,t}\circ\lb{S_\I^\s}{e\bowtie0?,\,t})X^\s))
    \\[3pt]
    \lb{S_\I^\s}{s_1;\,s_2,\,t} (R^\s,\Omega,I^\s)\deq(\lb{S_\I^\s}{s_2,t}\circ\lb{S_\I^\s}{s_1,t})(R^\s,\Omega,I^\s)
    \\
    \\
    \text{where: }\\
    \quad
    \apply(t,R^\s,I^\s,e)\deq\\
    \quad
    \qquad
    \letin{\fa{Y\in\V}V_Y^\s}{\cup^\s_\RR\;\setst{I^\s(t',Y)}{t'\neq t}}\\[3pt]
    \quad
    \qquad
    \letin{\fa{Y\in\V}e_Y}{
      \begin{cases}
        Y & \text{if $V_Y^\s=\bot^\s_\RR$}\\
        \asexpr(V_Y^\s\cup^\s_\RR\get(Y,R^\s)) & \text{if $V_Y^\s\neq\bot^\s_\RR$}
      \end{cases}}\\
    \qquad\quad
    e[\fa{Y\in\V}Y\mapsto e_Y]
  \end{array}$
  \caption{Abstract semantics of statements with interference.}
  \label{fig:absinterfstat}
\end{figure}

Finally, an abstraction of the interference fixpoint
(\ref{eq:intersem}) is computed by iteration on abstract interferences,
using the widening $\widen_\I$ to ensure termination, which provides
the abstract semantics $\lbp{P^\s_\I}$ of our program:
\begin{equation}
  \label{eq:absintersem}
  \begin{array}{l}
    \lbp{P_\I^\s}\deq\Omega,\text{ where }
    (\Omega,-)\deq\\
    \quad
    \lim\lbd{(\Omega,I^\s)}
      \letin{\fa{t\in\T}(-,\Omega'_t,I^\s_t{}')}{\lb{S_\I^\s}{\body_t,\,t}(\E^\s_0,\Omega,I^\s)}\\
      \qquad
      (\bigcup\;\setst{\Omega'_t}{t\in\T},\;
      I^\s\,\widen_\I\,\bigcup^\s_\I\;\setst{I^\s_t{}'}{t\in\T})
  \end{array}
\end{equation}
where $\lim F^\s$ denotes the limit of the iterates of $F^\s$ starting
from $(\emptyset,\bot^\s_\I)$.
The following theorem states the soundness of the analysis:
\begin{thm}
  \label{thm:soundproc}
  $\lbp{P_\I}\subseteq\lbp{P_\I^\s}.$
\end{thm}
\proof In Appendix~\ref{proof:soundproc}.\qed

The obtained analysis remains flow-sensitive and can be relational 
within each thread, provided that $\E^\s$ is relational.
However, interferences are abstracted in a flow-insensitive and 
non-relational way. This was already the case for the
concrete interferences in $\lbp{P_\I}$ and it is not related to the choice
of abstract domains.
The analysis is expressed as an outer iteration that completely 
re-analyzes each thread until the abstract interferences stabilize.
Thus, it can be implemented easily on top of an existing non-parallel 
analyzer.
Compared to a non-parallel program analysis, the cost is multiplied by
the number of outer iterations required to stabilize interferences.
Thankfully, our preliminary experimental results suggest that this number 
remains very low in practice --- 5 for our benchmark in Sec.~\ref{sec:result}.
In any case, the overall cost is not related to the (combinatorial) number 
of possible interleavings, but rather to the amount of abstract interferences $I^\s$,
i.e., of actual communications between the threads.
It is thus always possible to speed up the convergence of interferences or,
conversely, improve the precision at the expense of speed, by
adapting the widening $\widen_\RR$.

\medskip

In this article, we focus on analyzing systems composed of a fixed, finite
number of threads.
The finiteness of $\T$ is necessary for the computation 
of $\lbp{P_\I^\s}$ in (\ref{eq:absintersem}) to be effective.
However, it is actually possible to relax this hypothesis and allow an unbounded
number of instances of some threads to run in parallel.
For this, it is sufficient to consider {\em self-interferences}, i.e., 
replace the condition $t'\neq t$ in the definition
$\lb{E_\I}{X}(t,\rho,I)$ in Fig.~\ref{fig:interfexprsem} (for the concrete
semantics) and $\apply(t,R^\s,I^\s,e)$ in Fig.~\ref{fig:absinterfstat} 
(for the abstract semantics) with $t'\neq t\vee t\in\T'$, 
where $\T'\subseteq\T$ denotes the subset of threads that can have several 
instances.
The resulting analysis is necessarily uniform, i.e., 
it cannot distinguish different instances of the same thread nor express 
properties about the number of running instances ---
it is abstracted statically in a domain of two values: ``one''
($t\notin\T'$)  and ``two or more'' ($t\in\T'$).
In order to analyze actual programs spawning an unbounded number of threads, 
a non-uniform analysis (such as performed by Feret \cite{feret:getcol00} in 
the context of the $\pi-$calculus) may be necessary to achieve a
sufficient precision, but this is not the purpose of the present article.

\subsection{Weakly Consistent Memory Semantics \texorpdfstring{$\lbp{P'_*}$}{P'*}}
\label{sec:weaksem}

We now review the various parallel semantics we proposed in the preceding 
sections and discuss their adequacy to describe actual models of 
parallel executions.

\medskip

It appears that our first semantics, the concrete interleaving semantics 
$\lbp{P_*}$ of Sec.~\ref{sec:interleavesem}, while simple, is not realistic.
A first issue is that, as noted by Reynolds in \cite{reynolds-FSTTCS04}, such a 
semantics requires choosing a level of granularity, i.e., some basic set of 
operations that are assumed to be atomic and cannot be interrupted
by another thread.
In our case, we assumed assignments and guards (i.e., primitive statements) 
to be atomic.
In contrast, an actual system may
schedule a thread within an assignment and cause, for instance,
$x$ to be 1 at the end of the program in Fig.~\ref{fig:weakex2}
instead of the expected value 2.
A second issue, noted by Lamport in \cite{lamport-ACM78}, is that the
latency of loads and stores in a shared memory may break the
sequential consistency in true multiprocessor systems: threads running
on different processors may not always agree on the value of a shared variable.
For instance, in the mutual exclusion algorithm of Fig.~\ref{fig:weakex1}, 
the thread $t_2$ may still see the value 0 in $\mathrm{flag1}$ even after the thread
$t_1$ has entered its critical section, causing $t_2$ to also enter its
critical section, as the effect of the assignment $\mathrm{flag1}\leftarrow 1$
is propagated asynchronously and takes some time to be acknowledged by $t_2$.
Moreover, Lamport noted in \cite{lamport-TC79} that reordering
of independent loads and stores in one thread by the processor can also
break sequential consistency --- for instance performing the load from 
$\mathrm{flag2}$ before the store into $\mathrm{flag1}$, instead of after,
in the thread $t_1$ in Fig.~\ref{fig:weakex1}.
More recently, it has been observed by Manson et al.~\cite{manson-al-POPL05} 
that optimizations in modern compilers have the same
ill-effect, even on mono-processor systems: program transformations that are
perfectly safe on a thread considered in isolation 
(for instance, reordering the independent assignment 
$\mathrm{flag1}\leftarrow 1$ and test $\mathrm{flag2}=0$ in $t_1$)
can cause non-sequentially-consistent behaviors to appear.
In this section, we show that the interference semantics correctly handles
these issues by proving that it is invariant under a ``reasonable'' class of 
program transformations. This is a consequence of its coarse, flow-insensitive
and non-relational modeling of thread communications.

\medskip

Acceptable program transformations of a thread are defined with respect to 
the path-based semantics $\lbpx{\Pi}$ of Sec.~\ref{sec:pathsem}.
A transformation of a thread $t$ is acceptable if it gives rise to a set
$\pi'(t)\subseteq\Pi$ of control paths such that every path $p'\in\pi'(t)$ 
can be obtained from a path $p\in\pi(\body_t)$
by a sequence of elementary transformations described  below
in Def.~\ref{def:transform}.
Elementary transformations are denoted $q\becomes q'$, where $q$ and $q'$
are sequences of primitive statements.
This notation indicates that any occurrence of $q$ in a path 
of a thread can be replaced with
$q'$, whatever the context appearing before and after $q$.
The transformations in Def.~\ref{def:transform} try to account for
widespread compiler and hardware
optimizations, but are restricted to transformations that do not
generate new errors nor new interferences.\footnote{The environments at the 
end of the thread after transformations may be different, but this does not
pose a problem as environments are not observable in our semantics:
$\lbp{P_*}\subseteq\L$ (\ref{eq:interleavesem}).}
This ensures that an interference-based analysis of the original program is 
sound with respect to the transformed one, which is formalized below in
Thm.~\ref{thm:weakinterf}.

The elementary transformations of Def.~\ref{def:transform} require 
some side-conditions to hold in order to be acceptable.
They use the following notions.
We say that a variable $X\in\V$ is {\em fresh} if it does not occur in any 
thread, and {\em local} if it occurs only in the currently transformed 
thread.
We denote by $s[e'/e]$ the statement $s$ where some, but not necessarily all,
occurrences of the expression $e$ may be changed into $e'$.
The set of variables appearing in the expression $e$ is denoted $\var(e)$,
while the set of variables modified by the statement $s$ is
$\lval(s)$. Thus, $\lval(X\leftarrow e)=\sset{X}$ while 
$\lval(e\bowtie0?)=\emptyset$.
The predicate $\nonblock(e)$ holds if evaluating the expression $e$ cannot 
block the program --- as would, e.g., an expression with a definite run-time 
error, such as 1/0 --- i.e., 
$\nonblock(e)\diff\fa{\rho\in\E}V_\rho\neq\emptyset$ where 
$(V_\rho,-)=\lb{E}{e}\rho$.
We say that $e$ is {\em deterministic} if, moreover, $\fa{\rho\in\E}|V_\rho|=1$.
Finally, $\noerror(e)$ holds if evaluating $e$ is always error-free, i.e.,
$\noerror(e)\diff\fa{\rho\in\E}\Omega_\rho=\emptyset$ where
$(-,\Omega_\rho)=\lb{E}{e}\rho$.
We are now ready to state our transformations:
\begin{defi}[Elementary path transformations]
\label{def:transform}
~\par
\begin{enumerate}[(1)]
\item Redundant store elimination:
$X\leftarrow e_1\cdot X\leftarrow e_2\becomes X\leftarrow e_2$,
when $X\notin\var(e_2)$ and $\nonblock(e_1)$.
\item Identity store elimination: $X\leftarrow X\becomes\epsilon$.
\item Reordering assignments:
$X_1\leftarrow e_1\cdot X_2\leftarrow e_2\becomes X_2\leftarrow e_2\cdot X_1\leftarrow e_1$,
when $X_1\notin\var(e_2)$, $X_2\notin\var(e_1)$, $X_1\neq X_2$, and $\nonblock(e_1)$.
\item Reordering guards:
$e_1\bowtie 0?\cdot e_2\bowtie' 0?\becomes e_2\bowtie' 0?\cdot e_1\bowtie 0?$,
when $\noerror(e_2)$.
\item Reordering guards before assignments:
$X_1\leftarrow e_1\cdot e_2\bowtie 0?\becomes e_2\bowtie 0?\cdot X_1\leftarrow e_1$,
when $X_1\notin\var(e_2)$ and either $\nonblock(e_1)$ or $\noerror(e_2)$.
\item Reordering assignments before guards:
$e_1\bowtie 0?\cdot X_2\leftarrow e_2\becomes X_2\leftarrow e_2\cdot e_1\bowtie0?$,
when $X_2\notin\var(e_1)$, $X_2$ is local, and $\noerror(e_2)$.
\item Assignment propagation: $X\leftarrow e\cdot  s\becomes X\leftarrow e\cdot  s[e/X]$,
when $X\notin\var(e)$, $\var(e)$ are local, and $e$ is deterministic.
\item Sub-expression elimination:
$s_1\cdot \ldots\cdot s_n\becomes X\leftarrow e\cdot s_1[X/e]\cdot \ldots\cdot s_n[X/e]$,
when $X$ is fresh, $\fa{i}\var(e)\cap\lval(s_i)=\emptyset$,
and $\noerror(e)$.
\item Expression simplification: 
$s\becomes s[e'/e]$,
when $\fa{\rho\in\E}\lb{E}{e}\rho\sqsupseteq\lb{E}{e'}\rho$
and $\var(e)$ and $\var(e')$ are local.\footnote{The original expression 
simplification rule from \cite{mine:esop11} required a much stronger
side-condition:
$\lb{E_\I}{e}\br(t,\rho,I)\sqsupseteq\lb{E_\I}{e'}(t,\rho,I)$ for all
$\rho$ and $I$, which actually implied that $e$ and $e'$ were variable-free.
We propose here a more permissive side-condition allowing local
variables to appear in $e$ and $e'$.}
\end{enumerate}
\end{defi}
These simple rules, used in combination, allow modeling large classes of
classic program transformations as well as distributed memories.
Store latency can be simulated using rules 7 and 3.
Breaking a statement into several ones is possible with rules 7 and 8.
As a consequence, the rules can expose preemption points within statements,
which makes primitive statements no longer atomic.
Global optimizations, such as constant propagation and folding, can be 
achieved using rules 7 and 9.
Rules 1--6 allow peephole optimizations.
Additionally, transformations that do not change the set of control
paths, such as loop unrolling, are naturally
supported.

Given the set of transformed control paths $\pi'(t)$ for each thread
$t\in\T$, the set of transformed parallel control paths $\pi'_*$ is defined, 
similarly to (\ref{eq:interleave}), as:
\begin{equation}
  \label{eq:transinterleavesem}
  \pi'_*\deq\setst{p\in\Pi_*}{\fa{t\in\T}\proj_t(p)\in\pi'(t)}\\
\end{equation}
and the semantics $\lbp{P'_*}$ of the parallel program is, similarly to
(\ref{eq:interleavesem}):
\begin{equation}
  \label{eq:transinterleavesem2}
    \lbp{P'_*}\deq\Omega,\text{ where }(-,\Omega)=\lbx{\Pi_*}{\pi'_*}(\E_0,\emptyset)\enspace.
\end{equation}
Any original thread $\pi(\body_t)$ being a special case of transformed
thread $\pi'(t)$ (considering the identity transformation), we have
$\lbp{P_*}\subseteq\lbp{P'_*}$.
The following theorem extends Thm.~\ref{thm:interf} to transformed programs:
\begin{thm}
  \label{thm:weakinterf}
  $\lbp{P'_*}\subseteq\lbp{P_\I}$.
\end{thm}
\proof In Appendix~\ref{proof:weakinterf}.\qed
An immediate consequence of Thms.~\ref{thm:soundproc} and \ref{thm:weakinterf}
is the soundness of the abstract semantics $\lbp{P^\s_\I}$ with respect
to the concrete semantics of the transformed program $\lbp{P'_*}$, i.e.,
$\lbp{P'_*}\subseteq\lbp{P^\s_\I}$.
Note that, in general, $\lbp{P'_*}\neq\lbp{P_\I}$.
The two semantics may coincide, as for instance in the program of 
Fig.~\ref{fig:weakex1}. In that case:
$\lbp{P_*}\subsetneq\lbp{P'_*}=\lbp{P_\I}$.
However, in the case of Fig.~\ref{fig:weakex2}, $y$ can take any positive 
value according to the interference semantics $\lbp{P_\I}$
(as explained in Sec.~\ref{sec:interfersem}), while the interleaving semantics
after program transformation $\lbp{P'_*}$ only allows the values 1 and 2;
we have $\lbp{P_*}=\lbp{P'_*}\subsetneq\lbp{P_\I}$.

\medskip

Theorem~\ref{thm:weakinterf} holds for our ``reasonable'' 
collection of program transformations, but may not hold when considering
other, ``unreasonable'' ones.
For instance, in Fig.~\ref{fig:weakex1},
$\mathrm{flag1}\leftarrow 1$ should not be replaced
by a misguided prefetching optimizer
with $\mathrm{flag1}\leftarrow 42; \mathrm{flag1}\leftarrow 1$ in thread $t_1$.
This would create a spurious interference causing the value 42 to be possibly 
seen by thread $t_2$.
If there is no other reason for $t_2$ to see the value 42, such as a 
previous or future assignment of 42 into $\mathrm{flag1}$ by $t_1$, it would
create an execution outside those considered by the interference semantics
and invalidate Thm.~\ref{thm:weakinterf}.
Such ``out-of-thin-air'' values are explicitly forbidden by the Java
semantics as described by Manson et al.~\cite{manson-al-POPL05}. See
also \cite{saraswat-al-PPOPP07} for an in-depth discussion of out-of-thin-air
values.
Another example of invalid transformation is the reordering of assignments
$X_1\leftarrow e_1 \cdot X_2\leftarrow e_2\becomes X_2\leftarrow e_2 \cdot X_1\leftarrow e_1$
when $e_1$ may block the program, e.g., due to a division by zero
$X_1\leftarrow 1/0$. 
Indeed, the transformed program could expose errors in $e_2$ that cannot occur
in the original program because they are masked by the previous error in
$X_1\leftarrow e_1$.
This case is explicitly forbidden by the
$\nonblock(e_1)$ side condition in Def.~\ref{def:transform}.(3).
The proof in Appendix~\ref{proof:weakinterf} contains more examples of
transformations that become invalid when side-conditions are not respected.

Definition~\ref{def:transform} is not exhaustive.
It could be extended with other ``reasonable'' transformations, and some
restrictive side-conditions might be relaxed in future work without breaking
Thm.~\ref{thm:weakinterf}.
It is also possible to enrich Def.~\ref{def:transform} with new transformations
that do not respect Thm.~\ref{thm:weakinterf} as is, and then adapt the
interference semantics to retrieve a similar theorem. 
For instance, we could allow speculative stores of some special value,
such as zero, which only requires adding an interference $(t,X,0)$ for each
thread $t$ and each variable $X$ modified by $t$.
As another example, we could consider some memory writes to be non-atomic,
such as 64-bit writes on 32-bit computers, which requires adding
interferences that expose partially assigned values.

Finally, it would be tempting to, dually, reduce the number of allowed program
transformations, and enforce a stronger memory consistency.
For instance, we could replace Def.~\ref{def:transform} with a model of an
actual multiprocessor, such as the intel x86 architecture model proposed
by Sewell et al. in \cite{sewell-al:jacm10}, 
which is far less permissive and thus ensures many more properties.
We would obtain a more precise interleaving semantics $\lbp{P'_*}$,
closer to the sequentially consistent one $\lbp{P_*}$.
However, this would not mechanically improve the result of our static
analysis $\lbp{P^\s_\I}$, as it is actually an abstraction of the
concrete interference semantics $\lbp{P_\I}$, itself an incomplete 
abstraction of $\lbp{P_*}$.
Our choice of an interference semantics was not initially 
motivated by the modeling of weakly consistent memories (although this is
an important side effect), but rather by the construction of an effective and
efficient static analyzer.
Effectively translating a refinement of the memory model at the level of an
interference-based analysis without sacrificing the efficiency remains a 
challenging future work.

\section{Multi-threaded Programs With a Real-Time Scheduler}
\label{sec:sched}

We now extend the language and semantics of the preceding section
with explicit synchronization primitives.
These can be used to enforce mutual exclusion and construct critical
sections, avoiding the pitfalls of weakly consistent memories.
We also extend the semantics with a real-time scheduler taking thread
priorities into account, which provides an alternate way of implementing
synchronization.

\subsection{Priorities and Synchronization Primitives}
\label{sec:syncprim}

We first describe the syntactic additions to our language and introduce
informally the change in semantics.

\medskip

We denote by $\M$ a finite, fixed set of mutual exclusion locks, 
so-called {\em mutexes}.
The original language of Fig.~\ref{fig:syntax} is enriched with 
primitives to control mutexes and scheduling as follows:
\begin{equation}
  \label{eq:schedsyntax}
  \begin{array}{lcl@{\qquad}l}
    \mi{stat} & ::= & \sy{lock}(m) & \com{mutex locking, $m\in\M$}\\
    &|& \sy{unlock}(m) & \com{mutex unlocking, $m\in\M$}\\
    &|& X\leftarrow \sy{islocked}(m) & \com{mutex testing, $X\in\V,\,m\in\M$}\\
    &|& \sy{yield} & \com{thread pause}\\
  \end{array}
\end{equation}

The primitives $\sy{lock}(m)$ and $\sy{unlock}(m)$ respectively acquire
and release the mutex $m\in\M$.
The primitive $X\leftarrow \sy{islocked}(m)$ is used to test the status of the 
mutex $m$: it stores 1 into $X$ if $m$ is acquired by some thread, 
and 0 if it is free.
The primitive $\sy{yield}$ is used to voluntarily relinquish the control
to the scheduler.
The definition of control paths $\pi(s)$ from (\ref{eq:path}) is extended
by stating that $\pi(s)\deq\sset{s}$ for these statements, i.e., they
are primitive statements.
We also assume that threads have fixed, distinct priorities.
As only the ordering of priorities is significant,
we denote threads in $\T$ simply by integers ranging from 1 to $|\T|$, being 
understood that thread $t$ has a strictly higher priority than thread $t'$
when $t>t'$.

\medskip

To keep our semantics simple, we assume that acquiring a mutex for a thread 
already owning it is a no-op, as well as releasing a mutex it does not hold.
Our primitive mutexes can serve as the basis to implement more complex 
ones found in actual implementations.
For instance, mutexes that generate a run-time error or return an error
code when locked twice by the same thread can be implemented
using an extra program variable for each mutex / thread pair that 
stores whether the thread has already locked that mutex.
Likewise, recursive mutexes can be implemented by making these variables
count the number of times each thread has locked each mutex.
Finally, locking with a timeout can be modeled as a non-deterministic 
conditional that either locks the mutex, or yields and returns an error code.

\medskip

Our scheduling model is that of real-time processes, used noticeably
in embedded systems.
Example operating systems using this model include those obeying the ARINC 653 
standard \cite{ARINC} (used in avionics), as well as the real-time extension of 
the POSIX threads standard \cite{posix-threads}.
Hard guarantees about the execution time of services, although an important
feature of real-time systems, are not the purpose of this article as we
abstract physical time away.
We are interested in another feature: the strict interpretation of thread 
priorities when deciding which thread to schedule.
More precisely: a thread that is not blocked waiting for some resource can 
never be preempted by a lower priority thread.
This is unlike schedulers found in desktop computers (for instance,
vanilla POSIX threads \cite{posix-threads} without the real-time extension)
where even lower priority threads always get to run, 
preempting higher priority ones if necessary.
Moreover, we consider in this section that only a single thread can execute 
at a given time --- this was not required in Sec.~\ref{sec:shared}.
This is the case, for instance, when all the threads share a single processor.
In this model, the unblocked thread with the highest priority is always
the only one to run.
All threads start unblocked and may block, either by locking a mutex
that is already locked by another thread, or by yielding voluntarily, 
which allows lower priority threads to run.
Yielding denotes blocking for a non-deterministic amount of time, which is
useful to model timers of arbitrary duration
and waiting for external resources.
A lower priority thread can be preempted when unlocking a mutex
if a higher priority thread is waiting for this mutex.
It can also be preempted at any point by a yielding higher priority thread
that wakes up non-deterministically.
Thus, a preempted thread can be made to wait at an arbitrary program point,
and not necessarily at a synchronization statement.
The scheduling is dynamic and the number of possible thread interleavings 
authorized by the scheduler remains very large, despite being controlled 
by strict priorities.

This scheduling model is precise enough to take into account fine
mutual exclusion properties that would not hold if we considered arbitrary 
preemption or true parallel executions on concurrent processors.
For instance, in Fig.~\ref{fig:schedul}, the high priority thread avoids
a call to $\sy{lock}$ by testing with $\sy{islocked}$
whether the low priority
thread acquired the lock and, if not, executes its critical section and 
modifies $Y$ and $Z$, confident
that the low priority thread cannot execute and enter its
critical section before the high priority thread explicitly $\sy{yield}$s.

\begin{figure}[t]
  \centering
  \begin{tabular}{l@{\quad}|@{\quad}l}
    \quad\underline{low priority}\quad & 
    \quad\underline{high priority}\quad\\[7pt]
    $\sy{lock}(m);$ & $X\leftarrow\sy{islocked}(m);$\\
    $Y\leftarrow 1;$ & $\sy{if}\;X=0\;\sy{then}$\\
    $Z\leftarrow 1;$ & \quad $Z\leftarrow 2;$\\
    $T\leftarrow Y-Z$; & \quad $Y\leftarrow 2;$\\
    $\sy{unlock}(m)$ & \quad $\sy{yield}$\\
    \end{tabular}
  \caption{Using priorities to ensure mutual exclusion.}
  \label{fig:schedul}
\end{figure}

\subsection{Concrete Scheduled Interleaving Semantics \texorpdfstring{$\lbp{P_\H}$}{P-H}}
\label{sec:schedinterleavesem}

We now refine the various semantics of Sec.~\ref{sec:shared} to take scheduling into
account, starting with the concrete interleaving semantics $\lbp{P_*}$ of 
Sec.~\ref{sec:interleavesem}.
In this case, it is sufficient to redefine the semantics of primitive 
statements.
This new semantics will, in particular, exclude interleavings that do 
not respect mutual exclusion or priorities, and thus, we observe fewer 
behaviors.
This is materialized by the dotted $\subseteq$ arrow in Fig.~\ref{fig:sumsem}
between $\lbp{P_*}$ and the refined semantics $\lbp{P_\H}$ we are about to 
present.\footnote{Note that Fig.~\ref{fig:sumsem}
states that each concrete semantics without scheduler abstracts the 
corresponding concrete semantics with scheduler, but states nothing about
abstract semantics. Abstract semantics are generally incomparable due to 
the use of non-monotonic abstractions and widenings.}

\medskip

We define a domain of scheduler states $\H$ as follows:
\begin{equation}
  \H\deq(\T\rightarrow\setst{\ready,\yield,\wait(m)}{m\in\M})\times(\T\rightarrow \P(\M))\enspace.
\end{equation}
A scheduler state $(b,l)\in\H$ is a pair, where the function $b$ associates
to each thread whether it is ready (i.e., it is not blocked, and runs if
no higher priority thread is also ready), yielding (i.e., it is blocked at a
$\sy{yield}$ statement), or waiting for some mutex $m$ (i.e., it is blocked at 
a $\sy{lock}(m)$ statement). The function $l$ associates to each thread 
the set of all the mutexes it holds.
A program state is now a pair $((b,l),\rho)$ composed of a 
scheduler state $(b,l)\in\H$ and an environment $\rho\in\E$.
The semantic domain $\D\deq\P(\E)\times\P(\L)$ from
(\ref{eq:d}) is thus replaced with $\D_\H$ defined as:
\begin{equation}
  \D_\H\deq\P(\H\times\E)\times\P(\L)
\end{equation}
with the associated pairwise join $\sqcup_\H$.

The semantics $\lb{S_\H}{s,t}$ of a primitive statement $s$ executed 
by a thread $t\in\T$ is described in Fig.~\ref{fig:schedinterleavestatsem}.
It is decomposed into three steps: $\enabled_t$, $\lb{S^\dag_\H}{s,t}$, and
$\sched$, the first and the last steps being independent from the choice of
statement $s$.
Firstly, the function $\enabled_t$ filters states to keep only those where the 
thread $t$ can actually run, i.e., $t$ is the highest priority thread which
is ready.
Secondly, the function $\lb{S^\dag_\H}{s,t}$ handles the statement-specific
semantics.
For $\sy{yield}$, $\sy{lock}$, and $\sy{unlock}$ statements, this
consists in updating the scheduler part of each state.
For $\sy{lock}$ statements, the thread enters a wait state until the mutex 
is available. Actually acquiring the mutex is performed by the
following $\sched$ step if the mutex is immediately available, and by a 
later $\sched$ step following the unlocking of the mutex by its owner
thread otherwise.
The $\sy{islocked}$ statement updates each environment according to its
paired scheduler state.
The other primitive statements, assignments and guards, are not related to 
scheduling;
their semantics is defined by applying the regular, mono-threaded semantics
$\lb{S}{s}$ from Fig.~\ref{fig:statstructsem} to the environment part,
leaving the scheduler state unchanged.
Thirdly, the function $\sched$ updates the scheduler state by
waking up yielding threads non-deterministically, and
giving any newly available mutex to the highest priority thread waiting for it,
if any.

\begin{figure}
  \centering
  $\begin{array}{l}
    \underline{\lb{S_\H}{s,t}:\D_\H\stackrel{\sqcup_\H}{\longrightarrow}\D_\H}
    \\[4pt]
    \lb{S_\H}{s,t}\deq\sched \circ\lb{S^\dag_\H}{s,t}\circ\enabled_t
    \\[6pt]
    \text{where:}\\[3pt]
    \enabled_t(R,\Omega)\deq
    (\setst{((b,l),\rho)\in R}{b(t)=\ready\wedge\fa{t'>t}b(t')\neq\ready},\Omega)
    \\[5pt]
    \lb{S^\dag_\H}{\sy{yield},t}(R,\Omega)\deq
    (\setst{((b[t\mapsto\yield],l),\rho)}{((b,l),\rho)\in R},\Omega)
    \\
    \lb{S^\dag_\H}{\sy{lock}(m),t}(R,\Omega)\deq
    (\setst{((b[t\mapsto\wait(m)],l),\rho)}{((b,l),\rho)\in R},\Omega)
    \\
    \lb{S^\dag_\H}{\sy{unlock}(m),t}(R,\Omega)\deq
    (\setst{((b,l[t\mapsto l(t)\setminus\sset{m}]),\rho)}{((b,l),\rho)\in R},\Omega)
    \\
    \lb{S^\dag_\H}{X\leftarrow\sy{islocked}(m),t}(R,\Omega)\deq\\
    \qquad
    (\setst{((b,l),\rho[X\mapsto 0])}{((b,l),\rho)\in R,\,\fa{t'\in\T}m\notin l(t')}\;\cup\\
    \qquad\;\,
    \setst{((b,l),\rho[X\mapsto 1])}{((b,l),\rho)\in R,\,\ex{t'\in\T}m\in l(t')},\Omega)
    \\[5pt]
    \text{for all other primitive statements }s\in\set{X\leftarrow e, e\bowtie 0?}:\\
    \lb{S^\dag_\H}{s,t}(R,\Omega)\deq\\
    \qquad(\setst{((b,l),\rho')}{\ex{\rho}((b,l),\rho)\in R,\,(R',-)=\lb{S}{s}(\sset{\rho},\Omega),\,\rho'\in R'},\Omega')\\
    \qquad\text{where }(-,\Omega')=\lb{S}{s}(\setst{\rho}{(-,\rho)\in R},\Omega)
    \\[5pt]
    \sched(R,\Omega)\deq(\setst{((b',l'),\rho)}{((b,l),\rho)\in R},\Omega)\\
    \qquad\text{s.t. }\fa{t}\\
    \qquad\quad
    \begin{array}[t]{@{}l}
      \text{if }b(t)=\wait(m)\wedge
      (m\in l(t)\vee
      (\fa{t'}m\notin l(t')\wedge
      \fa{t'>t}b(t')\neq\wait(m)))\\
      \text{then }b'(t)=\ready\,\wedge\,l'(t)=l(t)\cup\sset{m}\\
      \text{else }l'(t)=l(t) \wedge (b'(t)=b(t)\vee (b'(t)=\ready\,\wedge\,b(t)=\yield))
    \end{array}
  \end{array}$
  \caption{Concrete semantics of primitive statements with a scheduler.}
  \label{fig:schedinterleavestatsem}
\end{figure}

The semantics $\lbx{\Pi_\H}{P}\in\D_\H\stackrel{\sqcup_\H}{\longrightarrow}\D_\H$
of a set $P\subseteq\Pi_*$ of parallel control
paths then becomes, similarly to (\ref{eq:interleavepathsem}):
\begin{equation}
  \label{eq:schedinterleavesem}
  \begin{array}{l}
    \lbx{\Pi_\H}{P}(R,\Omega)\deq\\
    \quad\bigsqcup_\H\;\setst{(\lb{S_\H}{s_n,t_n}\circ\cdots\circ\lb{S_\H}{s_1,t_1})(R,\Omega)}{(s_1,t_1)\cdot\ldots\cdot(s_n,t_n)\in P}\\[6pt]
  \end{array}
\end{equation}
and the semantics $\lbp{P_\H}$ of the program is, similarly to 
(\ref{eq:interleavesem}):
\begin{equation}
  \label{eq:schedinterleavesem2}
  \lbp{P_\H}\deq\Omega,\text{ where }(-,\Omega)=\lbx{\Pi_\H}{\pi_*}(\{h_0\}\times\E_0,\,\emptyset)
\end{equation}
where $\pi_*$ is the set of parallel control paths of the program, defined
in (\ref{eq:interleave}), and
$h_0\deq(\lbd{t}\ready,\lbd{t}\emptyset)$ denotes the initial
scheduler state (all the threads are ready and hold no mutex).
As in Sec.~\ref{sec:interleavesem}, 
many parallel control paths in $\pi_*$ are unfeasible, i.e., return an empty
set of environments, some of which are now ruled out by the
$\enabled_t$ function because they do not obey the real-time scheduling
policy or do not ensure the mutual exclusion enforced by locks.
Nevertheless, errors from a feasible prefix of an unfeasible path are
still taken into account.
This includes, in particular, the errors that occur before a dead-lock.

\subsection{Scheduled Weakly Consistent Memory Semantics \texorpdfstring{$\lbp{P'_\H}$}{P'-H}}
\label{sec:schedweaksem}

As was the case for the interleaving semantics without a scheduler 
(Sec.~\ref{sec:interleavesem}),
the scheduled interleaving semantics does not take into account the effect
of a weakly consistent memory.
Recall that a lack of memory consistency can be caused by the underlying
hardware memory model of a multi-processor, by compiler optimisations, or by
non-atomic primitive statements.
While we can disregard the hardware issues when considering mono-processor 
systems (i.e., everywhere in Sec.~\ref{sec:sched} except 
Sec.~\ref{sec:multi})
the other issues remain, and so, we must consider their interaction
with the scheduler.
Thus, we now briefly present a weakly consistent memory semantics for programs 
with a scheduler. 
The interference semantics designed in Secs.~\ref{sec:schedinterfersem} 
and \ref{sec:schedabs} will be sound with respect to this semantics.
\medskip

In addition to restricting the interleaving of threads, synchronization
primitives also have an effect when considering weakly consistent memory 
semantics: they enforce some form of sequential consistency at a coarse 
granularity level.
More precisely, the compiler and processor handle synchronization statements
specially, introducing the necessary flushes into memory and register reloads,
and refraining from optimizing across them.

Recall that the weakly consistent semantics $\lbp{P'_*}$ of 
Sec.~\ref{sec:weaksem} is based on the interleaving semantics $\lbp{P_*}$ 
of Sec.~\ref{sec:interleavesem} applied to transformed threads $\pi'(t)$,
which are obtained by transforming the paths in $\pi(\body_t)$
using elementary path transformations $q\becomes q'$ from 
Def.~\ref{def:transform}.
To take synchronization into account, we use the same definition of
transformed threads $\pi'(t)$, but restrict it to transformations
$q\becomes q'$ that do not contain any synchronization primitive.
For instance, we forbid the application of sub-expression elimination 
(Def.~\ref{def:transform}.(8)) on the following path:
$\sy{lock}(m)\cdot Y\leftarrow e\becomes X\leftarrow e\cdot \sy{lock}(m) \cdot Y\leftarrow X$.
However, if $q$ and $q'$ do not contain any synchronization primitive,
and $q\becomes q'$, then it is
legal to replace $q$ with $q'$ in a path containing synchronization primitives
before and after $q$.
For instance, the transformation 
$\sy{lock}(m)\cdot Y\leftarrow e\becomes \sy{lock}(m) \cdot X\leftarrow e\cdot Y\leftarrow X$
is acceptable.
The scheduled weakly consistent memory semantics is then, based on
(\ref{eq:schedinterleavesem2}):
\begin{equation}
  \label{eq:schedweaksem}
  \lbp{P'_\H}\deq\Omega,\text{ where }(-,\Omega)=\lbx{\Pi_\H}{\pi'_*}(\{h_0\}\times\E_0,\,\emptyset)
\end{equation}
where $\pi'_*$ is defined, as before (\ref{eq:transinterleavesem}),
as the interleavings of control paths from all $\pi'(t),\,t\in\T$.

\subsection{Concrete Scheduled Interference Semantics \texorpdfstring{$\lbp{P_\C}$}{P-C}}
\label{sec:schedinterfersem}

We now provide a structured version $\lbp{P_\C}$ of the scheduled 
interleaving semantics $\lbp{P_\H}$.
Similarly to the interference abstraction $\lbp{P_\I}$ from
Sec.~\ref{sec:interfersem} of the non-scheduled interleaving semantics 
$\lbp{P_*}$, it is based on a notion of 
interference, it is sound with respect to both the interleaving semantics
$\lbp{P_\H}$ and its weakly consistent version $\lbp{P'_\H}$,
but it is not complete with respect to either of them.
The main changes with respect to the interference abstraction $\lbp{P_\I}$ are:
a notion of scheduler configuration (recording
some information about the state of mutexes), a partitioning of interferences
and environments with respect to configurations, and a distinction 
between well synchronized thread communications and data-races.
As our semantics is rather complex, we first present it graphically on examples
before describing it in formal terms.

\subsubsection{Interferences}
In the non-scheduled semantics $\lbp{P_\I}$ (Sec.~\ref{sec:interfersem}),
any interference $(t,X,v)$, i.e., any write by a thread $t$ of a value $v$ into 
a variable $X$, could influence any read from the same variable $X$ in another 
thread $t'\neq t$.
While this is also a sound abstraction of the semantics with
a scheduler, the precision can be improved by refining our notion of 
interference and exploiting mutual exclusion properties enforced by the 
scheduler.
\medskip

\begin{figure}[t]
  \centering
  \begin{tabular}{c}
    \includegraphics[height=3.5cm]{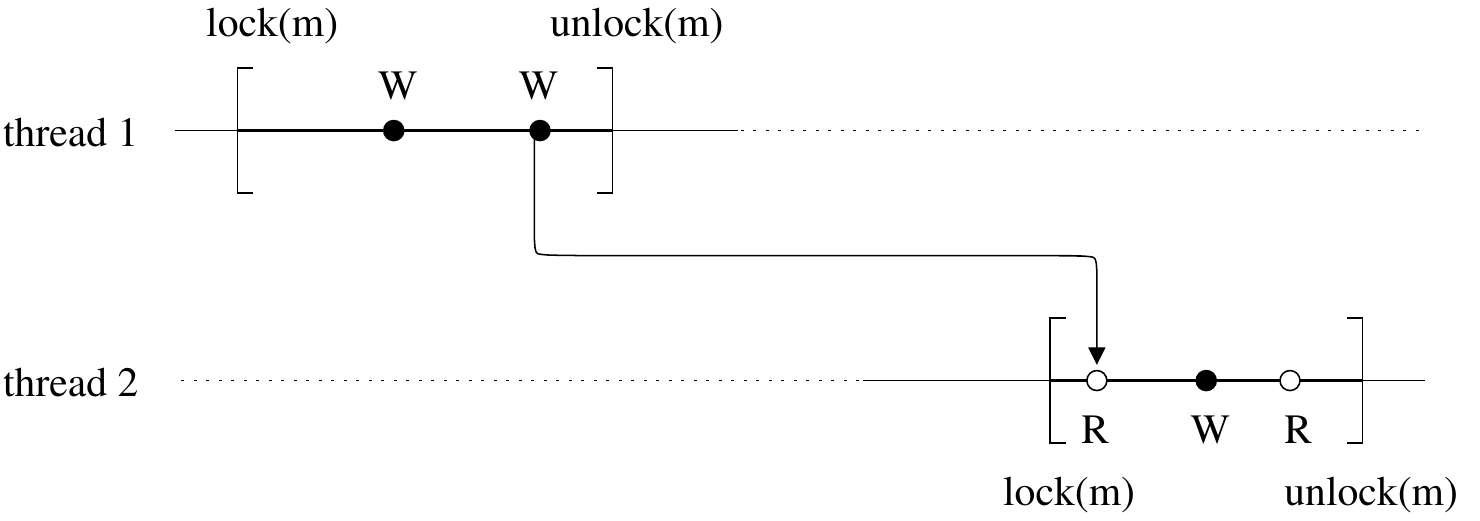}\\
    (a) Well synchronized communication.
    \\\\\\
    \includegraphics[height=3.5cm]{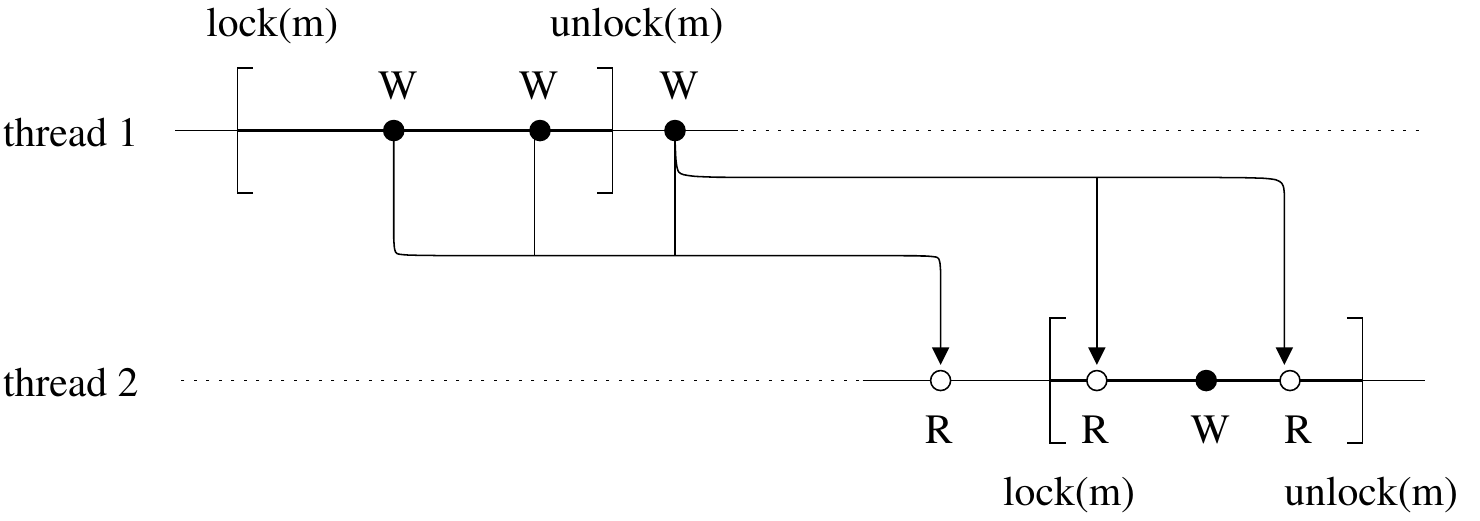}\\
    (b) Weakly consistent communications.
  \end{tabular}
  \caption{Well synchronized versus weakly consistent communications.}
  \label{fig:inter}
\end{figure}

Good programming practice dictates that all read and write accesses to a
given shared variable should be protected by a common mutex.
This is exemplified in Fig.~\ref{fig:inter}.(a) where $W$ and $R$ denote
respectively a write to and a read from a variable $X$, and all reads and 
writes are protected by a mutex $m$.
In this example, thread~1 writes twice to $X$ while holding $m$. Thus, when
thread~2 locks $m$ and reads $X$, it can see the second value written
by thread~1, but never the first one, which is necessarily overwritten before
thread~2 acquires $m$.
Likewise, after thread~2 locks $m$ and overwrites $X$, while it still holds
$m$ it can only read back the value it has written and not any value written 
by thread~1.
Thus, a single interference from thread~1 can effect thread~2, and at only
one read position; we call this read / write pair a 
``well synchronized communication.''
Well synchronized communications are flow-sensitive
(the order of writes and reads matters), and so, differ significantly
from the interferences of Sec.~\ref{sec:interfersem}.
In practice, we model such communications by recording at the point
$\sy{unlock}(m)$ in thread~1 the current value of all the variables that
are modified while $m$ is locked, and import these values in the
environments at the point $\sy{lock}(m)$ in thread~2.

Accesses are not always protected by mutexes, though.
Consider, for instance, the example in Fig.~\ref{fig:inter}.(b), where
$X$ may additionally be modified by thread~1 and read by thread~2 outside
the critical sections defined by mutex $m$.
In addition to the well synchronized communication of Fig.~\ref{fig:inter}.(a),
which is omitted for clarity in Fig.~\ref{fig:inter}.(b),
we consider that a write from thread~1 effects a read from thread~2
if either operation is performed while $m$ is not locked.
These read / write pairs correspond to data-races, and
neither the compiler nor the hardware is required to enforce 
memory consistency. We call these pairs ``weakly consistent communications.''
In practice, these are handled in a way similar to the interferences in
Sec.~\ref{sec:interfersem}: the values thread~1 can write into $X$ are
remembered in a flow-insensitive interference set, and the semantics
of expressions is modified so that, when reading $X$ in thread~2, 
either the thread's value for $X$ or a value from the interference set
is used. 
We also remember the set of mutexes that threads hold during
each read and each write, so that we
can discard communications that cannot occur due to mutual exclusion.
For instance, in Fig.~\ref{fig:inter}.(b), there is no communication of any 
kind between the first write in thread~1 and the second read in thread~2.
The example also shows that well synchronized and weakly consistent 
communications can mix freely: there is no weakly consistent
communication between the second write in thread~1 and the second read in
thread~2 due to mutual exclusion (both threads hold the mutex $m$); however,
there is a well synchronized communication --- shown in 
Fig.~\ref{fig:inter}.(a).

\begin{figure}[t]
  \centering
  \begin{tabular}{c}
    \includegraphics[height=5cm]{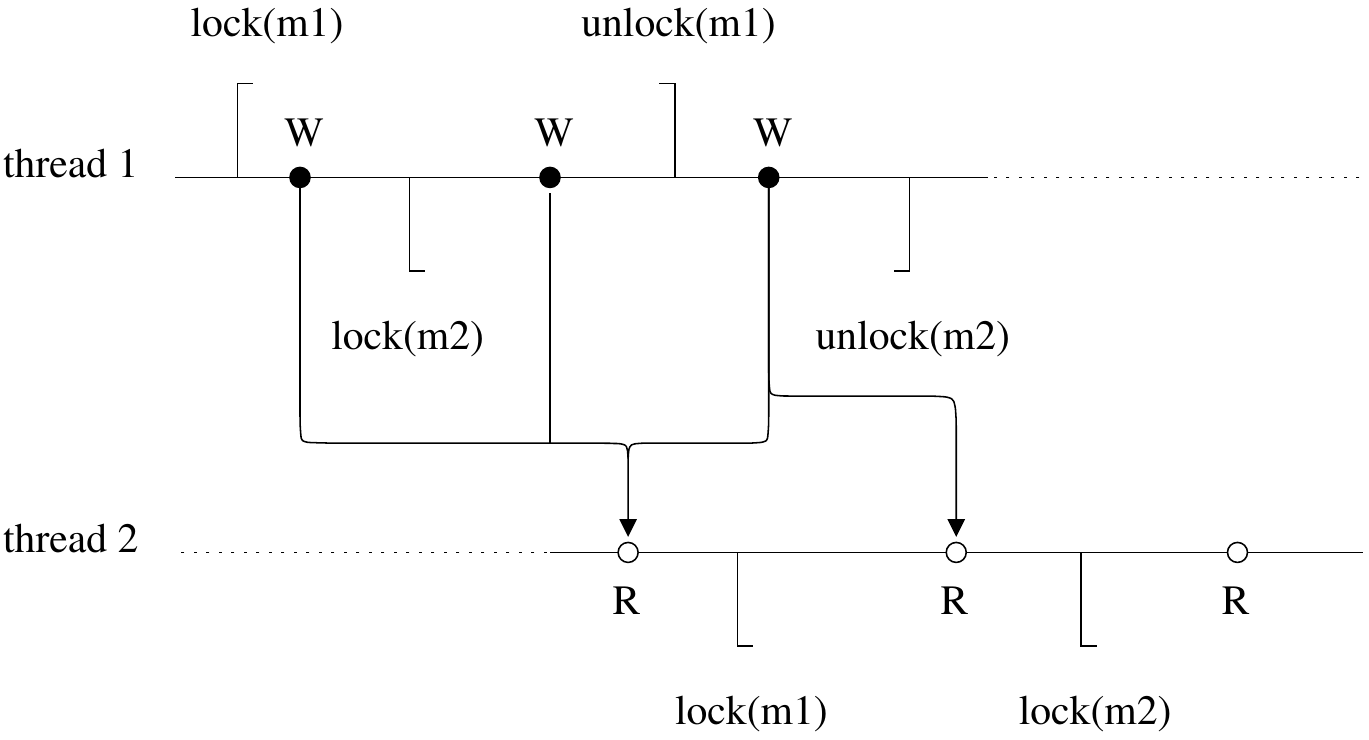}\\
    (a) Weakly consistent communications.
    \\\\\\
    \includegraphics[height=5cm]{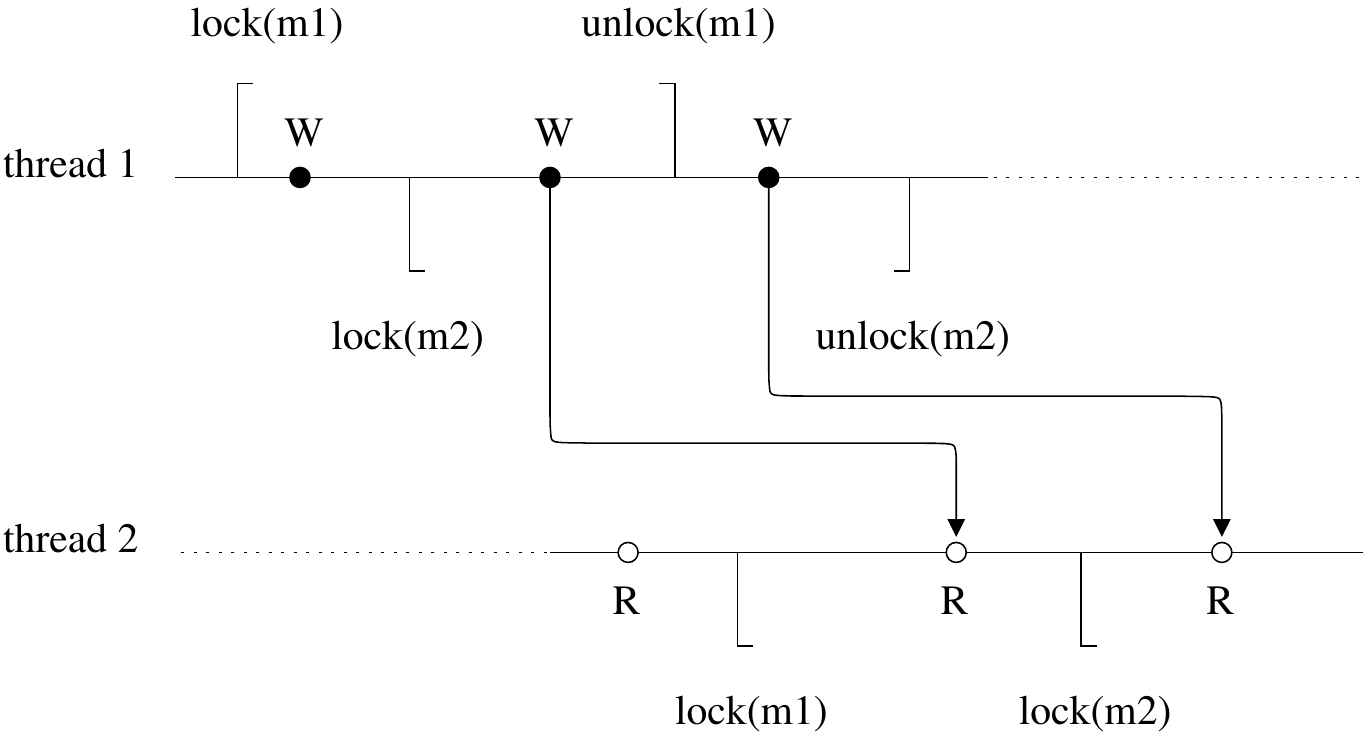}\\
    (b) Well synchronized communications.
    \\\\\\
    \includegraphics[height=5cm]{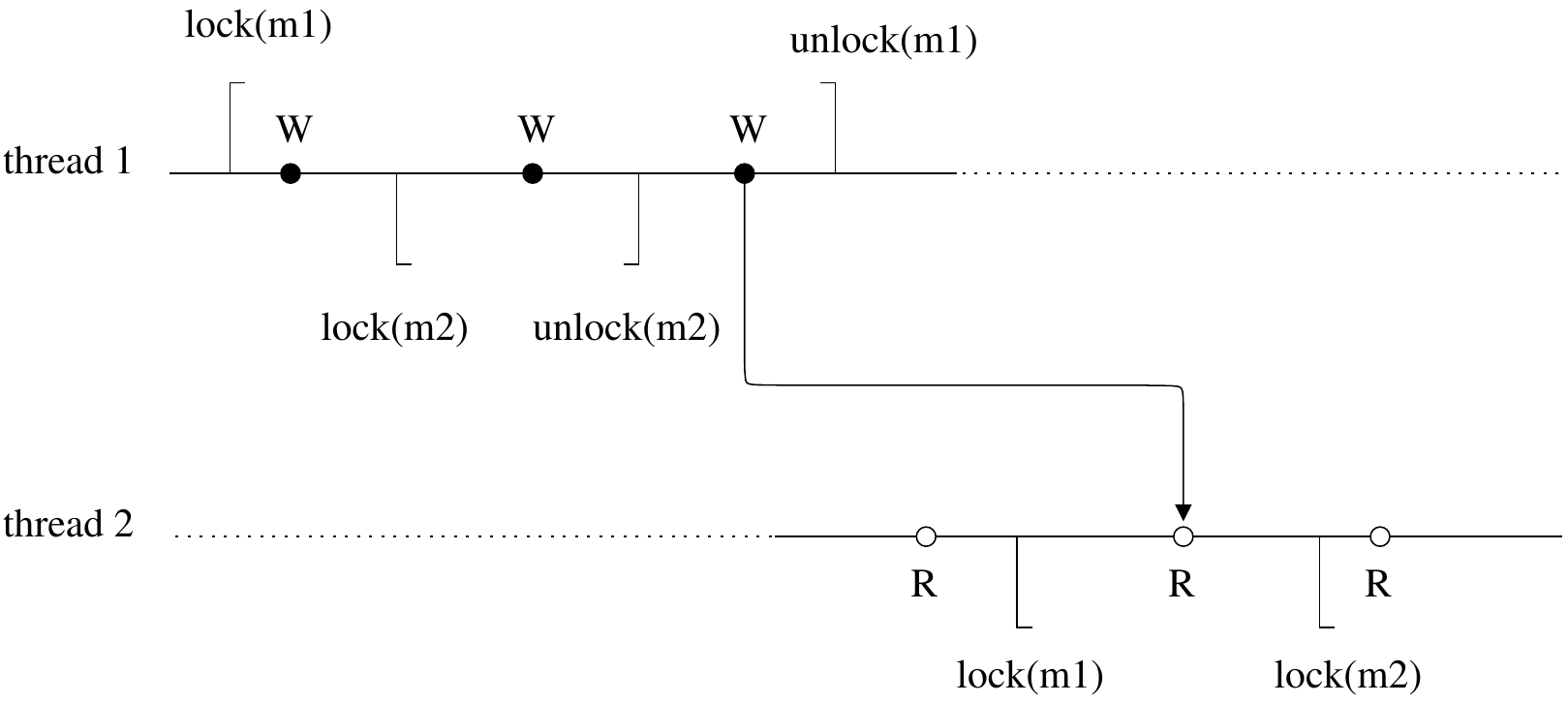}\\
    (c) Well synchronized communications.
  \end{tabular}
  \caption{Well synchronized and weakly consistent communications with two locks.}
  \label{fig:inter2}
\end{figure}

Figure~\ref{fig:inter2} illustrates the communications in the case of several
mutexes: $m1$ and $m2$.
In Fig.~\ref{fig:inter2}.(a), weakly consistent communications only occur 
between write / read pairs when the involved threads have not locked a
common mutex.
For instance, the first write by thread~1 is tagged with the set of locked 
mutexes $\sset{m1}$, and so, can only influence the first read by thread~2
(tagged with $\emptyset$) and not the following two (tagged respectively
with $\sset{m1}$ and $\sset{m1,m2}$).
Likewise, the second write, tagged with $\sset{m1,m2}$, only influences the
first read. However, the third write, tagged with only $\sset{m2}$,
influences the two first reads (thread~2 does not hold the mutex $m2$ there).
In Fig.~\ref{fig:inter2}.(b), well synchronized communications import, 
as before, at a lock of mutex $m1$ (resp.~$m2$) in thread~2, the 
last value written  by thread~1 before unlocking the same 
mutex $m1$ (resp.~$m2$).
The well synchronized communication in Fig.~\ref{fig:inter2}.(c) is 
more interesting.
In that case, thread~1 unlocks $m2$ before $m1$, instead of after.
As expected, when thread~2 locks $m1$, it imports the last (third) value 
written by thread~1, just before unlocking $m1$.
We note, however, that the second write in thread~1 does not influence
thread~2 while thread~2 holds mutex $m1$, as the value is always
over-written by thread~1 before unlocking $m1$.
We model this by importing, when locking a mutex $m$ in thread~2, only
the values written by thread~1 while it does not hold a common mutex 
(in addition to $m$) with thread~2.
Thus, when locking $m2$ while still holding the mutex $m1$, thread
2 does not import the second value written by thread~1
because thread~1 also holds $m1$ during this write.

\subsubsection{Interference partitioning}
To differentiate between well synchronized and weakly consistent communications,
and to avoid considering communications between parts of threads that are in
mutual exclusion, we partition interferences with respect to a
thread-local view of scheduler configurations.
The (finite) set $\C$ of configurations is defined as:
\begin{equation}
\label{eq:schedconf}
\C\deq\P(\M)\times\P(\M)\times\setst{\weak,\sync(m)}{m\in\M}\enspace.
\end{equation}
In a configuration $(l,u,s)\in\C$, the first component $l\subseteq\M$
denotes the exact set of mutexes locked by the thread creating the
interference, which is
useful to decide which reads will be affected by the interference.
The second component $u\subseteq\M$ denotes a set of mutexes that
are known to be locked by no thread (either current or not). This information
is inferred by the semantics of $\sy{islocked}$ statements and can be
exploited to detect extra mutual exclusion properties that further limit
the set of reads affected by an interference
(as in the example in Fig.~\ref{fig:schedul}).
The last component, $s$, allows distinguishing between 
weakly consistent and well synchronized communications:
$\weak$ denotes an interference that generates weakly
consistent communications, while $\sync(m)$ denotes an interference that 
generates
well synchronized communications for critical sections protected by the mutex 
$m$.
These two kinds of interferences are naturally called, respectively,
weakly consistent and well synchronized interferences.
The partitioned domain of interferences is then:
\begin{equation}
  \label{eq:schedinterf}
  \I\deq \T\times\C\times\V\times\R
\end{equation}
which enriches the definition of $\I$ from Sec.~\ref{sec:interfersem} with a 
scheduler configuration in $\C$.
The interference $(t,c,X,v)\in\I$ indicates that the thread $t\in\T$ can write
the value $v\in\R$ into the variable $X\in\V$ and the scheduler is in 
the configuration $c\in\C$ at the time of the write.

\subsubsection{Environment partitioning}
When computing program states in our semantics, environments are also
partitioned with respect to scheduler configurations in order to track
some information on the current state of mutexes.
Thus, our program states associate an environment $\rho\in\E$ and a
configuration in $(l,u,s)\in\C$, where the configuration $(l,u,s)$ indicates 
the set of mutexes $l$ held by the thread in that state, as well as the set 
of mutexes $u$ that are known to be held by no thread; the $s$ component 
is not used and always set by convention to $\weak$.
The semantic domain is now:
\begin{equation}
  \label{eq:schedstate}
  \D_\C\deq\P(\C\times\E)\times\P(\L)\times\P(\I)
\end{equation}
partially ordered by pointwise set inclusion.
We denote by $\sqcup_\C$ the associated pointwise join.
While regular statements (such as assignments and tests) update the 
environment part of each state, synchronization primitives update the
scheduler part of the state.

The use of pairs of environments and scheduler configurations allows
representing relationships between the value of a variable and the state of
a mutex,
which is important for the precise modeling of the $\sy{islocked}$ primitive
in code similar to that of Fig.~\ref{fig:schedul}.
For instance, after the statement $X\leftarrow\sy{islocked}(m)$, all states
$((l,u,s),\rho)$ satisfy $\rho(X)=0 \Longrightarrow m\in u$.
Thus, when the high thread enters the ``then'' branch of the subsequent 
$X=0$ test, we know that $m$ is not locked by any thread and we can disregard
the interferences generated by the low thread while holding $m$.

\subsubsection{Semantics}
\label{sec:schedinterfersemdetail}
We now describe in details the semantics of expressions and statements.
It is presented fully formally in Figs.~\ref{fig:schedinterfexprsem} 
and \ref{fig:schedinterferstatsem}.

\begin{figure}
  \centering
  $\begin{array}{l}
    \underline{\lb{E_\C}{e}:(\T\times\C\times\E\times\P(\I))\rightarrow(\P(\R)\times\P(\L))}
    \\[4pt]
    \lb{E_\C}{X}(t,c,\rho,I)\deq
    (\set{\rho(X)}\cup\setst{v}{\ex{(t',c',X,v)\in I}t\neq t'\wedge\excl(c,c')},\,\emptyset)
    \\[3pt]
    \lb{E_\C}{[c_1,c_2]}(t,c,\rho,I)\deq(\setst{c\in\R}{c_1\leq c\leq c_2},\,\emptyset)
    \\[3pt]
    \lb{E_\C}{-_\ell\,e}(t,c,\rho,I)\deq
    \letin{(V,\Omega)}{\lb{E_\C}{e}(t,c,\rho,I)}
    (\setst{-x}{x\in V},\,\Omega)
    \\[3pt]
    \lb{E_\C}{e_1\diamond_\ell e_2}(t,c,\rho,I)\deq\\
    \qquad\letin{(V_1,\Omega_1)}{\lb{E_\C}{e_1}(t,c,\rho,I)}\\
    \qquad\letin{(V_2,\Omega_2)}{\lb{E_\C}{e_2}(t,c,\rho,I)}\\
    \qquad(\setst{x_1\diamond x_2}{x_1\in V_1,\,x_2\in V_2,\,\diamond\neq /\vee x_2\neq 0},\\
    \qquad\;\Omega_1\cup \Omega_2\cup\setst{\ell}{\diamond=/\wedge 0\in V_2})\\
    \text{where }\diamond\in\set{+,-,\times,/}
    \\
    \\
    \text{where:}\\
    \excl((l,u,s),(l',u',s'))\diff l\cap l'=u\cap l'=u'\cap l=\emptyset\wedge s=s'=\weak
  \end{array}$
  \caption{Concrete scheduled semantics of expressions with interference.}
  \label{fig:schedinterfexprsem}
\end{figure}

The semantics $\lb{E_\C}{e}(t,c,\rho,I)$ of an expression $e$ is presented in 
Fig.~\ref{fig:schedinterfexprsem}.
It is similar to the non-scheduled semantics
$\lb{E_\I}{e}(t,\rho,I)$ of Fig.~\ref{fig:interfexprsem}, except that it
has an extra argument: the current configuration $c\in\C$ 
$(\ref{eq:schedconf})$ of the thread evaluating the expression.
The other arguments are: the thread $t\in\T$ evaluating the expression,
the environment $\rho\in\E$ in which it is evaluated, and a set 
$I\subseteq\I$ $(\ref{eq:schedinterf})$ of interferences from the whole program.
Interferences are applied when reading a variable $\lb{E_\C}{X}$.
Only weakly consistent interferences are handled in expressions --- well
synchronized interferences are handled in the semantics of synchronization
primitives, presented below. Moreover, we consider only interferences 
with configurations 
that are not in mutual exclusion with the current configuration $c$.
Mutual exclusion is enforced by the predicate $\excl$, which states that,
in two scheduler configurations $(l,u,\weak)$ and $(l',u',\weak)$
for distinct threads, no mutex can be
locked by both threads ($l\cap l'=\emptyset$), and no thread
can lock a mutex which is assumed to be free by the other one
($l\cap u'=l'\cap u=\emptyset$).
The semantics of other expression constructs remains the same, 
passing recursively the arguments  $t$, $c$, and $I$ unused and unchanged.

\begin{figure}
  \centering
  $\begin{array}{l}
    \underline{\lb{S_\C}{s,t}:\D_\C\stackrel{\sqcup_\C}{\longrightarrow}\D_\C}
    \\[4pt]
    \lb{S_\C}{X\leftarrow e,\,t}(R,\Omega,I)\deq\\
    \qquad
    (\emptyset,\Omega,I)\sqcup_\C
    \underset{(c,\rho)\in R}{\bigsqcup_\C}\;
    \begin{array}[t]{l}
      \letin{(V,\Omega')}{\lb{E_\C}{e}(t,c,\rho,I)}\\
      (\setst{(c,\rho[X\mapsto v])}{v\in V},\,\Omega',\,\setst{(t,c,X,v)}{v\in V})
      \\[3pt]
    \end{array}
    \\
    \lb{S_\C}{e\bowtie 0?,\,t}(R,\Omega,I)\deq\\
    \qquad
    (\emptyset,\Omega,I)\sqcup_\C
    \underset{(c,\rho)\in R}{\bigsqcup_\C}\;
    \begin{array}[t]{l}
      \letin{(V,\Omega')}{\lb{E_\C}{e}(t,c,\rho,I)}\\
      (\setst{(c,\rho)}{\ex{v\in V}v\bowtie 0},\,\Omega',\,\emptyset)
    \end{array}
    \\[3pt]
    \lb{S_\C}{\sy{if}\;e\bowtie0\;\sy{then}\;s,\,t}(R,\Omega,I)\deq\\
    \qquad
    (\lb{S_\C}{s,t}\circ\lb{S_\C}{e\bowtie0?,\,t})(R,\Omega,I)\sqcup_\C\lb{S_\C}{e\not\bowtie0?,\,t}(R,\Omega,I)
    \\[3pt]
    \lb{S_\C}{\sy{while}\;e\bowtie0\;\sy{do}\;s,\,t}(R,\Omega,I)\deq\\
    \qquad
    \lb{S_\C}{e\not\bowtie0?,\,t}(\lfp\lbd{X}(R,\Omega,I)\sqcup_\C(\lb{S_\C}{s,t}\circ\lb{S_\C}{e\bowtie0?,\,t})X)
    \\[3pt]
    \lb{S_\C}{s_1;\,s_2,\,t}(R,\Omega,I)\deq (\lb{S_\C}{s_2,t}\circ\lb{S_\C}{s_1,t})(R,\Omega,I)
    \\[3pt]
    \lb{S_\C}{\sy{lock}(m),\,t}(R,\Omega,I)\deq\\
    \qquad
    (\setst{((l\cup\sset{m},\emptyset,s),\rho')}{((l,-,s),\rho)\in R,\,
      \rho'\in\funin(t,l,\emptyset,m,\rho,I)},\\
    \qquad
    \;\Omega,\,I\cup\bigcup\,\setst{\funout(t,l,\emptyset,m',\rho,I)}{\ex{u}((l,u,-),\rho)\in R \wedge m'\in u})
    \\[3pt]
    \lb{S_\C}{\sy{unlock}(m),\,t}(R,\Omega,I)\deq\\
    \qquad
    (\setst{((l\setminus\sset{m},u,s),\rho)}{((l,u,s),\rho)\in R},\,\\
    \qquad
    \;\Omega,\,I\cup\bigcup\,\setst{\funout(t,l\setminus\sset{m},u,m,\rho,I)}{((l,u,-),\rho)\in R})
    \\[3pt]
    \lb{S_\C}{\sy{yield},\,t}(R,\Omega,I)\deq\\
    \qquad
    (\setst{((l,\emptyset,s),\rho)}{((l,-,s),\rho)\in R},\,\\
    \qquad
    \;\Omega,\,I\cup\bigcup\,\setst{\funout(t,l,\emptyset,m',\rho,I)}{\ex{u}((l,u,-),\rho)\in R \wedge m'\in u})
    \\[3pt]
    \lb{S_\C}{X\leftarrow\sy{islocked}(m),\,t}(R,\Omega,I)\deq\\
    \qquad\text{if no thread $t'>t$ locks $m$, then:}\\
    \qquad\quad
    (\setst{((l,u\cup \sset{m},s),\rho'[X\mapsto 0])}{((l,u,s),\rho)\in R,\,
      \rho'\in\funin(t,l,u,m,\rho,I)}\,\cup\\
    \qquad\quad
    \;\setst{((l,u\setminus\sset{m},s),\rho[X\mapsto 1])}{((l,u,s),\rho)\in R},\\
    \qquad\quad
    \;\Omega,\,
    I\cup\setst{(t,c,X,v)}{v\in\sset{0,1},\,(c,-)\in R})\\
    \qquad\text{otherwise:}\\
    \qquad\quad
    \lb{S_\C}{X\leftarrow[0,1],\,t}(R,\Omega,I)
    \\
    \\
    \text{where:}
    \\
    \quad
    \funin(t,l,u,m,\rho,I)\deq\\
    \quad\qquad
    \setst{\begin{array}[t]{@{}l}
        \rho'}{\fa{X\in\V}\rho'(X)=\rho(X)\vee
        (\ex{t',l',u'}
        (t',(l',u',\sync(m)),X,\rho'(X))\in I\\
        \wedge \, t\neq t' \wedge l\cap l'=l\cap u'=l'\cap u=\emptyset)}
    \end{array}
    \\[3pt]
    \quad
    \funout(t,l,u,m,\rho,I)\deq\\
    \quad\qquad
    \setst{(t,(l,u,\sync(m)),X,\rho(X))}{\ex{l'}(t,(l',-,\weak),X,-)\in I \wedge m\in l'}
  \end{array}$
  \caption{Concrete scheduled semantics of statements with interference.}
  \label{fig:schedinterferstatsem}
\end{figure}

\medskip

We now turn to the semantics $\lb{S_\C}{s,t}(R,\Omega,I)$ of a statement
$s$ executed by a thread $t$, which is defined in 
Fig.~\ref{fig:schedinterferstatsem}.
It takes as first argument a set $R$ of states which are now pairs
consisting of an environment $\rho\in\E$ and a scheduler configuration 
$c\in\C$, i.e., $R\subseteq \C\times\E$.
The other arguments are, as in the non-scheduled semantics of
Fig.~\ref{fig:interfstatsem}, a set of run-time errors $\Omega\subseteq\L$
to enrich, and a set of interferences $I\subseteq\I$ to use and enrich.
The semantics of assignments and tests in Fig.~\ref{fig:schedinterferstatsem}
is similar to the non-scheduled case (Fig.~\ref{fig:interfstatsem}).
The scheduler configuration associated with each input environment is 
simply passed as argument to the expression semantics $\lbp{E_\C}$ in order to 
select precisely which weakly relational interferences to apply 
(through $\excl$),
but it is otherwise left unmodified in the output.
Additionally, assignments $X\leftarrow e$ generate weakly consistent 
interferences, which store in $I$ the current thread $t$ and the scheduler 
configuration $c$ of its state, in addition to the modified variable $X$ and 
its new value.

The semantics of non-primitive statements remains the same as in previous
semantics by structural induction on the syntax of statements
(e.g., Fig.~\ref{fig:interfstatsem}).

\smallskip

The main point of note is thus the semantics of synchronization primitives.
It updates the scheduler configuration and takes care of
well synchronized interferences.

Let us explain first how the scheduler part $(l,u,s)\in\C$ of a state
$((l,u,s),\rho)\in R$ is updated.
Firstly, the set $l$ of mutexes held by the current thread is updated by
the primitives $\sy{lock}(m)$ and $\sy{unlock}(m)$ by respectively
adding $m$ to and removing $m$ from $l$.
Secondly, the set of mutexes $u$ that are known to be free in the system
is updated by $X\leftarrow \sy{islocked}(m)$.
Generally, no information on the state of the mutex is known {\em a priori}.
Each input state thus spawns two output states: one where $m$ is free
($u\in m$), and one where $m$ is not free ($u\notin m$).
In the first state, $X$ is set to $0$ while, in the second state, it is
set to $1$. As a consequence, although the primitive cannot actually infer 
whether the mutex is free or not, it nevertheless keeps the relationship 
between the value of $X$ and the fact that $m$ is free.
Inferring this relation is sufficient to analyze precisely the code in 
Fig.~\ref{fig:schedul}.
It is important to note that the information in $m\in u$ is transient as,
when a context switch occurs, another thread $t'$ can run and lock $m$,
thus invaliding the assumption by thread $t$ that no thread has locked $m$.
We distinguish two scenarios, depending on whether $t'$ has higher priority
than $t$ or not.
When $t'<t$, the thread $t'$ has lower priority and 
cannot preempt $t$ at an arbitrary point due to the real-time
nature of the scheduler. Instead, $t'$ must wait until $t$ performs a
blocking operation (i.e., calls a $\sy{lock}$ or $\sy{yield}$ primitive)
to get the opportunity to lock $m$.
This case is handled by having all our blocking primitives reset
the $u$ component to $\emptyset$.
When $t'>t$, the thread 
$t'$ can preempt $t$ at arbitrary points, including  just after 
the $\sy{islocked}$ primitive, and so, we can never safely assume 
that $m\in u$.
If this scenario is possible, $X\leftarrow\sy{islocked}(m)$ is 
modeled as $X\leftarrow[0,1]$, without updating $u$.
To decide which transfer function to use for $\sy{islocked}$, we need to
know the set of all mutexes than can be locked by each thread.
It is quite easy to enrich our semantics to track this information but, as it
is cumbersome, we did not include this in Fig.~\ref{fig:schedinterferstatsem} 
--- one way is to add
a new component $M:\T\rightarrow\P(\M)$ to the domain $\I$ of interferences, 
in which we remember
the set of arguments $m$ of each $\sy{lock}(m)$ encountered by each thread;
then, we check that $\not\ex{t'>t}m\in M(t')$ before applying the
precise transfer function for $X\leftarrow\sy{islocked}(m)$ in thread $t$.

We now discuss how synchronization primitives handle well synchronized
interferences.
We use two auxiliary functions, $\funin(t,l,u,m,\rho,I)$ and 
$\funout(t,l,u,m,\rho,I)$, that model respectively entering and exiting a 
critical section protected by a mutex $m\in\M$ in a thread $t\in\T$.
The arguments $l,u\subseteq \M$ reflect the scheduler configuration when
the primitive is called, i.e., they are respectively the set of mutexes
held by thread $t$ and those assumed to be free in the system.
The function $\funout(t,l,u,m,\rho,I)$ {\em collects\/}
a set of well synchronized 
interferences from an environment $\rho\in\E$.
These are constructed from the current value $\rho(X)$ of the variables 
$X$ that have been modified while the mutex $m$ was held.
Such information can be tracked precisely in the semantics by adding
another component in $\C\rightarrow\P(\V)$ to our program states $R$
but, for the sake of simplicity, the semantics we present simply extracts
this information from the interferences in $I$: we consider all the 
variables that have some weakly interference by thread $t$ in a configuration
where it holds $m$ ($m\in l$).
This may actually over-approximate the set of variables we seek as it 
includes 
variables that have been modified in previous critical sections protected by 
the same mutex $m$, but not in the current critical section.\footnote{Our prototype performs
the same over-approximation for the sake of keeping the analysis
simple, and we did not find any practical occurrence where this resulted
in a loss of precision. We explain this by remaking that critical sections 
delimited by the same mutex tend to protect the same set of modified 
variables.}
Given a variable $X$, the interference we store is then
$(t,(l,u,\sync(m)),X,\rho(X))$.
The function $\funin(t,l,u,m,\rho,I)$ {\em applies\/} well synchronized 
interferences from $I$ to an environment $\rho$: it returns all the
environments $\rho'$ that can be obtained from $\rho$ by setting one or 
several variables to their interference value.
It only considers well synchronized interferences with configuration
$\sync(m)$ and from threads $t'\neq t$.
Moreover, it uses a test similar to that of $\excl$ to avoid applying 
interferences that cannot occur due to mutual exclusion, by comparing the 
current state of mutexes ($l$ and $u$) to their state when the interference
was stored.

The function pair $\funin$ / $\funout$ is actually used to implement 
{\em two\/} kinds of critical sections.
A first kind stems from the use of $\sy{lock}(m)$ and $\sy{unlock}(m)$ 
statements, which naturally delimit a critical section protected by $m$.
Additionally, whenever a mutex $m$ is added to the $u$ scheduler component
by a primitive $X\leftarrow\sy{islocked}(m)$, we also enter a critical
section protected by $m$. Thus, $\funin$ is called
for mutex $m$, and $\excl$ ensures that weakly synchronized interferences 
where $m$ is locked are no longer applied.
Such critical sections end when $m$ leaves $u$, that is, whenever the thread
executes a blocking primitive: $\sy{lock}$ or $\sy{yield}$.
These primitives call $\funout$ for every mutex currently in $u$,
and reset $u$ to $\emptyset$ in the program state.

\medskip

Finally, we turn to the semantics $\lbp{P_\C}$ of a program, which has the 
same fixpoint form as $\lbp{P_\I}$ (\ref{eq:intersem}):
\begin{equation}
  \label{eq:interschedsem}
  \begin{array}{l}
    \lbp{P_\C}\deq\Omega,\text{ where }
    (\Omega,-)\deq
    \lfp\lbd{(\Omega,I)}\\
      \qquad
      \bigsqcup_{t\in\T}\;
      \letin{(-,\Omega',I')}{\lb{S_\C}{\body_t,\,t}(\{c_0\}\times\E_0,\Omega,I)}
      (\Omega',I')
  \end{array}
\end{equation}
where the initial configuration is $c_0\deq(\emptyset,\emptyset,\weak)\in\C$.
This semantics is sound with respect to that of Secs.~\ref{sec:schedinterleavesem}--\ref{sec:schedweaksem}:
\begin{thm}
  \label{thm:soundschedinterfer}
  $\lbp{P_\H}\subseteq\lbp{P_\C}$ and
  $\lbp{P'_\H}\subseteq\lbp{P_\C}.$
\end{thm}
\proof In Appendix~\ref{proof:soundschedinterfer}.\qed

\subsubsection{Multiprocessors and non-real-time systems}
\label{sec:multi}
The only part of our semantics that exploits the fact that only one
thread can execute at a given time is the semantics of
$X\leftarrow\sy{islocked}(m)$. It assumes that, after the current thread has
performed the test, the state of the mutex $m$ cannot change until the current 
thread calls a blocking primitive ($\sy{lock}$ or $\sy{yield}$) --- unless 
some higher priority thread can also lock the mutex $m$.
Thus, in order to obtain a semantics that is also sound for truly parallel 
or non-real-time systems, it is sufficient to interpret all statements
$X\leftarrow\sy{islocked}(m)$ as $X\leftarrow[0,1]$.

While more general, this semantics is less precise when analyzing
a system that is known to be mono-processor and real-time.
For instance, this semantics cannot prove that the two threads in 
Fig.~\ref{fig:schedul} are in mutual exclusion and that, as a result,
$T=0$ at the end of the program. 
It finds instead $T\in\set{-1,0,1}$, which is less precise.
As our target application (Sec.~\ref{sec:result}) is mono-processor
and real-time, we will not discuss this more general but less precise
semantics further.

\subsubsection{Detecting data-races}

\begin{figure}
  \centering
  $\begin{array}{l}
    \underline{\lb{E_\race}{e}:(\T\times\C\times\P(\I))\rightarrow\P(\T\times\T\times\V)}
    \\[4pt]
    \lb{E_\race}{X}(t,c,I)\deq
    \setst{(t,t',X)}{\ex{c'}(t',c',X,-)\in I\wedge t\neq t' \wedge \excl(c,c')}\\[3pt]
    \lb{E_\race}{[c_1,c_2]}(t,c,I)\deq\emptyset
    \\[3pt]
    \lb{E_\race}{-_\ell\,e}(t,c,I)\deq\lb{E_\race}{e}(t,c,I)
    \\[3pt]
    \lb{E_\race}{e_1\diamond_\ell e_2}(t,c,I)\deq
    \lb{E_\race}{e_1}(t,c,I)\cup\lb{E_\race}{e_2}(t,c,I)\\
    \text{where }\diamond\in\set{+,-,\times,/}
  \end{array}$
  \caption{Read / write data-race detection.}
  \label{fig:racesem}
\end{figure}

In our semantics, data-races silently cause weakly consistent interferences
but are otherwise not reported.
It is easy to modify the semantics to output them.
Write / write data-races can be directly extracted from the computed set of
interferences $I$ gathered by the least fixpoint in (\ref{eq:interschedsem})
as follows:
\begin{equation*}
  \setst{(t,t',X)\in\T\times\T\times\V}{\ex{c,c'}(t,c,X,-)\in I\wedge (t',c',X,-)\in I\wedge t\neq t' \wedge\excl(c,c')}
\end{equation*}
is a set where each element $(t,t',X)$ indicates that threads
$t$ and $t'$ may both write into $X$ at the same time.
Read / write data-races cannot be similarly extracted from $I$ as the set of
interferences does not remember which variables are read from, 
only which ones are written to.
A simple solution is to instrument the semantics of expressions so that,
during expression evaluation, it gathers the set of read variables that
are affected by an interference.
This is performed, for instance, by $\lbp{E_\race}$ presented in
Fig.~\ref{fig:racesem}.
This function has the same arguments as $\lbp{E_\C}$, except that no environment
$\rho$ is needed, and it outputs a set of data-races $(t,t',X)$ instead
of environments and errors.

\subsubsection{Precision}
The interference abstraction we use in $\lbp{P_\C}$ is sound be not complete
with respect to the interleaving-based semantics $\lbp{P_\H}$.
In addition to the incompleteness already discussed in 
Sec.~\ref{sec:interfersem}, some loss of precision comes from the handling 
of well synchronized accesses.
A main limitation is that such accesses are handled in a
non-relational way, hence $\lbp{P_\C}$ cannot represent relations enforced
at the boundary of critical sections but broken within, while
$\lbp{P_\H}$ can.
For instance, in Fig.~\ref{fig:schedul}, we cannot prove that
$Y=Z$ holds outside critical sections, but only that $Y,Z\in\sset{1,2}$.
This shows in particular that even programs without data-races have 
behaviors  in $\lbp{P_\C}$ outside the sequentially consistent ones.
However, we can prove that the assignment into $T$ is free from interference,
and so, that $T=0$.
By contrast, the interference semantics $\lbp{P_\I}$ of
Sec.~\ref{sec:interfersem} ignores synchronization and would output 
$T\in\set{-1,0,1}$, which is less precise.

\begin{figure}[t]
  \centering
  \begin{tabular}{l@{\quad}|@{\quad}l}
    \multicolumn{2}{c}{$\E_0:X=Y=5$}\\[4pt]
    \quad\underline{thread $t_1$}\quad & 
    \quad\underline{thread $t_2$}\quad\\[7pt]
    $\sy{while}\ 0=0\ \sy{do}$ & $\sy{while}\ 0=0\ \sy{do}$\\
    \quad $\sy{lock}(m);$ & \quad $\sy{lock}(m);$\\
    \quad $\sy{if}\ X >0\ \sy{then}$ & \quad $\sy{if}\ X <10\ \sy{then}$\\
    \qquad $X\leftarrow X-1;$ & \qquad $X\leftarrow X+1;$\\
    \qquad $Y\leftarrow Y-1;$ & \qquad $Y\leftarrow Y+1;$\\
    \quad $\sy{unlock}(m)$ & \quad $\sy{unlock}(m)$\\
    \end{tabular}
  \caption{Imprecisely analyzed program due to the lack of relational lock invariant.}
  \label{fig:imprecise1}
\end{figure}

\begin{figure}[t]
  \centering
  \begin{tabular}{l@{\quad}|@{\quad}l}
    \multicolumn{2}{c}{$\E_0:X=Y=0$}\\[4pt]
    \quad\underline{high thread}\quad & 
    \quad\underline{low thread}\quad\\[7pt]
    $X\leftarrow 1;$ & $B\leftarrow 1/X;$\\
    $A\leftarrow 1/(Y-1);$ & $Y\leftarrow 1$\\
    $\sy{yield}$\\
    \end{tabular}
  \caption{Imprecisely analyzed program due to the lack of inter-thread flow-sensitivity.}
  \label{fig:imprecise2}
\end{figure}

Figure~\ref{fig:imprecise1} presents another example where the lack of
relational interference results in a loss of precision.
This example implements an abstract producer / consumer system, where
a variable $X$ counts the number of resources, thread $t_1$ consumes
resources ($X\leftarrow X-1$) if available ($X>0$), and thread $t_2$
generates resources ($X\leftarrow X+1$) if there is still room for
resources ($X<10$).
Our interference semantics can prove that $X$ is always bounded in $[0,10]$.
However, it cannot provide an upper bound on the variable $Y$.
Actually, $Y$ is also bounded by $[0,10]$ as it mirrors $X$.
Proving this would require inferring a relational lock invariant:
$X=Y$.

Finally, Fig.~\ref{fig:imprecise2} presents an example where the lack
of inter-thread flow sensitivity results in a loss of precision.
In this example, the high priority thread always executes first,
until it reaches $\sy{yield}$, at which point it allows the lower priority
thread to execute.
To prove that the expression $1/(Y-1)$ does not perform an error, it is
necessary to prove that it is executed before the low thread stores $1$
into $Y$.
Likewise, to prove that the expression $1/X$ does not perform an error
in the low thread, it is necessary to prove that it is executed after the 
high thread stores $1$ into $X$. 
With respect to flow sensitivity,
our semantics is only able to express that an event is performed before
another one within the same thread (intra-thread flow sensitivity) and that 
a thread communication between a pair of locations cannot occur 
(mutual exclusion), but it is not
able to express that an event in a thread is performed before another
one in another thread (inter-thread flow sensitivity).

\subsection{Abstract Scheduled Interference Semantics \texorpdfstring{$\lbp{P_\C^\s}$}{P\#-C}}
\label{sec:schedabs}

We now abstract the interference semantics with scheduler $\lbp{P_\C}$ from the
preceding section in order to construct an effective static analyzer.
We reuse the ideas from the abstraction $\lbp{P^\s_\I}$ of $\lbp{P_\I}$
in Sec.~\ref{sec:sharedabs}.
The main difference is that we track precisely scheduler configurations in 
$\C$ (\ref{eq:schedconf}), and we partition abstract environments and 
interferences with respect to them.

\medskip

\begin{figure}
  \centering
  $\begin{array}{l}
    \I^\s\deq(\T\times \C\times\V)\rightarrow\RR^\s
    \\[3pt]
    \gamma_\I:\I^\s\rightarrow\P(\I)\\
    \qquad\text{s.t. }\gamma_\I(I^\s)\deq\setst{(t,c,X,v)}{t\in\T,\,c\in\C,\,X\in\V,\,v\in\gamma_\RR(I^\s(t,c,X))}
    \\[3pt]
    \bot^\s_\I\deq\lbd{(t,c,X)}\bot^\s_\RR
    \\[3pt]
    I_1^\s\cup^\s_\I I_2^\s\deq\lbd{(t,c,X)}I_1^\s(t,c,X)\cup^\s_\RR I_2^\s(t,c,X)
    \\[3pt]
    I_1^\s\widen_\I I_2^\s\deq\lbd{(t,c,X)}I_1^\s(t,c,X)\widen_\RR I_2^\s(t,c,X)
  \end{array}$
  \caption{Abstract domain of scheduled interferences $\I^\s$, derived from $\RR^\s$.}
  \label{fig:absinterfscheddomain}
\end{figure}

\begin{figure}
  \centering
  $\begin{array}{l}
    \D^\s_\C\deq(\C\rightarrow\E^\s)\times\P(\L)\times\I^\s
    \\[3pt]
    \gamma:\D^\s_\C\rightarrow\D_\C\\
    \qquad\text{s.t. }\gamma(R^\s,\Omega,I^\s) \deq 
    (\setst{(c,\rho)}{c\in\C,\,\rho\in\gamma_\E(R^\s(c))},\,\Omega,\,\gamma_\I(I^\s))
    \\[3pt]
    (R^\s_1,\Omega_1,I^\s_1)\cup^\s (R^\s_2,\Omega_2,I^\s_2)\deq
    (\lbd{c}R^\s_1(c)\cup^\s_\E R^\s_2(c),\Omega_1\cup\Omega_2,I^\s_1 \cup^\s_\I I^\s_2)
    \\[3pt]
    (R^\s_1,\Omega_1,I^\s_1)\widen (R^\s_2,\Omega_2,I^\s_2)\deq
    (\lbd{c}R^\s_1(c)\widen_\E R^\s_2(c),\Omega_1\cup\Omega_2,I^\s_1 \widen_\I I^\s_2)
  \end{array}$
  \caption{Abstract domain of statements $\D^\s_\C$, derived from $\E^\s$ and $\I^\s$.}
  \label{fig:absschedomain}
\end{figure}

As in Sec.~\ref{sec:sharedabs}, we assume that an abstract domain $\E^\s$ of
environment sets $\P(\E)$ is given 
(with signature in Fig.~\ref{fig:absdomain}), 
as well as an abstract domain $\RR^\s$ of real sets $\P(\R)$
(with signature in Fig.~\ref{fig:absrealdomain}).
The abstract domain of interferences $\I^\s$, abstracting $\I$ 
(\ref{eq:schedinterf}), is obtained by partitioning
$\RR^\s$ with respect to $\T$ and $\V$, similarly to the interference
domain of Fig.~\ref{fig:absinterfdomain}, but also $\C$, as shown in
Fig.~\ref{fig:absinterfscheddomain}.
As $\V$, $\T$, and $\C$ are all finite, a map from $\T\times \C\times\V$ to
$\RR^\s$ can indeed be represented in memory, and the join $\cup^\s_\I$ and
widening $\widen_\I$ can be computed pointwise.
Moreover, abstract environments $\E^\s$ are also partitioned with respect to
$\C$.
Hence, the abstract semantic domain $\D^\s_\C$ abstracting
$\D_\C$ $(\ref{eq:schedstate})$ becomes:
\begin{equation}
  \D^\s_\C\deq(\C\rightarrow\E^\s)\times\P(\L)\times\I^\s\enspace.
\end{equation}
It is presented in Fig.~\ref{fig:absschedomain}.

\begin{figure}
  \centering
  $\begin{array}{l}
    \underline{\lb{S_\C^\s}{s,t}:\D^\s_\C\rightarrow\D^\s_\C}
    \\[4pt]
    \lb{S_\C^\s}{X\leftarrow e,\,t}(R^\s,\Omega,I^\s)\deq\\
    \qquad
    \letin{\fa{c\in\C}(R^\s_c,\Omega_c)}{\lb{S^\s}{X\leftarrow\apply(t,c,R^\s,I^\s,e)}(R^\s(c),\Omega)}\\
    \qquad
    (\lbd{c}{R^\s_c},\;\bigcup_{c\in\C}\,\Omega_c,\;I^\s[\fa{c\in\C}(t,c,X)\mapsto I^\s(t,c,X)\cup^\s_\RR\get(X,R^\s_c)])
    \\[3pt]
    \lb{S_\C^\s}{e\bowtie 0?,\,t}(R^\s,\Omega,I^\s)\deq\\
    \qquad
    \letin{\fa{c\in\C}(R^\s_c,\Omega_c)}{\lb{S^\s}{\apply(t,c,R^\s,I^\s,e)\bowtie0?}(R^\s(c),\Omega)}\\
    \qquad
    (\lbd{c}{R^\s_c},\;\bigcup_{c\in\C}\;\Omega_c,\,I^\s)
    \\[3pt]
    \lb{S_\C^\s}{\sy{if}\;e\bowtie0\;\sy{then}\;s,\,t} (R^\s,\Omega,I^\s)\deq\\
    \qquad
    (\lb{S_\C^\s}{s,t}\circ\lb{S_\C^\s}{e\bowtie0?,\,t})(R^\s,\Omega,I^\s)\;\cup^\s\;\lb{S_\C^\s}{e\not\bowtie0?,\,t} (R^\s,\Omega,I^\s)
    \\[3pt]
    \lb{S_\C^\s}{\sy{while}\;e\bowtie0\;\sy{do}\;s,\,t} (R^\s,\Omega,I^\s)\deq\\
    \qquad
    \lb{S_\C^\s}{e\not\bowtie0?,\,t}(\lim\lbd{X^\s}X^\s\widen((R^\s,\Omega,I^\s)\;\cup^\s\;(\lb{S_\C^\s}{s,\,t}\circ\lb{S_\C^\s}{e\bowtie0?,\,t})X^\s))
    \\[3pt]
    \lb{S_\C^\s}{s_1;\,s_2,\,t} (R^\s,\Omega,I^\s)\deq(\lb{S_\C^\s}{s_2,t}\circ\lb{S_\C^\s}{s_1,t})(R^\s,\Omega,I^\s)
    \\
    \\
    \text{where: }\\
    \quad
    \apply(t,c,R^\s,I^\s,e)\deq\\
    \quad\qquad
    \letin{\fa{Y\in\V}V_Y^\s}{\cup^\s_\RR\;\setst{I^\s(t',c',Y)}{t'\neq t\wedge \excl(c,c')}}\\[3pt]
    \quad\qquad
    \letin{\fa{Y\in\V}e_Y}{
      \begin{cases}
        Y & \text{if $V_Y^\s=\bot^\s_\RR$}\\
        \asexpr(V_Y^\s\cup^\s_\RR\get(Y,R^\s(c))) & \text{if $V_Y^\s\neq\bot^\s_\RR$}
      \end{cases}}\\
    \quad\qquad
    e[\fa{Y\in\V}Y\mapsto e_Y]
  \end{array}$
  \caption{Abstract scheduled semantics of statements with interference.}
  \label{fig:absschedstat}
\end{figure}

 \begin{figure}
  \centering
  $\begin{array}{l}
    \lb{S^\s_\C}{\sy{lock}(m),\,t}(R^\s,\Omega,I^\s)\deq\\\quad
    (\lbd{(l,u,s)}
    \bigcup^\s_\E
    \begin{array}[t]{l}
      \{\,\funin^\s(t,l',\emptyset,m,R^\s(l',u',s),I^\s)\\
      |\;l=l'\cup\sset{m} \wedge u=\emptyset \wedge u'\subseteq\M \wedge s=\weak\,\},
    \end{array}
    \\\quad
    \;\Omega,
    \;I^\s\cup^\s_\I \bigcup^\s_\I\;
    \setst{\funout^\s(t,l,\emptyset,m',R^\s(l,u,s),I^\s)}
    {l,u\subseteq \M \wedge m'\in u \wedge s=\weak})
    \\[3pt]
    \lb{S^\s_\C}{\sy{unlock}(m),\,t}(R^\s,\Omega,I^\s)\deq\\\quad
    (\lbd{(l,u,s)}
    \bigcup^\s_\E\;
    \setst{R^\s(l',u',s)}
    {l=l'\setminus\sset{m}\wedge u=u' \wedge s=\weak},
    \\\quad
    \;\Omega,
    \;I^\s\cup^\s_\I \bigcup^\s_\I\;
    \setst{\funout^\s(t,l\setminus\sset{m},u,m,R^\s(l,u,s),I^\s)}
    {l,u\subseteq \M \wedge s=\weak})
    \\[3pt]
    \lb{S^\s_\C}{\sy{yield},\,t}(R^\s,\Omega,I^\s)\deq\\\quad
    (\lbd{(l,u,s)}
    \bigcup^\s_\E\;
    \setst{R^\s(l',u',s)}
    {l=l'\wedge u=\emptyset \wedge u'\subseteq\M \wedge s=\weak},
    \\\quad
    \;\Omega,
    \;I^\s\cup^\s_\I \bigcup^\s_\I\;
    \setst{\funout^\s(t,l,\emptyset,m',R^\s(l,u,s),I^\s)}
    {l,u\subseteq \M \wedge m'\in u \wedge s=\weak})
    \\[3pt]
    \lb{S^\s_\C}{X\leftarrow\sy{islocked}(m),\,t}(R^\s,\Omega,I^\s)\deq\\
    \quad\letin{(R^\s{}',-,I^\s{}')}
    {\lb{S^\s_\C}{X\leftarrow[0,1],\,t}(R^\s,\Omega,I^\s)}\\
    \quad \text{if no thread $t'>t$ locks $m$, then:}\\
    \qquad
    (\lbd{(l,u,s)}
    \bigcup^\s_\E
    \begin{array}[t]{l}
      \{\;\letin{(V^\s,-)}{
        \\\qquad
        \begin{cases}
          \lb{S^\s}{X\leftarrow 0}(\funin^\s(t,l',u',m,R^\s(l',u',s),I^\s),\emptyset)
          & \text{if }m\in u
          \\
          \lb{S^\s}{X\leftarrow 1}(R^\s(l',u',s),\emptyset)
          & \text{if }m\notin u
        \end{cases}\\\;\;
      }
      V^\s
      \\
      \;| \;l=l'\wedge u\setminus\sset{m}=u'\setminus\sset{m} \wedge s=\weak
      \;\},
    \end{array}
    \\
    \qquad\;\Omega,I^\s{}')\\
    \quad\text{otherwise: }\\
    \qquad(R^\s{}',\Omega,I^\s{}')
    \\
    \\
    \text{where:}
    \\
    \quad
    \funin^\s(t,l,u,m,V^\s,I^\s)\deq\\\quad\quad
    V^\s \cup^\s_\E \bigcup^\s_\E
    \begin{array}[t]{l}
      \{ 
      \begin{array}[t]{l}
        \letin{X^\s}{I^\s(t',(l',u',\sync(m)),X)}\\
        \letin{(V^\s{}',-)}{\lb{S^\s}{X\leftarrow \asexpr(X^\s)}(V^\s,\emptyset)}\\
        V^\s{}'
      \end{array}
      \\
      \;|\;X\in \V \wedge t\neq t' \wedge l\cap l'=l\cap u'=l'\cap u =\emptyset
      \;\}
      \\[3pt]
    \end{array}
    \\
    \quad
    \funout^\s(t,l,u,m,V^\s,I^\s)\deq\\\qquad
    \lbd{(t',c,X)}
    \begin{cases}
      \get(X,V^\s)& 
      \text{if } t=t' \wedge c=(l,u,\sync(m)),\\
      & \ex{c'=(l',-,\weak)}m\in l' \wedge I^\s(t,c',X)\neq\bot^\s_\RR\\
      \bot^\s_\RR & 
      \text{otherwise}
    \end{cases}
  \end{array}$
  \caption{Abstract scheduled semantics with interference of synchronization primitives.}
  \label{fig:absschedstat2}
\end{figure}

Sound abstract transfer functions $\lbp{S^\s_\C}$, derived from those in 
$\E^\s$ ($\lbp{S^\s}$),
are presented in Figs.~\ref{fig:absschedstat}--\ref{fig:absschedstat2}.

Assignments and tests in Fig.~\ref{fig:absschedstat} 
are very similar to the non-scheduled case $\lbp{S^\s_\I}$ 
(Fig.~\ref{fig:absinterfstat}) with two differences.
Firstly, $\lbp{S^\s}$ is applied pointwise to each abstract environment
$R^\s(c)\in\E^\s$, $c\in\C$.
New interferences due to assignments are also considered pointwise.
Secondly, the $\apply$ function now takes as extra argument a configuration $c$,
and then only considers interferences from configurations $c'$ not in mutual
exclusion with $c$. This is defined through the same function $\excl$ we used in the
concrete semantics (Fig.~\ref{fig:schedinterfexprsem}).
The semantics of non-primitive statements is the same as for previous
semantics, by structural induction on the syntax of statements.

The semantics of synchronization primitives is presented in
Fig.~\ref{fig:absschedstat2}.
It uses the functions $\funin^\s$ and $\funout^\s$ which abstract, 
respectively, the functions $\funin$ and $\funout$ presented in
Fig.~\ref{fig:schedinterferstatsem}.
As their concrete versions, $\funin^\s$ and $\funout^\s$ take as arguments a 
current thread $t\in\T$, a mutex $m\in\M$ protecting a critical section, 
and sets of mutexes $l,u\subseteq\M$ describing the current scheduling 
configuration.
Moreover, they take as arguments
an abstract set of interferences $I^\s\in\I^\s$ instead
of a concrete set, and an abstract set of environments $V^\s\in\E^\s$ instead
of a single concrete one.
The function $\funout^\s$ uses $\get$ (Fig.~\ref{fig:absrealdomain})
to extract abstract sets of 
variable values from abstract environments and construct new abstract 
well synchronized interferences.
The function $\funin^\s$ applies these interferences to an abstract 
environment by converting them to an expression (using $\asexpr$) and
updating the value of variables (using an assignment in $\lbp{S^\s}$).
Additionally, the semantics of synchronization primitives models updating
the scheduler configuration from $c'=(l',u',\weak)$ to $c=(l,u,\weak)$
by moving abstract environments from $R^\s(c')$ into $R^\s(c)$;
when partitions are collapsed, all the abstract environments mapped to the
same configuration $c$ are merged into $R^\s(c)$ using $\cup^\s_\E$.
Finally, the abstract analysis $\lbp{P_\C^\s}$ computes a fixpoint with
widening over abstract interferences, which is similar to 
(\ref{eq:absintersem}):
\begin{equation}
  \label{eq:absschedintersem}
  \begin{array}{l}
    \lbp{P_\C^\s}\deq\Omega,\text{ where }
    (\Omega,-)\deq\\
    \quad
    \lim\lbd{(\Omega,I^\s)}
      \letin{\fa{t\in\T}(-,\Omega'_t,I^\s_t{}')}{\lb{S_\C^\s}{\body_t,\,t}(R^\s_0,\Omega,I^\s)}\\
      \qquad
      (\bigcup\;\setst{\Omega'_t}{t\in\T},\;
      I^\s\,\widen_\I\,\bigcup^\s_\I\;\setst{I^\s_t{}'}{t\in\T})
  \end{array}
\end{equation}
where the partitioned initial abstract environment
$R^\s_0\in\C\rightarrow \E^\s$ is defined as:
\begin{equation*}
  R^\s_0\deq\lbd{c}
  \begin{cases}
    \E^\s_0 & \text{if }c=(\emptyset,\emptyset,\weak)\\
    \bot^\s_\E & \text{otherwise}
  \end{cases}
\end{equation*}
The resulting analysis is sound:
\begin{thm}
  \label{thm:soundsched}
  $\lbp{P_\C}\subseteq\lbp{P_\C^\s}.$
\end{thm}
\proof In Appendix~\ref{proof:soundsched}.\qed

Due to partitioning, $\lbp{P_\C^\s}$ is less efficient than $\lbp{P_\I^\s}$.
The abstract semantic functions for primitive statements, as well as
the join $\cup^\s$ and widening $\widen$, are performed pointwise on all
configurations $c\in\C$.
However, a clever implementation need not represent explicitly nor iterate over
partitions mapping a configuration to an empty environment $\bot^\s_\E$ or an 
empty interference $\bot^\s_\RR$.
The extra cost with respect to a non-scheduled analysis has thus a 
component that is linear in the number of non-$\bot^\s_\E$ environment
partitions and a component linear in the number of non-$\bot^\s_\RR$
interferences.
Thankfully, partitioned environments are extremely sparse:
Sec.~\ref{sec:result} shows that, in practice, at most program points,
$R^\s(c)=\bot^\s_\E$ except for a few configurations
(at most 4 in our benchmark).
Partitioned interferences are less sparse (52 in our benchmark)
because, being flow-insensitive,
they accumulate information for configurations reachable from any program 
point.
However, this is not problematic: as interferences are non-relational, a 
larger number of partitions can be stored and manipulated efficiently.

Thanks to partitioning, the precision of $\lbp{P_\C^\s}$ is 
much better than that of $\lbp{P_\I^\s}$ in the presence of locks and priorities.
For instance, $\lbp{P_\C^\s}$ using the interval domain discovers that 
$T=0$ in Fig.~\ref{fig:schedul},
while the analysis of Sec.~\ref{sec:sharedabs} would only discover that
$T\in[-1,1]$ due to spurious interferences from the high priority thread.

\section{Experimental Results}
\label{sec:result}

We implemented the abstract analysis of Sec.~\ref{sec:schedabs} in Thésée, 
our prototype analyzer based on the Astrée static analyzer 
\cite{blanchet-al-PLDI03}.
We first describe succinctly Astrée, then Thésée, and finally
our target application and its analysis by Thésée.

\subsection{The \texorpdfstring{Astrée}{Astree} Analyzer}

\begin{figure}
  \centering
  \includegraphics[width=7cm]{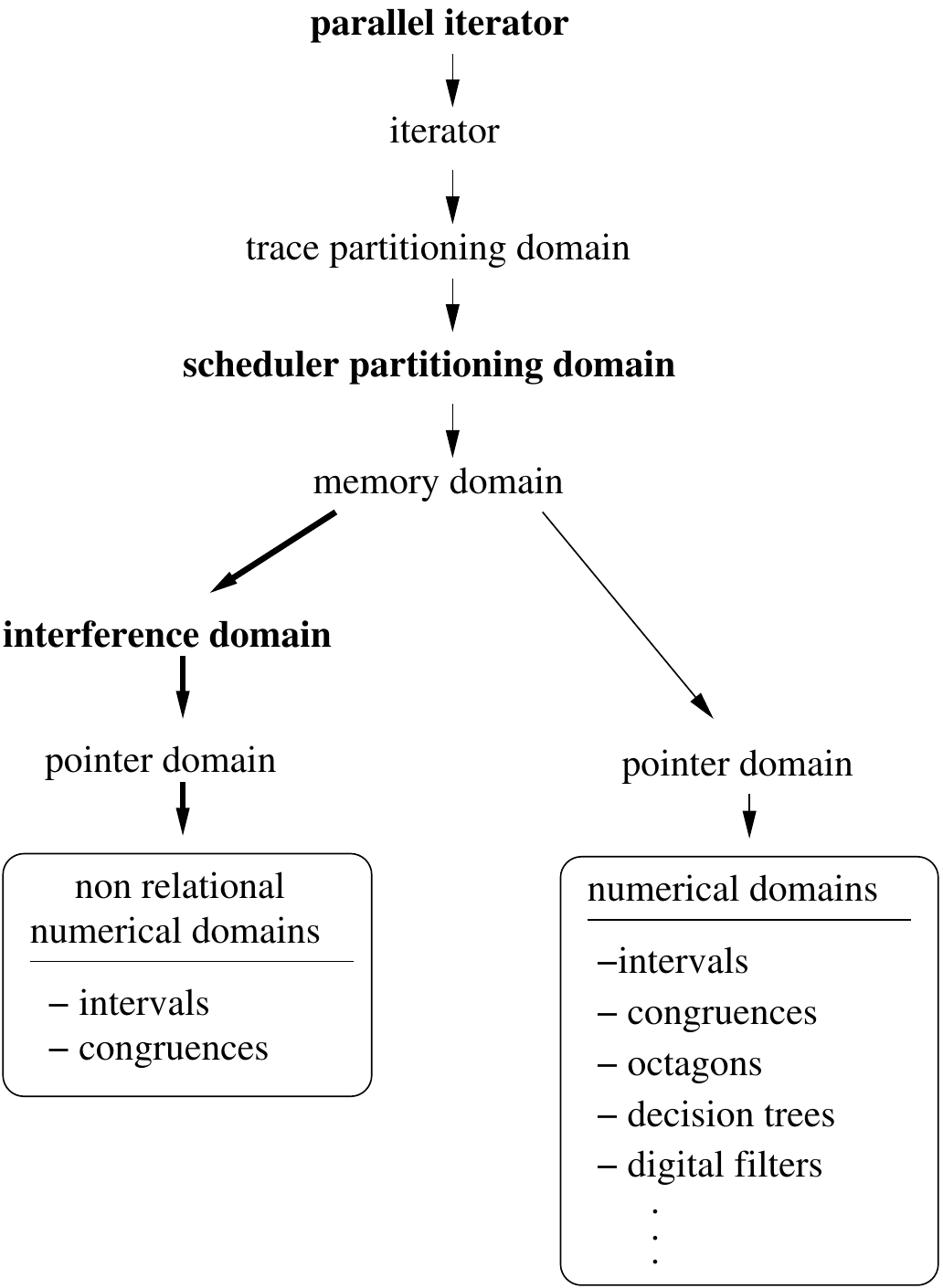}
  \caption{Hierarchy of abstractions in Astrée and Thésée.
Domains in boldface are specific to Thésée and not included in Astrée.}
  \label{fig:herarchy}
\end{figure}

Astrée is a static analyzer that checks for run-time errors in 
embedded C programs.
Astrée accepts a fairly large subset of C, excluding notably dynamic memory 
allocation and recursion, that are generally unused (or even forbidden)
in embedded code.
Moreover, Astrée does not analyze multi-threaded programs, which is the very
issue we address in the present article.

\medskip

The syntax and semantics assumed by Astrée are based on the
C99 norm \cite{c99}, supplemented with the IEEE 754-1985 norm 
for floating-point arithmetics \cite{ieee754}.
The C99 norm underspecifies many aspects of the semantics, leaving much
leeway to compiler implementations, including random undocumented and
unpredictable behaviors in case of an error such as an integer overflow.
A strictly conforming program would rely only on the semantics defined
in the norm. 
Few programs are strictly conforming; they rely instead on
additional, platform-specific semantic hypotheses. This is especially true in
the embedded software industry, where programs are designed for a specific,
well-controlled platform, and not for portability.
Thus, Astrée provides options to set platform-specific semantic features,
such as the bit-size and byte-ordering of data-types, and the
subsequent analysis is only sound with respect to these hypotheses.
The run-time errors checked by Astrée are: overflows in integer arithmetics 
and casts, integer divisions by zero, invalid bit-shifts, infinities and Not 
a Number floating-point values (caused by overflows, divisions by zero,
or invalid operations),
out-of-bound array accesses, invalid pointer arithmetics or dereferences
(including null, dangling, and misaligned pointers), and failure of 
user-defined assertions (specified in a syntax similar to the 
standard {\tt assert} function).

Astrée takes as input the C source after preprocessing by a standard
preprocessor and a configuration file describing the ranges of the
program inputs (such as memory-mapped sensors) if any.
It then runs fully automatically and outputs a list of alarms corresponding
to potential errors, and optionally program invariants for selected program
points and variables.
Astrée is sound in that it computes an over-approximation of all program 
traces, for all input scenarios.
Moreover, the analysis continues even for erroneous program traces if the
behavior after the error has a reasonable semantics. This is the case after
integer overflows, for instance, using a wrap-around semantics, but it is not
the case after dereferencing a dangling pointer, which has truly 
unpredictable results.
In all cases, when there is no alarm, or when all the alarms can be proved
by other means to be spurious, then the program is indeed proved free of
run-time error.

\medskip

Although Astrée accepts a large class of C programs, it cannot analyze most of
them precisely and efficiently.
It is specialized, by its choice of abstractions, towards control~/ command
aerospace code, for which it gives good results.
Thanks to a modular design, it can be adapted to other application domains
 by adding new abstractions.
Actually, the initial specialization towards control~/ command avionic software
\cite{blanchet-al-PLDI03} was achieved by incrementally adding new domains
and refining existing ones until all false alarms could be removed on a target
family of large control software from Airbus (up to 1 M lines)
\cite{delmas-souyris-sas07}.
The resulting analyzer achieved the zero false alarm goal in a few hours of
computation on a standard 2.66 GHz 64-bit intel server, and could be
deployed in an industrial context \cite{delmas-souyris-sas07}.
This specialization can be continued with limited effort, at least for related
application domains, as shown by our case study on space software 
\cite{mine:dasia09}.
Astrée is now a mature tool industrialized by AbsInt \cite{absint-web}.

Figure~\ref{fig:herarchy} presents the design of Astrée as a hierarchy of 
abstract domains --- we ignore for now boldface domains, which are 
specific to Thésée.
Actually, Astrée does not contain a single ``super-domain'' but rather
many small or medium-sized domains that focus on a specific kind of properties each, possess
a specific encoding of these properties and algorithms to manipulate them,
and can be easily plugged in and out.
One of the first domain included in Astrée was the simple interval domain
\cite{cc-POPL77} that expresses properties of the form $X\in[a,b]$ for every 
machine integer and floating-point variable $X\in\V$.
The interval domain is key as it is scalable, hence it can be 
applied to all variables at all program points. Moreover, it is able to express 
sufficient conditions for the absence of many kinds of errors, e.g., 
overflows.
Astrée also includes relational domains, such as the octagon domain
\cite{mine-HOSC06} able to infer relations of the form $\pm X \pm Y\leq c$.
Such relations are necessary at a few locations, for instance to infer
precise loop invariants, which then lead to tighter variable bounds. 
However, as the octagon domain
is less scalable, it is used only on a few variables, selected 
automatically by a syntactic heuristic.
Astrée also includes abstract domains specific to the target application domain,
such as a domain to handle digital filtering featured in many
control~/ command applications \cite{feret-ESOP04}.
The computations are performed in all the domains in parallel, and the
domains communicate information through a partially reduced product
\cite{cousot-al-ASIAN06}, so that they can improve each other in a controlled
way --- a fully reduced product, where all domains communicate all their
finds, would not scale up.
Additionally to numeric variables, the C language features pointers.
Pointer values are modeled in the concrete semantics of Astrée as
semi-symbolic pairs containing a variable name and an integer byte-offset.
The pointer abstract domain is actually a functor that adds support for 
pointers to
any (reduced product of) numerical abstract domain(s) by maintaining internally
for each pointer a set of pointed-to variables, and delegating the abstraction
of the offset to the underlying numerical domain(s) (associating a
synthetic integer variable to each offset).
Another functor, the memory domain, handles the decomposition of
aggregate variables (such as arrays and structures) into simpler scalar variables.
The decomposition is dynamic to account for the weak type system of C and
the frequent reinterpretation of the same memory portions as values of 
different types (due to union types and to type-punning).
Both functors are described in \cite{mine-LCTES06}.
Finally, a trace partitioning domain \cite{mauborgne-rival-ESOP05} adds a
limited amount (for efficiency) of path-sensitivity by maintaining at the
current control point several abstract states coming from execution traces
with a different history (such as which branches of $\sy{if}$ statements
were taken in the past).
The computation in these domains is driven by an iterator that traverses the
code by structural induction on its syntax, iterating loops with widening
and stepping into functions to achieve a fully flow- and context-sensitive
analysis.

\medskip

More information and pointers about Astrée can be found in 
\cite{bertrane-al-aiaa10}.

\subsection{The \texorpdfstring{Thésée}{Thesee} Analyzer}

Thésée is a prototype extension of Astrée that uses the abstract scheduled
interference semantics of Sec.~\ref{sec:schedabs} to support the analysis of
multi-threaded programs.
Thésée checks for the same classes of run-time errors as Astrée.
Additionally, it reports data-races, but ignores other parallel-related 
hazards, such as dead-locks and priority inversions, that are not described
in our concrete semantics.

\medskip

Thésée benefited directly from Astrée's numerous abstract domains and iteration
strategies targeting embedded C code.
Figure~\ref{fig:herarchy} presents the design of Thésée, where non-boldface
domains are inherited from Astrée and boldface ones have been added.

Firstly, the memory domain has been modified to compute the abstract 
interferences generated by the currently analyzed thread and apply the 
interferences from other threads.
We use the method of Fig.~\ref{fig:absschedstat}: the memory domain
dynamically modifies expressions to include interferences explicitly 
(e.g., replacing variables with intervals)
before passing the expressions to a stack of domains that are unaware of 
interferences.
Interferences are themselves stored and manipulated by a specific domain
which maintains abstract sets of values.
Non-relational abstractions from Astrée, such as intervals but also abstract 
pointer values, are directly exploited to represent abstract interferences.

Secondly, a scheduler partitioning domain has been added.
It maintains an abstraction of environments and of interferences for each
abstract scheduled configuration live at the current program point.
Then, for each configuration, it calls the underlying domain with
the abstract environment associated to this configuration, as well as the
abstract interferences that can effect this environment
(i.e., a join of interferences from all configurations not in mutual 
exclusion with the current one).
Additionally, the scheduler domain interprets directly all the instructions 
related to synchronization, which involves copying and
joining abstract environments from different configurations,
as described in Fig.~\ref{fig:absschedstat2}.

Finally, we introduced an additional, parallel iterator driving the whole 
analysis.
Following the execution model of the ARINC 653 specification, the
parallel iterator
first executes the main function as a regular single-threaded program and
collects the set of resources (threads, synchronization objects) it creates.
Then, as the program enters parallel mode, the iterator analyzes each
thread in sequence, and keeps re-analyzing them until their interferences are
stable (in parallel mode, no new thread may be created).

All these changes add up to approximately 10~K lines of code in the
100~K lines analyzer and did not require much structural change.

\subsection{Analyzed Application}

Thésée has been applied to the analysis of a large industrial program 
consisting of 1.7~M lines of unpreprocessed C code\footnote{After preprocessing and removal of comments, empty lines, and multiple definitions, the code is  2.1~M lines. The increase in size is due to the use of macros.}
and 15 threads, and running
under an ARINC 653 real-time OS \cite{ARINC}.
The analyzed program is quite complex as it mixes string formatting, list 
sorting,
network protocols (e.g., TFTP), and automatically generated synchronous logic.

The application performs system calls that must be properly modeled
by the analyzer.
To keep the analyzer simple, Thésée implements natively only low-level
primitives to declare and manipulate threads as well as simple mutexes
having the semantics described in Sec.~\ref{sec:syncprim}.
However, ARINC 653 objects have a more complex semantics.
The analyzed program is thus completed with a 2,500-line hand-written model 
of the ARINC 653 standard, designed specifically for the analysis with Thésée.
It implements all the system calls in C extended with Thésée primitives.
The model maps high-level ARINC 653 objects to lower-level Thésée ones.
For instance, ARINC processes\footnote{In the ARINC 653 \cite{ARINC} terminology, execution units in shared memory are called ``processes''; they correspond to POSIX threads and not to POSIX processes \cite{posix-threads}.}
have a name while Thésée threads only have
an integer identifier, so, the model keeps track of the correspondence between 
names and identifiers in C arrays and implements system calls to look up names
and identifiers. It also emulates the ARINC semantics using Thésée
primitives. For instance, a lock with a timeout is modeled as a 
non-deterministic test that either actually locks the mutex, or yields and 
returns an error code without locking the mutex.
An important feature of the program we analyze is that all potentially 
blocking calls have a finite timeout, so, by construction, no dead-lock nor 
unbounded priority inversion can occur.
This explains why we did not focus on detecting statically these issues
in the present article.

\subsection{Analysis Results}

\begin{figure}[t]
  \centering
  \includegraphics[width=12cm]{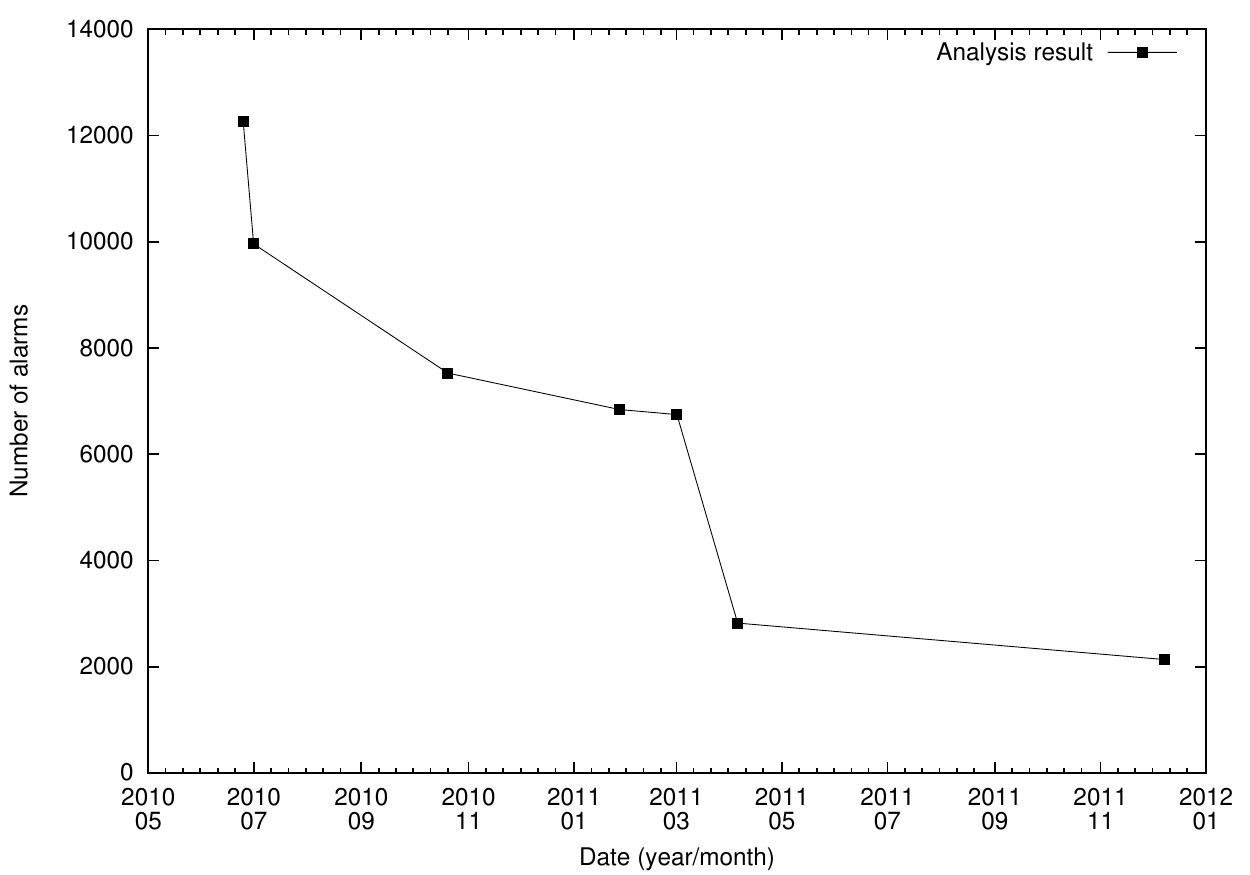}\\
  \caption{Evolution of the number of alarms in the analysis of our target 
application as we improved our prototype analyzer.}
  \label{fig:alarms}
\end{figure}

At the time of writing, the analysis with Thésée of this application
takes 27~h on our 2.66~GHz 64-bit intel server.
An important result is that only 5 iterations are required to stabilize 
abstract interferences.
Moreover, there is a maximum of 52 partitions for abstract interferences
and 4 partitions at most for abstract environments, so that the analysis fits 
in the 32~GB of memory of our server.
The analysis currently generates 2,136 alarms (slightly less than
one alarm per 800 lines of unpreprocessed code).

\begin{figure}
  \begin{tabular}{c@{\qquad}c}
    {\small\tt\begin{tabular}[b]{l}
        int clock, acc;\\
        \\
        void accum(int reset)\\
        \{\\
        \quad static int t0;\\
        \quad if (reset) \{\\
        \qquad acc = 0;\\
        \quad \}\\
        \quad  else \{\\
        \qquad acc += clock - t0;\\
        \quad \}\\
        \quad t0 = clock;\\
        \quad /* $0\leq acc \leq clock$ */\\
        \}\\
      \end{tabular}}
    &
    {\small\tt\begin{tabular}[b]{l}
        struct t \{\\
        \quad int id;\\
        \quad struct \{ char msg[23]; \} x[3];\\
        \} tab[12];\\
        \\
        char* end\_of\_msg(int ident, int pos)\\
        \{\\
        \quad int i;\\
        \quad struct t* p = tab;\\
        \quad for (i=0; i<12 \&\& p[i].id!=ident; i++);\\
        \quad char* m = p[i].x[pos].msg;\\
        \quad for (i=0; i<23 \&\& m[i]; i++);\\
        \quad /* $\mi{offset}(m+i)\in 4 + 292[0,11] + 96[0,2] + [0,22]$ */\\
        \quad return m+i;\\
        \}
      \end{tabular}}
    \\\\
    (a) & (b)
  \end{tabular}
  \caption{Program fragments that required an improvement in the analyzer prototype.}
  \label{fig:improvements}
\end{figure}

These figures have evolved before and during the writing of this article, as 
we improved the analysis.
Figure~\ref{fig:alarms} presents the evolution of the number of alarms on a
period of 18~months.
As our improvement effort focuses on optimizing the analysis precision, 
we do not present the 
detailed evolution of the analysis time (it oscillates between 14~h and 
28~h,\footnote{Intuitively, adding domains and refining their precision degrades
the efficiency. However, inferring tighter invariants can also reduce the 
number of loop iterations to reach a fixpoint, and so, improving the precision
may actually lower the overall analysis time.}
with a number of iterations between 4 and 7) nor the memory consumption 
(stable at a little under 30~GB).
The number of alarms started at 12,257 alarms mid-2010, as reported in
\cite[\S~VI]{bertrane-al-aiaa10}.
This high initial number can be explained by the lack  of specialization of 
the analyzer: the first versions of Thésée were naturally tuned for avionic 
control~/ command software as they inherited abstract domains $\E^\s$ and 
$\RR^\s$ from Astrée, but our target application for Thésée is {\em not\/}
limited to control~/ command processing.
To achieve our current results, we improved the numerical, pointer, and
memory domains in Thésée, and designed new ones.
We illustrate two of these improvements in Fig.~\ref{fig:improvements}.
A first example is the improvement of transfer functions of existing
domains.
For instance, the function {\tt accum} from Fig.~\ref{fig:improvements}.(a)
accumulates, in {\tt acc}, elapsed time as counted by a {\tt clock} variable 
(updated elsewhere), and we need to discover the invariant 
${\tt acc} \leq {\tt clock}$.
This requires analyzing precisely the incrementation of {\tt acc} in a
relational domain. As the octagon domain we use \cite{mine-HOSC06} 
only supported precisely assignments involving two variables, we added
new transfer functions for three-variable assignments (another solution, 
using the more powerful polyhedron domain \cite{ch:popl78}, did not
prove scalable enough).
A second example is an improvement of the pointer domain to precisely
track pointers traversing nested arrays and structures, as in 
Fig.~\ref{fig:improvements}.(b) where the precise location of the pointer
{\tt m+i} needs to be returned.
Our target application features similar complex data-structures and
their traversal extensively.
We thus added a non-relational integer domain for offsets of the form
$\alpha_0+\sum_i\alpha_i[\ell_i,h_i]$, where the values of $\alpha_i$, $\ell_i$, 
and $h_i$ are inferred dynamically.
Note that all these improvements concern only the abstract domain parameters;
neither the interference iterator nor the scheduler partitioning were
refined.

\medskip

Following the design-by-refinement used in Astrée 
\cite{blanchet-al-PLDI03}, we have focused on the analysis of a single
(albeit large and complex) industrial software and started refining the 
analyzer to lower the number of alarms, instead of multiplying 
shallower case studies. 
We plan to further improve the analysis precision in order to approach the 
zero false alarm goal. This is the objective of the AstréeA project 
\cite{astreea-web}, successor to Thésée.
The remaining 2,136 alarms can be categorized into three kinds.
Firstly, some alarms are, similarly to the ones described in 
Fig.~\ref{fig:improvements}, not related to parallelism but to the imprecision
of the parameter abstract domains.
An important class of properties currently not supported by Astrée nor
Thésée is that of memory shape properties \cite{chang-al:popl08}.
In the context of embedded software, dynamic memory allocation is disabled;
nevertheless, our target code features dynamic linked lists allocated
in large static arrays.
Another class of properties concerns the correct manipulation of
zero-terminated C strings.
A significant part of the remaining alarms may be removed by
designing new memory domains for these properties.
Secondly, some alarms can be explained by an imprecise abstraction of
thread interferences, similar to the imprecision observed in
Figs.~\ref{fig:imprecise1}--\ref{fig:imprecise2} 
(these examples were inspired from our target code).
Hence the need to extend our framework to support relational and 
flow-sensitive abstractions of interferences.
Thirdly, some alarms have simply not yet been fully investigated.
Although Thésée provides verbose information on the context of each alarm
as well as the thread-local and interference invariants, discovering the
origin of alarms is a challenging task on such a large code:
it often requires tracking the imprecision upstream and understanding the
interplay of thread interferences.

\section{Related Work}
\label{sec:relwork}

There are far too many works on the semantics and analysis of parallel
programs to provide a fair survey and comparison here. 
Instead, we focus on a few works that are either recent or provide a fruitful 
comparison with ours.

\subsection{Interferences}
The idea of attaching to each thread location a local invariant and
handling proofs of parallel programs similarly to that of sequential programs
dates back to the Hoare-style logic of Owicki and Gries 
\cite{owicki-gries-AI76} and the inductive assertion method of
Lamport \cite{lamport-TSE77,lamport-AI80}.
It has been well studied since; see \cite{roever-01} for a recent
account and survey.
The difference between the proofs of sequential programs and that of
parallel programs in this framework
is that the local invariants of each thread must be proved invariant by the
execution of all other threads --- i.e., a non-interference check.
These proof methods are studied from an Abstract Interpretation theoretical
point of view by Cousot and Cousot \cite{cc-APCT84}, which leads to two 
results: an expression of each method as a decomposition of the 
global invariant into thread-local invariants, and a framework to apply 
abstractions and derive effective and sound static analyzers.
When shifting from proof methods to inference methods, the non-interference
check naturally becomes an inference of interferences.
Our work is thus strongly inspired from \cite{cc-APCT84}:
it is based on an Owicki--Gries style decomposition of the global invariant
(although it is not only based on the control points of threads, but also
on a more complex scheduler state).
The thread-local and interference parts are then abstracted separately,
using a relational flow-sensitive analysis for the former and a coarser
non-relational flow-insensitive analysis for the later.
Our work is also similar to the recent static analysis of C programs with 
POSIX threads by Carré and Hymans \cite{carre-hymans-ARXIV09}:
both are based on Abstract Interpretation and interference computation, and 
both are
implemented by modifying existing static analyses of sequential programs.
Their analysis is more powerful than ours in that it handles dynamic thread creation and
the concrete semantics models interferences as relations (instead of actions),
but the subsequent abstraction leads to a non-relational analysis; moreover,
real-time scheduling is not considered.

\subsection{Data-Flow Analysis}
Fully flow-insensitive analyses, such as Steensgaard's points-to
analysis \cite{steensgaard-popl96}, naturally handle parallel programs.
To our knowledge, all such analyses are also non-relational.
These fast analyses are adequate for compiler optimization but,
unfortunately, the level of accuracy required to prove safety properties
demands the use of (at least partially) flow-sensitive and relational methods,
which we do.
By contrast, S\u{a}lcianu and Rinard \cite{salciano-al-ppopp01} proposed
a flow-sensitive pointer and escape analysis for parallel programs
which is more precise (and more costly), although it still targets 
program optimisation.
It uses a notion of interference to model the effect of threads and
method calls.

\subsection{Model Checking}
Model-checking also has a long history of verifying parallel systems,
including recently weak memory models \cite{atig-al-popl10}.
To prevent state explosion, Godefroid \cite{godefroid-phd}
introduced partial order reduction methods.
They limit the number of interleavings to consider, with no impact on 
soundness nor completeness.
Due to the emphasis on completeness, the remaining set of interleavings 
can still be high.
By contrast, we abstract the problem sufficiently so that no interleaving need 
to be considered at all, at the cost of completeness. 
Another way to reduce the complexity of model checking is the context bound
approach, as proposed by Qadeer et al.~\cite{qadeer-rehof-tacas05}.
As it is unsound, it may fail to find some run-time errors.
By contrast, our method takes into account all executions until completion.
In his PhD, Malkis \cite{malkis-phd} used abstract interpretation
to prove the equivalence of Owicki and Gries's proof method and
the more recent model-checking algorithm by Flanagan and Qadeer
\cite{flanagan-al-spin03}, and presented an improvement based on 
counterexample-guided abstract refinement, a method which, unlike ours, is not
guaranteed to converge in finite time.

\subsection{Weakly Consistent Memories}
Weakly consistent memory models have been studied originally for hardware ---
see \cite{adve-charachorloo-WRL95} for a tutorial.
Precise formal models are now available for popular architectures, 
such as the intel x86 model by Sewell et al.~\cite{sewell-al:jacm10},
either inferred from informal processor documentations or 
reverse-engineered through ``black-box'' testing
\cite{alglave-al-tacas11}.
Pugh \cite{pugh-JGI99} pioneered the use of weakly consistent memory models in 
programming 
language semantics in order to take into account hardware and compiler 
optimizations.
This culminated in the Java memory model of Manson et 
al.~\cite{manson-al-POPL05,java3}.

Weakly consistent memory models are now recognised as an important part
of language semantics and are increasingly supported by verification methods.
An example in model-checking is the work of Atig et 
al.~\cite{atig-al-popl10}.
An example in theorem proving is the extension by \v{S}ev\v{c}ík et 
al.~\cite{sevcik-popl11} of Leroy's formally proved C compiler
\cite{leroy-POPL06}.
Testing methods have also been proposed, such as that of Alglave et 
al.~\cite{alglave-al-tacas11}.
In the realm of static analysis, we can cite the
static analysis by Abstract Interpretation of the happens-before
memory model (at the core of the Java model)
designed by Ferrara \cite{ferrara-TAP08}.
Recently, Alglave and al. proposed in \cite{algave-aplas11} to lift
analyses that are only sound with respect to sequential consistency,
to analyses that are also sound
in weak memory models.
Their method is generic and uses a ``repair loop'' similar 
to our fixpoint of flow-insensitive interferences.

Memory models are often defined implicitly, by
restricting execution traces using global conditions, following
the approach chosen by the Java memory model \cite{java3}.
We chose instead a generative model based on local control path 
transformations, which is reminiscent of the approach by Saraswat et 
al.~\cite{saraswat-al-PPOPP07}.
We believe that it matches more closely classic software 
and hardware optimizations.
Note that we focus on models that are not only realistic, but 
also amenable to abstraction into an interference semantics.
The first condition ensures the soundness of the static analysis,
while the second one ensures its efficiency.

\subsection{Real-Time Scheduling}
Many analyses of parallel programs assume arbitrary preemption, either
implicitly at all program points (as in flow-insensitive analyses),
or explicitly at specified program points 
(as in context-bounded approaches \cite{qadeer-rehof-tacas05}), but few
analyses 
model and exploit the strict scheduling policy of real-time schedulers.
A notable exception is the work of Gamatié et al.~\cite{gamatie-RTAS03} 
on the modeling of systems under an ARINC 653 operating system.
As the considered systems are written in the SIGNAL language, their ARINC 653 
model is naturally also written in SIGNAL, while ours in written in C 
(extended with low-level primitives for parallelism, which were not necessary 
when modeling in SIGNAL as the language can naturally express parallelism).

\subsection{Further Comparison}
A detailed comparison between domain-aware static analyzers, such as
Astrée, and other verification methods, such as theorem proving and
model checking, is presented in \cite{mine-al:tase07}.
These arguments are still valid in the context of a parallel program analysis 
and not repeated here.
On the more specific topic of parallel program analysis, we refer the 
reader to the comprehensive survey by Rinard~\cite{rinard-sas01}.

\section{Conclusion}
\label{sec:conclusion}

We presented a static analysis by Abstract Interpretation to detect in a 
sound way run-time errors in embedded C
software featuring several threads, a shared memory with weak consistency,
mutual exclusion locks, thread priorities, and a real-time scheduler.
Our method is based on a notion of interferences and a partitioning with
respect to an abstraction of the scheduler state. 
It can be implemented on top of existing
analyzers for sequential programs, leveraging a growing library
of abstract domains.
Promising early experimental results on an
industrial code demonstrate the scalability of our approach.

\smallskip

A broad avenue for future work is to bridge the gap between the interleaving
semantics and its incomplete abstraction using interferences.
In particular, it seems important to abstract interferences due to well
synchronized accesses in a relational way (this is in particular needed to
remove some alarms remaining in our target application).
We also wish to add some support for abstractions that are (at least
partially) sensitive to the history of thread interleavings.
This would be useful to exploit more properties of real-time
schedulers, related for instance to the guaranteed ordering of some 
computations
(by contrast,  we focused in this article mainly on properties related 
to mutual exclusion).

Moreover, we wish to extend our framework to include more models of
parallel computations.
This includes support for alternate real-time operating systems with similar 
scheduling policies but manipulating different synchronization objects,
for instance the condition variables in real-time POSIX 
systems \cite{posix-threads}, or alternate priority schemes, such as
the priority ceiling protocol for mutexes.
Another example is the support for the OSEK/VDX and Autosar real-time 
embedded platforms widely used in the automotive industry.
We also wish to study more closely weak memory consistency semantics and,
in particular, how to design more precise or more general interference 
semantics, and abstract them efficiently.
Supporting atomic variables, recently included in the C and C++ languages,
may also trigger the need for a finer, field-sensitive handling of
weak memory consistency.

A long term goal is the analysis of other errors specifically 
related to parallelism, such as dead-locks, live-locks, and 
priority inversions.
In a real-time system, all system calls generally have a timeout in
order to respect hard deadlines.
Thus, interesting properties are actually quantitative: by construction,
unbounded priority inversions cannot occur, so, we wish to detect
bounded priority inversions.

On the practical side, we wish to improve our prototype analyzer to reduce
the number of false alarms on our target industrial code.
This requires some of the improvements to the parallel analysis framework 
proposed above (such as relational and flow-sensitive abstractions
for interferences), but also the design of new numerical, pointer, and memory
domains which are not specific to parallel programs.

\section*{Acknowledgement}
We wish to thank the ESOP'11 and LMCS anonymous reviewers as well as
David Pichardie for their helpful comments on several versions of this 
article.

\bibliographystyle{plain}
\bibliography{bib}

\appendix


\section{Proof of Theorems}
\label{sec:proof}

\def\thmproof#1#2{\subsection{Proof of Theorem~\ref{thm:#1}}\label{proof:#1} #2}


\thmproof{morphism}{$\fa{s\in\mi{stat}}\lb{S}{s}$ is well defined and a complete $\sqcup-$morphism.}
\proof 
  
  We prove both properties at the same time by structural induction on $s$.
  \begin{iteMize}{$\bullet$}
  \item 
    The case of assignments $X\leftarrow e$ and guards
    $e\bowtie 0?$ is straightforward.
  \item 
    The semantics of conditionals and sequences is well-defined as its
    components are well-defined by induction hypothesis.
    It is a complete $\sqcup-$morphism, by induction hypothesis and the fact
    that the composition and the join of complete $\sqcup-$morphisms are
    complete $\sqcup-$morphisms.
  \item 
    For a loop $\sy{while}\;e\bowtie0\;\sy{do}\;s$, consider
    $F$ and $G$ defined as:
    $$\begin{array}{l}
      F(X)\deq(\lb{S}{s}\circ\lb{S}{e\bowtie 0?})X\\
      G(X)\deq(R,\Omega)\sqcup F(X)\enspace.
    \end{array}$$
    By induction hypothesis, $F$, and so $G$, are complete $\sqcup-$morphisms.
    Thus, $G$ has a least fixpoint \cite{cc-PJM79}.
    We note that:
    $$\lb{S}{\sy{while}\;e\bowtie0\;\sy{do}\;s}(R,\Omega)=\lb{S}{e\not\bowtie0?}(\lfp G)$$
    which proves that the semantics of loops is well-defined.
    Moreover, according to \cite{cc-PJM79} again, the least fixpoint can
    be computed by countable Kleene iterations:
    $\lfp G=\bigsqcup_{i\in\N} G^i(\emptyset,\emptyset)$.
    We now prove by induction on $i$ that
    $G^i(\emptyset,\emptyset)=\bigsqcup_{k<i} F^k(R,\Omega)$.
    Indeed:
    $$\begin{array}{ll}
      &G^1(\emptyset,\emptyset)\\[2pt]
      =&(R,\Omega)\sqcup F(\emptyset,\emptyset)\\[2pt]
      =&(R,\Omega)\\[2pt]
      =&F^0(R,\Omega)
    \end{array}$$
    and
    $$\begin{array}{ll}
      &G^{i+1}(\emptyset,\emptyset)\\[2pt]
      =&G\left(\bigsqcup_{k<i} F^k(R,\Omega)\right)\\[2pt]
      =&(R,\Omega)\sqcup F\left(\bigsqcup_{k<i} F^k(R,\Omega)\right)\\[2pt]
      =&(R,\Omega)\sqcup \bigsqcup_{k<i} F^{k+1}(R,\Omega)\\[2pt]
      =&\bigsqcup_{k<i+1} F^k(R,\Omega)
    \end{array}$$
    because $F$ is a $\sqcup-$morphism.
    As a consequence, $\lfp G=\bigsqcup_{i\in\N} F^i(R,\Omega)$.
    Note that, $\fa{i\in\N}F^i$ is a complete $\sqcup-$morphism.
    Thus, the function $(R,\Omega)\mapsto\lfp G$ is also a complete
    $\sqcup-$morphism, and so is 
    $\lb{S}{\sy{while}\;e\bowtie0\;\sy{do}\;s}$.
    \qed
  \end{iteMize}


\thmproof{sound}{$\lbp{P}\subseteq\lbp{P^\s}.$}
\proof 
  We prove by structural induction on $s$ that:
  $$\fa{s,R^\s,\Omega}
  (\lb{S}{s}\circ\gamma)(R^\s,\Omega)\sqsubseteq(\gamma\circ\lb{S^\s}{s})(R^\s,\Omega)\enspace.$$
  \begin{iteMize}{$\bullet$}
  \item 
    The case of primitive statements $X\leftarrow e$ and $e\bowtie0?$
    holds by hypothesis: the primitive abstract functions provided by the 
    abstract domain are assumed to be sound.
  \item
    The case of sequences $s_1;\,s_2$ is settled by noting that
    the composition of sound abstractions is a sound abstraction.
  \item 
    The case of conditionals $\sy{if}\,e\bowtie0\,\sy{then}\,s$ is settled by
    noting additionally that $\cup^\s$ is a sound abstraction of $\sqcup$, 
    as  $\cup^\s_\E$ is a sound abstraction of $\cup$.
  \item 
    We now treat the case of loops.
    By defining:
    $$\begin{array}{lcl}
      F^\s(X)&\deq&(R^\s,\Omega)\cup^\s(\lb{S^\s}{s}\circ\lb{S^\s}{e\bowtie0?})(X)\\
      F(X)&\deq&(R,\Omega)\sqcup(\lb{S}{s}\circ\lb{S}{e\bowtie0?})(X)
    \end{array}$$
    we have
    $$\begin{array}{lcl}
      \lb{S^\s}{\sy{while}\;e\bowtie0\;\sy{do}\;s}(R^\s,\Omega)&=&\lb{S^\s}{e\not\bowtie0?}(\lim\lbd{X^\s}X^\s\widen F^\s(X^\s))\\
      \lb{S}{\sy{while}\;e\bowtie0\;\sy{do}\;s}(R,\Omega)&=&\lb{S}{e\not\bowtie0?}(\lfp F)\enspace.
    \end{array}$$
    By induction hypothesis, soundness of $\cup^\s$ and composition of
    sound functions, $F^\s$ is a sound abstraction of $F$.
    Assume now that $(R^\s{}',\Omega')$ is the limit of iterating
    $\lbd{X}X\widen F^\s(X)$ from $(\bot^\s_\E,\emptyset)$.
    Then, it is a fixpoint of $\lbd{X^\s}X^\s\widen F^\s(X^\s)$, hence
    $(R^\s{}',\Omega')=(R^\s{}',\Omega')\widen F^\s(R^\s{}',\Omega')$.
    Applying $\gamma$ and the soundness of $\widen$ and $F^\s$, we get:
    $$\begin{array}{ll}
      &\gamma(R^\s{}',\Omega')\\
      = &\gamma((R^\s{}',\Omega')\widen F^\s(R^\s{}',\Omega'))\\[2pt]
      \sqsupseteq & \gamma(F^\s(R^\s{}',\Omega'))\\[2pt]
      \sqsupseteq & F(\gamma(R^\s{}',\Omega'))\enspace.
    \end{array}$$
    Thus, $\gamma(R^\s{}',\Omega')$ is a post-fixpoint of $F$.
    As a consequence, it over-approximates $\lfp F$, which implies
    the soundness of the semantics of loops.
  \end{iteMize}
  We now apply the property we proved to $s=\body$, $R^\s=\E^\s_0$ and
  $\Omega=\emptyset$, and use the monotony of $\lb{S}{s}$ 
  and the fact that $\gamma(\E^\s_0)\supseteq\E_0$ to get:
  $$\begin{array}{ll}
    &\gamma(\lb{S^\s}{\body}(\E^\s_0,\emptyset))\\[2pt]
    \sqsupseteq & \lb{S}{\body}(\gamma(\E^\s_0,\emptyset))\\[2pt]
    \sqsupseteq & \lb{S}{\body}(\E_0,\emptyset)
  \end{array}$$
  which implies
  $\lbp{P^\s}\supseteq\lbp{P}$.
\qed


\thmproof{structpath}{$\fa{s\in\mi{stat}}\lbx{\Pi}{\pi(s)}=\lb{S}{s}.$}
\proof 
  The proof is by structural induction on $s$.
  \begin{iteMize}{$\bullet$}
  \item 
    The case of primitive statements is straightforward
    as $\pi(s)=\set{s}$.
  \item 
    For sequences, we use the induction hypothesis and the fact that
    $\lbpx{\Pi}$ is a morphism for path concatenation to get:
    $$\begin{array}{ll}
      & \lb{S}{s_1;s_2}\\[1pt]
      = & \lb{S}{s_2}\circ\lb{S}{s_1}\\[1pt]
      = & \lbx{\Pi}{\pi(s_2)}\circ\lb{\Pi}{\pi(s_1)}\\[1pt]
      = & \lbx{\Pi}{\pi(s_1)\cdot\pi(s_2)}\\[1pt]
      = & \lbx{\Pi}{\pi(s_1;s_2)}\enspace.
    \end{array}$$
  \item
    For conditionals, we also use the fact that $\lbpx{\Pi}$ is a morphism
    for $\cup$:
    $$\begin{array}{ll}
      & \lb{S}{\sy{if}\;e\bowtie0\;\sy{then}\;s}(R,\Omega)\\[2pt]
      = & (\lb{S}{s}\circ\lb{S}{e\bowtie0?})(R,\Omega)\sqcup\lb{S}{e\not\bowtie0?}(R,\Omega)\\[2pt]
      = & (\lbx{\Pi}{\pi(s)}\circ\lbx{\Pi}{\set{e\bowtie0?}})(R,\Omega)\sqcup\lbx{\Pi}{\set{e\not\bowtie0?}}(R,\Omega)\\[2pt]
      = & \lbx{\Pi}{((e\bowtie0?)\cdot\pi(s))\cup\set{e\not\bowtie0?}}(R,\Omega)\\[2pt]
      = & \lbx{\Pi}{\pi(\sy{if}\;e\bowtie0\;\sy{then}\;s)}(R,\Omega)\enspace.
    \end{array}$$
  \item 
    For loops $\sy{while}\;e\bowtie0\;\sy{do}\;s$, we define
    $F$ and $G$ as in proof~\ref{proof:morphism}, i.e., 
    $F(X)\deq(\lb{S}{s}\circ\lb{S}{e\bowtie0?})X$ and
    $G(X)\deq(R,\Omega)\sqcup F(X)$.
    Recall then that $\lfp G=\bigsqcup_{i\in\N} F^i(R,\Omega)$.
    By induction hypothesis and $\cdot-$morphism, we have:
    $$\begin{array}{ll}
      &F^i\\[1pt]
      =&(\lb{S}{s}\circ\lb{S}{e\bowtie0?})^i\\[1pt]
      =&(\lbx{\Pi}{\{e\bowtie0?\}\cdot\pi(s)})^i\\[1pt]
      =&\lbx{\Pi}{(\{e\bowtie0?\}\cdot\pi(s))^i}\enspace.
    \end{array}$$
    Let us now define the set of paths
    $P\deq\lfp\lbd{X}\sset{\epsilon}\cup(X\cdot\{e\bowtie0?\}\cdot\pi(s))$.
    By \cite{cc-PJM79}, $P=\bigcup_{i\in\N} (\{e\bowtie0?\}\cdot\pi(s))^i$.
    As a consequence:
    $$\lbx{\Pi}{P}=\bigsqcup_{i\in\N}\lbx{\Pi}{(\{e\bowtie0?\}\cdot\pi(s))^i}=\bigsqcup_{i\in\N}F^i$$
    and $\lbx{\Pi}{P}(R,\Omega)=\bigsqcup_{i\in\N}F^i(R,\Omega)=\lfp G$.\\[2pt]
    Finally:
    $$\begin{array}{ll}
      &\lb{S}{\sy{while}\;e\bowtie0\;\sy{do}\;s}(R,\Omega)\\[2pt]
      =&\lb{S}{e\not\bowtie0?}(\lfp G)\\[2pt]
      =&(\lbx{\Pi}{\set{e\not\bowtie0?}}\circ\lbx{\Pi}{P})(R,\Omega)\\[2pt]
      =&\lbx{\Pi}{P\cdot \{e\not\bowtie0?\}}(R,\Omega)\\[2pt]
      =&\lbx{\Pi}{\pi(\sy{while}\;e\bowtie0\;\sy{do}\;s)}(R,\Omega)
    \end{array}$$
    which concludes the proof.\qed
  \end{iteMize}


\thmproof{structpath2}{$\fa{t\in\T,s\in\mi{stat}}\lbx{\Pi_\I}{\pi(s),\,t}=\lb{S_\I}{s,\,t}.$}
\proof 
  The proof \ref{proof:structpath} of Thm.~\ref{thm:structpath} 
  only relies on the fact that the
  semantic functions $\lb{S}{s}$ are complete $\sqcup-$morphisms.
  As the functions $\lb{S_\I}{s,t}$ are complete $\sqcup_\I-$morphisms, the same
  proof holds.
\qed


\thmproof{interf}{$\lbp{P_*}\subseteq\lbp{P_\I}$.}
\proof 
  To ease the proof, we will use the notations $R(X)$, $\Omega(X)$, $I(X)$ to 
  denote the various components of a triplet $X=(R(X),\Omega(X),I(X))$ in the
  concrete domain $\D_\I=\P(\E)\times\P(\L)\times\P(\I)$,
  and $V(X)$, $\Omega(X)$ for pairs $X=(V(X),\Omega(X))$ in
  $\P(\R)\times\P(\L)$ output by the semantics of expressions.

  Let $(\Omega_\I,I_\I)$ be the fixpoint computed in (\ref{eq:intersem}), i.e.,
  $(\Omega_\I,I_\I)=\bigsqcup_{t\in\T}\,(\Omega'_t,I'_t)$ where
  $(-,\Omega'_t,I'_t)=\lb{S_\I}{\body_t,t}(\E_0,\Omega_\I,I_\I)$.
  Then, by definition, $\lbp{P_\I}=\Omega_\I$.
  We first prove the following properties that compare respectively the 
  set of errors and environments
  output by the interleaving semantics and the interference semantics
  of any path $p\in\pi_*$:
  $$\begin{array}{l@{\quad}l}
    \text{(i)}  & \Omega(\lbx{\Pi_*}{p}(\E_0,\emptyset))\subseteq\bigcup_{t\in\T}\;\Omega(\lbx{\Pi_\I}{\proj_t(p),\,t}(\E_0,\emptyset,I_\I)) \\[8pt]
    \text{(ii)}  & \fa{t\in\T,\,\rho\in R(\lbx{\Pi_*}{p}(\E_0,\emptyset))}
    \ex{\rho'\in R(\lbx{\Pi_\I}{\proj_t(p),\,t}(\E_0,\emptyset,I_\I))}\\[2pt]&
    \fa{X\in\V}(\rho(X)=\rho'(X)\vee\ex{t'\neq t}(t',X,\rho(X))\in I_\I)
  \end{array}$$
  The proof is by induction on the length of $p$.
  The case $p=\epsilon$ is straightforward: 
  the $\Omega$ component is $\emptyset$ on both sides of 
  (i), and we can take $\rho'=\rho$ in (ii) as the R component
  is $\E_0$ on both sides.
  Consider now $p=p'\cdot(s',t')$, i.e., $p$ is a path $p'$ followed by an 
  assignment or guard $s'$ from thread $t'$.
  Consider some $\rho\in R(\lbx{\Pi_*}{p'}(\E_0,\emptyset))$ and the expression
  $e'$ appearing in $s'$.
  We can apply the (ii) recurrence hypothesis to $p'$ and $t'$ which 
  returns some 
  $\rho'\in R(\lbx{\Pi_\I}{\proj_{t'}(p'),\,t'}(\E_0,\emptyset,I_\I))$.
  The fact that, for any $X\in\V$, either 
  $\rho(X)=\rho'(X)$ or $\ex{t''\neq t'}(t'',X,\rho(X))\in I_\I$
  implies, given the definition of $\lb{E_\I}{e'}$ (Fig.~\ref{fig:interfexprsem}),
  that:
  $$\begin{array}{l@{\quad}l}
    \text{(iii)} & \lb{E}{e'}\rho\sqsubseteq\lb{E_\I}{e'}(t',\rho',I_\I)
  \end{array}$$
  When considering, in particular, the error component $\Omega(\lb{E}{e'}\rho)$,
  (iii) allows extending the recurrence hypothesis (i) on $p'$ to also hold
  on $p$, i.e., executing $s'$ adds more errors on the right-hand side of (i) 
  than on the left-hand side.

  To prove (ii) on $p$, we consider two cases, depending on the kind of 
  statement $s'$:
  \begin{iteMize}{$\bullet$}
  \item 
    Assume that $s'$ is a guard, say $e'=0?$ (the proof is identical
    for other guards $e'\bowtie 0?$).
    Take any $t\in\T$ and $\rho\in R(\lbx{\Pi_*}{p}(\E_0,\emptyset))$.
    By definition of $\lb{S}{e'=0?}$, we have
    $\rho\in R(\lbx{\Pi_*}{p'}(\E_0,\emptyset))$ and $0\in V(\lb{E}{e'}\rho)$.
    Take any $\rho'$ that satisfies the recurrence hypothesis (ii) on $p'$ 
    for $t$ and $\rho$.
    We prove that it also satisfies (ii) on $p$ for $t$ and $\rho$.

    The case $t\neq t'$ is straightforward as, in this case,
    $\proj_t(p)=\proj_t(p')$.

    When $t=t'$, then $\proj_t(p)=\proj_t(p')\cdot (e'=0?)$.
    Recall property (iii):
    $\lb{E}{e'}\rho\sqsubseteq\lb{E_\I}{e'}(t',\rho',I_\I)$, which implies
    that  $0\in V(\lb{E_\I}{e'}(t',\rho',I_\I))$, and so, we have
    $\rho'\in R(\lbx{\Pi_\I}{\proj_t(p),\,t'}(\E_0,\emptyset,I_\I))$.
    \medskip

  \item 
    Assume that $s'$ is an assignment $X\leftarrow e'$.
    Take any $\rho\in R(\lbx{\Pi_*}{p}(\E_0,\emptyset))$ and 
    $t\in\T$.
    Then $\rho$ has the form $\rho_0[X\mapsto \rho(X)]$ for some 
    $\rho_0\in R(\lbx{\Pi_*}{p'}(\E_0,\emptyset))$ and $\rho(X)\in V(\lb{E}{e'}\rho_0)$.

    We first prove that $(t',X,\rho(X))\in I_\I$.
    Take $\rho_0'$ as defined by the recurrence hypothesis (ii) for $t'$,
    $p'$ and $\rho_0$.
    Then $\rho(X)\in V(\lb{E}{e'}\rho_0)\subseteq V(\lb{E_\I}{e'}(t',\rho_0',I_\I))$
    by property (iii).
    By definition of $\lb{S_\I}{X\leftarrow e',\,t'}$ (Fig.~\ref{fig:interfstatsem}),
    we have $(t',X,\rho(X))\in I(\lb{S_\I}{X\leftarrow e',\,t'}(\sset{\rho_0'},\emptyset,I_\I))$.
    We have $\rho'_0\in R(\lbx{\Pi_\I}{\proj_{t'}(p'),\,t'}(\E_0,\emptyset,I_\I))$
    by definition.
    Because $\proj_{t'}(p)=\proj_{t'}(p')\cdot (X\leftarrow e')\in\pi(\body_{t'})$,
    it follows that
    $(t',X,\rho(X))\in I(\lbx{\Pi_\I}{\pi(\body_{t'}),\,t'}(\E_0,\emptyset,I_\I))$.
    \\\noindent
    By Thm.~\ref{thm:structpath2}, we then have
    $(t',X,\rho(X))\in I(\lb{S_\I}{\body_{t'},\,t'}(\E_0,\emptyset,I_\I))=I'_{t'}$.
    Thus, $(t',X,\rho(X))\in I_\I$, as $I_\I$ satisfies
    $I_\I=\bigcup_{t\in\T} I'_t$.

    To prove (ii), we consider first the case where $t\neq t'$.
    Take $\rho_0'$ as defined by the recurrence hypothesis (ii) for $t$,
    $p'$ and $\rho_0$.
    As $t\neq t'$, $\proj_t(p)=\proj_t(p')$, and
    $\rho'_0\in R(\lbx{\Pi_\I}{\proj_t(p),\,t}(\E_0,\emptyset,I_\I))$.
    As $\rho$ and $\rho_0$ are equal except maybe on $X$, and we have
    $(t',X',\rho(X))\in I_\I$, then $\rho'_0$ satisfies (ii) for $t$, $p$, and
    $\rho$.

    We now consider the case where $t=t'$.
    Take $\rho_0'$ as defined by the recurrence hypothesis (ii) for $t$,
    $p'$ and $\rho_0$.
    We define $\rho'=\rho'_0[X\mapsto \rho(X)]$.
    The property (iii) implies $V(\lb{E}{e'}\rho_0)\subseteq V(\lb{E_\I}{e'}(t',\rho'_0,I_\I))$.
    We get $\rho'\in R(\lbx{\Pi_\I}{\proj_t(p')\cdot (X\leftarrow e'),\,t}\br(\E_0,\emptyset,I_\I))=R(\lbx{\Pi_\I}{\proj_t(p),\,t}(\E_0,\emptyset,I_\I))$.
    As $\rho$ and $\rho_0$ are equal except maybe on $X$,
    and $\rho'$ and $\rho'_0$ are also equal except maybe on $X$,
    and on $X$ we have $\rho(X)=\rho'(X)$, 
    then $\rho'$ satisfies (ii) for $t$, $p$ and $\rho$.
  \end{iteMize}
  The theorem then stems from applying property (i) to all $p\in\pi_*$ and using
  Thm.~\ref{thm:structpath2}:
  $$\begin{array}{ll}
    &\lbp{P_*}\\[2pt]
    =&\Omega(\lbx{\Pi_*}{\pi_*}(\E_0,\emptyset))\\[2pt]
    =&\bigcup_{p\in\pi_*}\;\Omega(\lbx{\Pi_*}{p}(\E_0,\emptyset))\\[2pt]
    \subseteq&\bigcup_{t\in\T,\,p\in\pi_*}\;\Omega(\lbx{\Pi_\I}{\proj_t(p),\,t}(\E_0,\emptyset,I_\I))\\[2pt]
    =&\bigcup_{t\in\T}\;\Omega(\lbx{\Pi_\I}{\pi(\body_t),\,t}(\E_0,\emptyset,I_\I))\\[2pt]
    =&\bigcup_{t\in\T}\;\Omega(\lb{S_\I}{\body_t,\,t}(\E_0,\emptyset,I_\I))\\[2pt]
    \subseteq&\bigcup_{t\in\T}\;\Omega(\lb{S_\I}{\body_t,\,t}(\E_0,\Omega_\I,I_\I))\\[2pt]
    =&\Omega_\I\\[2pt]
    =&\lbp{P_\I}\enspace.
  \end{array}$$
\qed


\thmproof{soundproc}{$\lbp{P_\I}\subseteq\lbp{P_\I^\s}.$}
\proof 
  We start by considering the semantics of expressions.
  We first note that, for any abstract environment $R^\s\in\E^\s$, 
  abstract interference  $I^\s\in\I^\s$, thread $t$, and expression $e$,
  if $\rho\in\gamma_\E(R^\s)$, then 
  $\lb{E_\I}{e}(t,\rho,\gamma_\I(I^\s))\subseteq\lb{E}{\apply(t,R^\s,I^\s,e)}\rho$, i.e., 
  $\apply$ can be used to over-approximate the semantics with interference
  $\lbp{E_\I}$ using the non-interfering semantics $\lbp{E}$ in the concrete.
  We prove this by structural induction on the syntax of expressions $e$. 
  We consider first the case of variables $e=X\in\V$, which is the
  interesting case.
  When $\gamma_\E(R^\s)$ has no interference for $X$, then
  $\apply(t,R^\s,I^\s,X)=X$ and:
  $$\lb{E}{\apply(t,R^\s,I^\s,X)}\rho
  =\lb{E}{X}\rho\\[2pt]=\sset{\rho(X)}\\[2pt] 
  =\lb{E_\I}{X}(t,\rho,\gamma_\I(I^\s))\enspace.$$
  In case of interferences on $X$, we have
  $\apply(t,R^\s,I^\s,X)=\asexpr(V^\s_X\cup^\s_\RR\get(X,R^\s))$, which is an 
  interval  $[l,h]$ containing, by definition of $\asexpr$ and soundness of 
  $\get$, both $\rho(X)$ and 
  $\setst{v}{\ex{t'\neq t}(t',X,v)\in\gamma_\I(I^\s)}$.
  Thus, $\lb{E}{\apply(t,R^\s,I^\s,X)}\rho=[l,h]\subseteq\lb{E}{X}\rho$.
  For other expressions, we note that $\lbp{E_\I}$ and $\lbp{E}$ are defined
  in similar way, and so, the proof stems from the induction hypothesis and
  the monotony of $\lbp{E_\I}$ and $\lbp{E}$.

  To prove the soundness of primitive statements $\lb{S^\s_\I}{s}$,
  we combine the above result with the soundness of $\lbp{S^\s}$ with
  respect to $\lbp{S}$.
  We immediately get that
  $R((\lb{S_\I}{s}\circ\gamma)(R^\s,\Omega,I^\s))\subseteq
   R((\gamma\circ\lb{S^\s_\I}{s})(R^\s,\Omega,I^\s))$
  for an assignment or test $s$, and likewise for the $\Omega$ component.
  (We reuse the notations $R(X)$, $\Omega(X)$, $I(X)$ from the
  proof \ref{proof:interf}.)
  The $I$ component is unchanged by $\lb{S_\I}{s}$ and $\lb{S^\s_\I}{s}$ 
  when $s$ is a test.
  For the $I$ component after an assignment, we remark that 
  $I((\lb{S_\I}{X\leftarrow e}\circ\gamma)(R^\s,\Omega,I))$ can be written as
  $$\gamma_\I(I)\cup\setst{(t,X,\rho(X))}{\rho\in R((\lb{S_\I}{X\leftarrow e}\circ\gamma)(R^\s,\Omega,I))}\enspace.$$
  We then reuse the fact that
  $R((\lb{S_\I}{s}\circ\gamma)(R^\s,\Omega,I^\s))\subseteq
  R((\gamma\circ\lb{S^\s_\I}{s})(R^\s,\Omega,I^\s))$
  to conclude the proof of primitive statements.

  The case of non-primitive statements is easier as it is mostly unchanged between
  $\lbp{S}$ and $\lbp{S_\I}$ (hence, between $\lbp{S^\s}$ and $\lbp{S^\s_\I}$). 
  Hence, the proof in \ref{proof:sound} of Thm.~\ref{thm:sound} still applies.

  As a consequence, we have:
  $$\fa{t\in\T}(\lb{S_\I}{\body_t,\,t}\circ\gamma)(R^\s,\Omega,I^\s)\sqsubseteq_\I(\gamma\circ\lb{S_\I^\s}{\body_t,\,t})(R^\s,\Omega,I^\s)\enspace.$$

  Consider now the solutions $(\Omega_\I,I_\I)$ and
  $(\Omega^\s_\I,I^\s_\I)$ of the fixpoints (\ref{eq:intersem}) and
  (\ref{eq:absintersem}).
  We have:
  $$\begin{array}{rl}
    \multicolumn{2}{l}{(\Omega_\I,I_\I)=\lfp F}\\[2pt]
    \text{where}&F(\Omega,I)=\bigsqcup_{t\in\T}\;(\Omega'_t,I'_t)\\
    \text{and}&(-,\Omega'_t,I'_t)=\lb{S_\I}{\body_t,t}(\E_0,\Omega,I)\\
  \end{array}$$
  Likewise:
  $$\begin{array}{rl}
    \multicolumn{2}{l}{(\Omega^\s_\I,I^\s_\I)=\lim F^\s}\\[2pt]
    \text{where}&F^\s(\Omega^\s,I^\s)=(\Omega^\s,I^\s)\widen\bigsqcup_{t\in\T}^\s\;(\Omega^\s_t{}',I^\s_t{}')\\
    \text{and}&(-,\Omega^\s_t{}',I^\s_t{}')=\lb{S_\I^\s}{\body_t,\,t}(\E^\s_0,\Omega^\s,I^\s)\\[2pt]
    \text{defining}&(\Omega^\s_1,I^\s_1)\sqcup^\s(\Omega^\s_2,I^\s_2)\deq(\Omega^\s_1\cup\Omega^\s_2,I^\s_1\cup^\s_\I I^\s_2)\\
    \text{and}&(\Omega^\s_1,I^\s_1)\widen(\Omega^\s_2,I^\s_2)\deq(\Omega^\s_2,I^\s_1\widen_\I I^\s_2)\\
  \end{array}$$
  By soundness of $\cup^\s_\I$, $\widen_\I$, $\lbp{S_\I^\s}$, and $\E^\s_0$,
  we get that $F^\s$ is a sound abstraction of $F$.
  The same fixpoint transfer property that was used for loops in
  proof~\ref{proof:sound} can be used here to prove that
  $\lim F^\s$ is a sound abstraction of $\lfp F$.
  As a consequence, we have $\Omega^\s_\I\supseteq\Omega_\I$, and so,
  $\lbp{P^\s_\I}\supseteq\lbp{P_\I}$.
\qed


\thmproof{weakinterf}{$\lbp{P'_*}\subseteq\lbp{P_\I}$.}
\proof 
  We reuse the notations $R(X)$, $\Omega(X)$, $I(X)$, $V(X)$ from
  proof~\ref{proof:interf}.
  Consider a path $p$ from a thread $t$ that gives, under an elementary 
  transformation from Def.~\ref{def:transform}, a path $p'$.
  Let us denote by $\V_f$ the subset of fresh variables (i..e, that do not 
  appear anywhere in the program, except maybe in $p'$),
  and by $\V_l(t)$ the subset of variables that are local to $t$ (i.e., 
  that do not appear anywhere, except maybe in thread $t$).

  We consider triples $(R,\Omega,I)$ such that the
  interferences $I$ are consistent
  with the fresh and local variables, i.e.,
  if $(t,X,v)\in I$, then $X\notin\V_f$ and $X\in\V_l(t')\Longrightarrow t=t'$.
  We prove that, for such triples, the following holds:
  $$\begin{array}{l@{\quad}l}
    \text{(i)}   & \Omega(\lbx{\Pi_\I}{p',t}(R,\Omega,I)) \subseteq \Omega(\lbx{\Pi_\I}{p,t}(R,\Omega,I)) \\[8pt]
    \text{(ii)}  & \fa{(t',X,v)\in I(\lbx{\Pi_\I}{p',t}(R,\Omega,I))}\\[2pt]&
    (t',X,v)\in I(\lbx{\Pi_\I}{p,t}(R,\Omega,I))
    \text{ or }(t=t'\wedge X\in\V_l(t))\\[8pt]
    \text{(iii)} & \fa{\rho'\in R(\lbx{\Pi_\I}{p',t}(R,\Omega,I))}
    \ex{\rho\in R(\lbx{\Pi_\I}{p,t}(R,\Omega,I))}\\[2pt]
    &\fa{X\in\V}\rho(X)=\rho'(X)\text{ or }X\in \V_f\\
  \end{array}$$
  i.e., the transformation does not add any error (i), it can only add 
  interferences on local variables (ii), and change environments on
  fresh variables (iii).
  We now prove (i)--(iii) for each case in Def.~\ref{def:transform}.
  \begin{enumerate}[(1)]
  \item Redundant store elimination:
    $X\leftarrow e_1\cdot X\leftarrow e_2\becomes X\leftarrow e_2$,
    where $X\notin\var(e_2)$ and $\nonblock(e_1)$.

    \noindent
    We note:
    \begin{iteMize}{$-$}
    \item[--] $(R_i,\Omega_i,I_i)=\lbx{\Pi_\I}{X\leftarrow e_i,\,t}(R,\Omega,I)$ for $i=1,2$, and
    \item[--] $(R_{1;2},\Omega_{1;2},I_{1;2})=\lbx{\Pi_\I}{X\leftarrow e_1\cdot X\leftarrow e_2,\,t}(R,\Omega,I)$.
    \end{iteMize}
    As $X\notin\var(e_2)$, $\lb{E_\I}{e_2}(t,\rho,I)=\lb{E_\I}{e_2}(t,\rho[X\mapsto v],I)$ for any $\rho$ and $v$.
    Moreover, as $\nonblock(e_1)$:
    $$\fa{\rho\in R}\ex{v\in V(\lb{E_\I}{e_1}(t,\rho,I))}\rho[X\mapsto v]\in R_1\enspace.$$
    This implies $R_{1;2}=R_2$, and so, (iii).
    Moreover, $\Omega_{1;2}=\Omega_1\cup\Omega_2\supseteq\Omega_2$, and so, (i).
    Likewise, $I_{1;2}=I_1\cup I_2\supseteq I_2$, and so, (ii).
    
    Note that the hypothesis $\nonblock(e_1)$ is important.
    Otherwise we could allow $X\leftarrow 1/_\ell\,0\cdot X\leftarrow 1/_{\ell'}\,0 \becomes X\leftarrow 1/_{\ell'} 0$,
    where the error $\ell'$ is in the transformed program but not in the original
    one (here, $\Omega_{1;2}=\Omega_1\not\supseteq\Omega_2$).
    \medskip

  \item Identity store elimination $X\leftarrow X\becomes\epsilon$.
    
    \noindent
    We have:
    $$\begin{array}{lcl}
      \Omega(\lbx{\Pi_\I}{X\leftarrow X,\,t}(R,\Omega,I))&=&\Omega\\
      I(\lbx{\Pi_\I}{X\leftarrow X,\,t}(R,\Omega,I))&\supseteq& I\\
      R(\lbx{\Pi_\I}{X\leftarrow X,\,t}(R,\Omega,I))&\supseteq& R\enspace.
    \end{array}$$
    In the two last inequalities, the converse inequality 
    does not necessarily hold.
    Indeed, $X\leftarrow X$ creates interferences that may be observed
    by other thread.
    Moreover $V(\lb{E_\I}{X}(t,\rho,I))$ is not necessarily $\set{\rho(X)}$, 
    but may contain interferences from other threads.
    This shows in particular that the converse transformation, identity
    insertion, is not acceptable as it introduces interferences.
    \medskip

  \item Reordering assignments:
    $X_1\leftarrow e_1\cdot X_2\leftarrow e_2\becomes X_2\leftarrow e_2\cdot X_1\leftarrow e_1$,
    where $X_1\notin\var(e_2)$, $X_2\notin\var(e_1)$, $X_1\neq X_2$, and $\nonblock(e_1)$.

    \noindent
    We note:
    \begin{iteMize}{$-$}
    \item[--] $(R_i,\Omega_i,I_i)=\lbx{\Pi_\I}{X_i\leftarrow e_i,\,t}(R,\Omega,I)$
      for $i=1,2$, and
    \item[--] 
      $(R_{1;2},\Omega_{1;2},I_{1;2})=\lbx{\Pi_\I}{X_1\leftarrow e_1\cdot X_2\leftarrow e_2,\,t}(R,\Omega,I)$, and
    \item[--] 
      $(R_{2;1},\Omega_{2;1},I_{2;1})=\lbx{\Pi_\I}{X_2\leftarrow e_2\cdot X_1\leftarrow e_1,t}(R,\Omega,I)$.
    \end{iteMize}
    \smallskip
    As $X_2\notin\var(e_1)$, $X_1\notin\var(e_2)$, and $X_1\neq X_2$, we have
    $$\begin{array}{l}
      \fa{\rho,v}\lb{E_\I}{e_1}(t,\rho,I)=\lb{E_\I}{e_1}(t,\rho[X_2\mapsto v],I)\text{  and }\\
      \fa{\rho,v}\lb{E_\I}{e_2}(t,\rho,I)=\lb{E_\I}{e_2}(t,\rho[X_1\mapsto v],I)\enspace.
    \end{array}$$
    This implies $R_{1;2}=R_{2;1}$, and so, (iii).

    As $\nonblock(e_1)$, we have $\fa{\rho\in R}\ex{v}\rho[X_1\mapsto v]\in R_1$.
    This implies $\Omega_2=\Omega(\lbx{\Pi_\I}{X_2\leftarrow e_2,\,t}(R_1,\Omega,I))$, and so
    $\Omega_{1;2}=\Omega_1\cup\Omega_2$.
    Likewise, $I_{1;2}=I_1\cup I_2$.
    Moreover, $\Omega_{2;1}\subseteq \Omega_1\cup\Omega_2=\Omega_{1;2}$ and 
    $I_{2;1}\subseteq I_1\cup I_2=I_{1;2}$.
    Thus (i) and (ii) hold.

    As in (1), the $\nonblock(e_1)$ hypothesis is important so that
    errors in $e_2$ masked by $X_1\leftarrow e_1$ in the original program
    do not appear in the transformed program.
    \medskip

  \item Reordering guards:
    $e_1\bowtie 0?\cdot e_2\bowtie' 0?\becomes e_2\bowtie' 0?\cdot e_1\bowtie 0?$,
    where $\noerror(e_2)$.

    \noindent
    We use the same notations as in the preceding proof.
    We have $I_{2;1}=I_{1;2}=I$, which proves (ii).
    Consider $\rho\in R$, then either $\rho\in R_1\cap R_2$, in which case
    $\rho\in R_{1;2}$ and $\rho\in R_{2;1}$, or $\rho\notin R_1\cap R_2$, in
    which case $\rho\notin R_{1;2}$ and $\rho\notin R_{2;1}$.
    In both cases, $R_{1;2}=R_{2;1}$, which proves (iii).
    We have
    $\Omega_2\subseteq\Omega_{2;1}\subseteq \Omega_1\cup\Omega_2$ and
    $\Omega_1\subseteq\Omega_{1;2}\subseteq \Omega_1\cup\Omega_2$.
    But, as $\noerror(e_2)$, $\Omega_2=\emptyset$, which implies
    $\Omega_{2;1}\subseteq\Omega_1\subseteq\Omega_{1;2}$, and so, (i).

    The property $\noerror(e_2)$ is important.
    Consider otherwise the case where $e_1\bowtie 0?$ filters out an 
    environment leading to an error in $e_2$.
    The error would appear in the transformed program but not in the
    original one.
    \medskip

  \item Reordering guards before assignments:
    $X_1\leftarrow e_1\cdot e_2\bowtie 0?\becomes e_2\bowtie 0?\cdot X_1\leftarrow e_1$,
    where $X_1\notin\var(e_2)$ and either $\nonblock(e_1)$ or $\noerror(e_2)$.
    
    \noindent
    We use the same notations as in the preceding proofs.
    As $X_1\notin\var(e_2)$, we have
    $$\fa{\rho,v}
    \lb{E_\I}{e_2}(t,\rho,I)=\lb{E_\I}{e_2}(t,\rho[X_1\mapsto v],I)\enspace.$$
    This implies $R_{1;2}=R_{2;1}$, and so, (iii).

    Moreover, we have $I_{1;2}=I_1$ and $I_{2;1}\subseteq I_1$, thus (ii) holds.

    For (i), consider first the case $\nonblock(e_1)$.
    We have $\fa{\rho\in R}\ex{v}\rho[X_1\mapsto v]\in R_1$.
    This implies $\Omega_2=\Omega(\lbx{\Pi_\I}{e_2\bowtie 0?,\,t}(R_1,\Omega,I))$, and so
    $\Omega_{1;2}=\Omega_1\cup\Omega_2$.
    As $\Omega_2\subseteq\Omega_{2;1}\subseteq \Omega_1\cup\Omega_2$,
    (i) holds.
    Consider now the case $\noerror(e_2)$.
    We have $\Omega_{1;2}=\Omega_1$ and
    $\Omega_{2;1}\subseteq\Omega_1$, and so, (i) holds.
    \medskip

  \item Reordering assignments before guards:
    $e_1\bowtie 0?\cdot X_2\leftarrow e_2\becomes X_2\leftarrow e_2\cdot e_1\bowtie0?$,
    where $X_2\notin\var(e_1)$, $X_2\in\V_l(t)$, and $\noerror(e_2)$.

    \noindent
    As $X_2\notin\var(e_1)$, we have
    $$\fa{\rho,v}
    \lb{E_\I}{e_1}(t,\rho,I)=\lb{E_\I}{e_1}(t,\rho[X_2\mapsto v],I)$$
    and thus, using the same notations as before,
    $R_{1;2}=R_{2;1}$, and so, (iii).

    Because, $\noerror(e_2)$, we have
    $\Omega_{2;1}\subseteq\Omega_1\cup\Omega_2=\Omega_1$ and $\Omega_{1;2}=\Omega_1$, and so $\Omega_{2;1}\subseteq\Omega_{1;2}$ (i).

    Unlike the preceding proofs, we now have $I_{1;2}\subseteq I_{2;1}=I_2$
    and generally $I_{1;2}\not\supseteq I_{2;1}$.
    However, as $X_2\in\V_l(t)$, we have
    $I_{2;1}\setminus I_{1;2}\subseteq \sset{t}\times\V_l(t)\times\R$, i.e.,
    the only interferences added by the transformation concern the variable 
    $X_2$, local to $t$.
    This is sufficient to ensure (ii).
    \medskip

  \item Assignment propagation: $X\leftarrow e\cdot s\becomes X\leftarrow e\cdot s[e/X]$,
    where $X\notin\var(e)$, $\var(e)\subseteq\V_l(t)$, and $e$ is deterministic.

    \noindent
    Let us note :
    \begin{iteMize}{$-$}
    \item[--] $(R_1,\Omega_1,I_1)=\lbx{\Pi_\I}{X\leftarrow e,\,t}(R,\Omega,I)$,\smallskip
    \item[--] $(R_{1;2},\Omega_{1;2},I_{1;2})=\lbx{\Pi_\I}{s,\,t}(R_1,\Omega_1,I_1)$,\smallskip
    \item[--] $(R'_{1;2},\Omega'_{1;2},I'_{1;2})=\lbx{\Pi_\I}{s[e/X],\,t}(R_1,\Omega_1,I_1)$.\smallskip
    \end{iteMize}
    Take $\rho_1\in R_1$, then there exists $\rho\in R$ such that
    $\rho_1=\rho[X\mapsto \rho_1(X)]$ and 
    $\rho_1(X)\in V(\lb{E_\I}{e}(t,\rho,I))$.
    As $e$ is deterministic and $X\notin\var(e)$,
    $V(\lb{E}{X}\rho_1)=V(\lb{E}{e}\rho)=\sset{\rho_1(X)}$.
    Additionally, as $\var(e)\subseteq\V_l(t)$, there is no interference
    in $I$ for variables in $e$ from other threads, and so,
    $V(\lb{E_\I}{X}(t,\rho_1,I_1))=V(\lb{E_\I}{e}(t,\rho,I))=\sset{\rho_1(X)}$.
    As a consequence, $R_{1;2}=R'_{1;2}$, which proves (iii).
    Moreover, $I_{1;2}=I'_{1;2}$, hence (ii) holds.
    Finally, we have $\bigcup\,\setst{\Omega(\lb{E_\I}{e}(t,\rho,I))}{\rho\in R}\subseteq \Omega_1$ but also $\bigcup\,\setst{\Omega(\lb{E_\I}{e}(t,\rho_1,I_1))}{\rho_1\in R_1}\subseteq\Omega_1$, i.e.,
    any error from the evaluation of $e$ 
    while executing $s[e/X]$ in $R_1$ was already present during the 
    evaluation of $e$ while executing $X\leftarrow e$ in $R$.
    Thus, $\Omega_{1;2}=\Omega'_{1;2}$, and (i) holds.

    The fact that $e$ is deterministic is important.
    Otherwise, consider $X\leftarrow[0,1]\cdot Y\leftarrow X+X \becomes X\leftarrow[0,1]\cdot Y\leftarrow [0,1]+[0,1]$.
    Then $Y\in\sset{0,2}$ on the left of $\becomes$
    while $Y\in\sset{0,1,2}$ on the right of $\becomes$:
    the transformed program has more behaviors than the original one.
    Likewise, $\var(e)\subseteq\V_l(t)$ ensures that $e$ may not evaluate
    to a different value due to interferences from other threads.
    \medskip

  \item Sub-expression elimination:
    $s_1\cdot \ldots \cdot s_n\becomes X\leftarrow e\cdot s_1[X/e]\cdot \ldots\cdot s_n[X/e]$,
    where $X\in\V_f$, $\var(e)\cap\lval(s_i)=\emptyset$,
    and $\noerror(e)$.

    \noindent
    Let us note: $$(R',\Omega',I')=\lbx{\Pi_\I}{X\leftarrow e,\,t}(R,\Omega,I)\enspace.$$
    Consider $\rho'\in R'$.
    Then $\rho'=\rho[X\mapsto \rho'(X)]$ for some $\rho\in R$ and
    $\rho'(X)\in V(\lb{E_\I}{e}(t,\rho,I))$.
    As $X$ does not appear in $s_i$ (being fresh), and noting:
    $$(R'_i,\Omega'_i,I'_i)=\lbx{\Pi_\I}{s_1[X/e]\cdot \ldots\cdot s_i[X/e],\,t}(\set{\rho'},\Omega',I')$$
    we get: $$\fa{i,\,\rho'_i\in R'_i}V(\lb{E}{X}\rho'_i)=\sset{\rho'(X)}$$ and,
    as  $X\in\V_f$, there is no interference on $X$, and
    $$\fa{\rho'_i\in R'_i}V(\lb{E_\I}{X}(t,\rho'_i,I'_i))=\sset{\rho'(X)}\enspace.$$
    As $\var(e)\cap\lval(s_i)=\emptyset$, and noting
    $$(R_i,\Omega_i,I_i)=\lbx{\Pi_\I}{s_1\cdot \ldots \cdot s_i,\,t}(\sset{\rho},\Omega,I)$$
    we get: $$\fa{i,\,\rho_i\in R_i}V(\lb{E_\I}{e}(t,\rho_i,I_i))=V(\lb{E_\I}{e}(t,\rho,I))\supseteq\sset{\rho'(X)}\enspace.$$
    As a consequence, $\fa{i,\,\rho'_i\in R'_i}\exists \rho_i\in R_i$ such
    that $\rho'_i=\rho_i[X\mapsto \rho'(X)]$.
    When $i=n$, this implies (iii).
    Another consequence is that
    $\fa{i}I'_i\subseteq I_i \cup \setst{(t,X,\rho'(X))}{\rho'\in R'}$.
    As $X\in\V_f$, this implies (ii) when $i=n$.
    Moreover, as $\noerror(e)$, $\Omega'=\Omega$.
    Note that: $$\fa{i}\Omega(\lbx{\Pi_\I}{s[X/e],\,t}(R'_i,\Omega'_i,I'_i))\subseteq\Omega(\lbx{\Pi_\I}{s,\,t}(R'_i,\Omega'_i,I'_i))\enspace.$$
    Thus, $\fa{i}\Omega'_i\subseteq\Omega_i$, which implies (i) when $i=n$.
    \medskip

  \item Expression simplification: $s\becomes s[e'/e]$,
    when $\fa{\rho}\lb{E}{e}\rho\sqsupseteq\lb{E}{e'}\rho$
    and $\var(e)\cup\var(e')\subseteq\V_l(t)$.

    \noindent
    As $\var(e)\subseteq\V_l(t)$, there is no interference in $I$ for
    variables in $e$ from other threads.
    We deduce that $\lb{E_\I}{e}(t,\rho,I)=\lb{E}{e}\rho$.
    Likewise, $\lb{E_\I}{e'}(t,\rho,I)=\lb{E}{e'}\rho$, and so,
    $\lb{E_\I}{e}(t,\rho,I)\sqsupseteq\lb{E_\I}{e'}(t,\rho,I)$.
    By monotony of $\lbp{S_\I}$, we get:
    $$\lbx{\Pi_\I}{s,\,t}(R,\Omega,I)\sqsupseteq \lbx{\Pi_\I}{s[e'/e],\,t}(R,\Omega,I)$$
    which implies (i)--(iii).
  \end{enumerate}
  We now prove that, if the pair $p,p'$ satisfies (i)--(iii),
  then so does the pair $p\cdot s,\,p'\cdot s$ for any primitive statement $s$ 
  executed by $t$.
  We consider a triple $(R_0,\Omega_0,I_0)$ and note
  $(R',\Omega',I')=\lbx{\Pi_\I}{p',t}(R_0,\Omega_0,I_0)$
  and $(R,\Omega,I)=\lbx{\Pi_\I}{p,t}(R_0,\Omega_0,I_0)$.
  Take any $\rho'\in R'$.
  By (iii) on the pair $p,p'$, there is some $\rho\in R$ that equals
  $\rho'$ except maybe for some variables that are free, and so, cannot occur 
  in $s$.
  Moreover, by (ii) on $p,p'$, $I$ contains $I'$ except maybe for some
  interferences that are from $t$, and so, cannot influence the expression 
  in $s$.
  So, by noting $e$ the expression appearing in $s$, 
  $\lb{E_\I}{e}(t,\rho',I')\sqsubseteq_\I\lb{E_\I}{e}(t,\rho,I)$.
  As a consequence, (i) and (ii) hold for the pair $p\cdot s,\,p'\cdot s$.
  Consider now $\rho'\in R(\lbx{\Pi_\I}{p'\cdot s,\,t}(R_0,\Omega_0,I_0))$.
  If $s$ is a guard $e=0?$ (the other guards $e\bowtie 0?$ being similar), 
  then $\rho'\in R'$ and, by (iii) for $p,p'$,
  $\exists \rho\in R$ equal to $\rho'$
  except for some free variables.
  As, $0\in V(\lb{E_\I}{e}(t,\rho',I'))\subseteq V(\lb{E_\I}{e}(t,\rho,I))$, 
  $\rho$ proves the property (iii) for  $p\cdot s,\,p'\cdot s$.
  If $s$ is an assignment $X\leftarrow e$ then $\rho'=\rho'_0[X\mapsto \rho'(X)]$ for
  some $\rho'_0\in R'$ and $\rho'(X)\in V(\lb{E_\I}{e}(t,\rho'_0,I'))$.
  By (iii) for $p,p'$, $\exists \rho_0\in R$ 
  equal to $\rho'_0$ except for some free variables.
  So, $\rho'(X)\in V(\lb{E_\I}{e}(t,\rho'_0,I'))\subseteq V(\lb{E_\I}{e}(t,\rho_0,I))$.
  Thus, $\rho_0[X\mapsto \rho'(X)]$ proves the property (iii) for  $p\cdot s,\,p'\cdot s$.

  The following two properties are much easier to prove.
  Firstly, if the pair $p,p'$ satisfies (i)--(iii),
  so does the pair $q\cdot p,\,q\cdot p'$ for any path $q$.
  This holds because (i)--(iii) are stated for all $R$, $\Omega$ and $I$.
  Secondly, if both pairs $p,p'$ and $p',p''$ satisfy (i)--(iii), 
  then so does the pair $p,p''$.
  This completes the proof that elementary path transformations can be applied 
  in a context with an arbitrary prefix and suffix, and that
  several transformations can be applied sequentially.

  We are now ready to prove the theorem.
  Consider the least fixpoint computed by the interference semantics 
  (\ref{eq:intersem}):
  $$\begin{array}{l}
   (\Omega_\I,I_\I)=\lfp F\\[2pt]
   \text{where }F(\Omega,I)=(\bigcup_{t\in\T}\,\Omega_t,\,\bigcup_{t\in\T}\,I_t)\\[2pt]
    \text{and }(-,\Omega_t,I_t)=\lb{S_\I}{\body_t,\,t}(\E_0,\Omega,I)\enspace.
  \end{array}$$
  By Thm.~\ref{thm:structpath2}, we have
  $(-,\Omega_t,I_t)=\lbx{\Pi_\I}{\pi(\body_t),\,t}(\E_0,\Omega,I)$.
  Given transformed threads $\pi'(t)$, consider also the
  fixpoint:
  $$\begin{array}{l}
    (\Omega'_\I,I'_\I)=\lfp F'\\[2pt]
    \text{where }F'(\Omega',I')=(\bigcup_{t\in\T}\,\Omega'_t,\,\bigcup_{t\in\T}\,I'_t\setminus\I_l)\\[2pt]
    \text{and }(-,\Omega'_t,I'_t)=\lbx{\Pi_\I}{\pi'(t),\,t}(\E_0,\Omega',I')
  \end{array}$$
  and $\I_l$ is a set of interferences we can ignore as they only affect
  local or fresh variables:
  $$\I_l=\setst{(t,X,v)}{t\in\T,\,v\in\R,\,X\in\V_f\cup\V_l(t)}\enspace.$$
  Then, given each path in $p'\in\pi'(t)$, we can apply properties (i) and (ii)
  to the pair $p$, $p'$, where $p$ is the path in $\pi(\body_t)$ that gives
  $p'$ after transformation.
  We get that $F'(X)\sqsubseteq_\I F(X)$.
  As a consequence, $\lfp F'\sqsubseteq_\I\lfp F$.
  The transformed semantics, however, is not exactly $\lfp F'$, but rather:
  $$\begin{array}{l}
     (\Omega''_t,I''_t)=\lfp F''\\[2pt]
     \text{where }F''(\Omega',I')=(\bigcup_{t\in\T}\,\Omega'_t,\bigcup_{t\in\T}\,I'_t)\\[2pt]
     \text{and }(\Omega'_t,I'_t)\text{ defined as before}
   \end{array}$$ 
  The difference lies in the extra interferences generated by
  the transformed program, which are all in $\I_l$.
  Such interferences, however, have no influence on the semantics of
  threads, as we have:
  $$\lbx{\Pi_\I}{\pi'(t),\,t}(\E_0,\Omega,I')=\lbx{\Pi_\I}{\pi'(t),\,t}(\E_0,\Omega,I'\setminus\I_l)\enspace.$$
  Indeed, any interference $(t',X,v)\in\I_l$ is either from thread
  $t$ and  then ignored 
  for the thread $t$ itself, or it is from another thread $t'\neq t$ in
  which case $X$ 
  is a local variable of $t$ or a free variable, which does not occur in $t'$.
  As a consequence, $I'_t=I''_t$ and $\Omega''_t=\Omega'_t$, and so,
  $\Omega''=\Omega'\subseteq\Omega$.
  Hence, the interference semantics of the original program contains all the 
  errors that can occur in any program obtained by acceptable thread
  transformations.
\qed


\thmproof{soundschedinterfer}{$\lbp{P_\H}\subseteq\lbp{P_\C}$ and $\lbp{P'_\H}\subseteq\lbp{P_\C}.$}
\proof
  We first prove $\lbp{P_\H}\subseteq\lbp{P_\C}$.

  In order to do so, we need to consider a path-based interference semantics
  $\lbx{\Pi_\C}{P,t}$ that, given a set of paths $P\subseteq\pi(\body_t)$ in a thread $t$,
  computes:
  \begin{center}$\lbx{\Pi_\C}{P,t}(R,\Omega,I)\deq\bigsqcup_\C\;\setst{(\lb{S_\C}{s_n,t}\circ\cdots\circ\lb{S_\C}{s_1,t})(R,\Omega,I)}{s_1\cdot \ldots\cdot s_n\in P}\enspace.$\end{center}
  Similarly to Thms.~\ref{thm:structpath}, \ref{thm:structpath2},
  the two semantics are equal:
  $$\fa{t\in\T,\,s\in\mi{stat}}\lbx{\Pi_\C}{\pi(s),\,t}=\lb{S_\C}{s,\,t}$$
  The proof is identical to that in \ref{proof:structpath} as
  the $\lbp{S_\C}$ functions are complete $\sqcup-$morphisms, and so, 
  we do not repeat it here.

  The rest of the proof that $\lbp{P_\H}\subseteq\lbp{P_\C}$ follows a similar
  structure as \ref{proof:interf}.
  Let $(\Omega_\C,I_\C)$ be the fixpoint computed in (\ref{eq:interschedsem}), 
  i.e.,
  $$\begin{array}{l}
    (\Omega_\C,I_\C)=\bigsqcup_\C\setst{(\Omega'_t,I'_t)}{t\in\T}\\
    \text{where }(-,\Omega'_t,I'_t)=\lb{S_\C}{\body_t,t}(\sset{c_0}\times\E_0,\Omega_\C,I_\C)\enspace.
  \end{array}$$
  We denote initial states for $\lbp{P}_\H$ and $\lbp{P}_\C$ as respectively
  $\E_{0h}\deq\sset{h_0}\times\E_0$ and $\E_{0c}\deq\sset{c_0}\times\E_0$.  
  Furthermore, we link scheduler states $h\in \H$
  and partitioning configurations $c\in \C$ in a thread $t$
  with the following abstraction $\alpha_t:\H\rightarrow\P(\C)$:
  $$
  \alpha_t(b,l)\deq
  \setst{(l(t),u,\weak)}{\fa{x\in u,t'\in\T}x\notin l(t')}
  \enspace.
  $$
  i.e., a configuration forgets the ready state $b(t)$ of the thread
  ($\ready$, $\yield$ing, or $\wait$ing for some mutex), 
  but remembers the exact set of mutexes $l(t)$
  held by the current thread, and optionally remembers the mutexes
  not held by any thread $u$.
  We prove the following properties by induction on the length of the path
  $p\in\pi_*$:
  $$\begin{array}{l@{\quad}l}
    \text{(i)}  & \Omega(\lbx{\Pi_\H}{p}(\E_{0h},\emptyset))\subseteq\bigcup_{t\in\T}\;\Omega(\lbx{\Pi_\C}{\proj_t(p),\,t}(\E_{0c},\emptyset,I_\C)) \\[8pt]
    \text{(ii)}  & \fa{t\in\T}\fa{(h,\rho)\in R(\lbx{\Pi_\H}{p}(\E_{0h},\emptyset))}
    \ex{(c,\rho')\in R(\lbx{\Pi_\C}{\proj_t(p),\,t}(\E_{0c},\emptyset,I_\C))}\\[2pt]&
    c\in\alpha_t(h)\wedge\\[2pt]
    &\fa{X\in\V}\rho(X)=\rho'(X)\text{ or }
    \ex{t',c'}t\neq t',\,\excl(c,c'),\,(t',c',X,\rho(X))\in I_\C.
  \end{array}$$

  The properties hold for $p=\epsilon$ as $\fa{t\in\T}$
  $$\begin{array}{lcl}
    \lbx{\Pi_\H}{\epsilon}(\E_{0h},\emptyset)&=&(\E_{0h},\emptyset)\\[2pt]
    \lbx{\Pi_\C}{\epsilon,t}(\E_{0c},\emptyset,I_\C)&=&(\E_{0c},\emptyset,I_\C)
  \end{array}$$
  and $c_0\in\alpha_t(h_0)$.

  Assume that the properties hold for $p'$ and consider $p=p'\cdot (s',t')$, i.e.,
  $p'$ followed by a primitive statement $s'$ executed by a thread $t'$.
  The case where $s'$ is an assignment or a guard is very similar to that of
  proof \ref{proof:interf}: we took care to update (ii) to reflect the
  change in the evaluation of variables in expressions
  $\lb{E_\C}{X}$ (in particular, the use of $\excl$ to determine which
  interferences from other threads can influence the current one, given their
  respective configuration).
  The effect of $\enabled_t$ in $\lb{S_\H}{s',t'}$ is to remove states from
  $R(\lbx{\Pi_\H}{p'}(\E_{0h},\emptyset))$, and so, it does not invalidate (ii).
  Moreover, as assignments and guards do not modify the scheduler
  state, the subsequent $\sched$ application has no effect.
  Consider now the case where $s'$ is a synchronization primitive.
  Then (i) holds as these primitives do not modify the error set.
  We now prove (ii).
  Given $(h,\rho)\in R(\lbx{\Pi_\H}{p}(\E_{0h},\emptyset))$, there
  is a corresponding 
  $(h_1,\rho_1)\in R(\lbx{\Pi_\H}{p'}(\E_{0h},\emptyset))$
  such that
  $(h,\rho)\in R(\lb{S_\H}{s',t'}(\sset{(h_1,\rho_1)},\emptyset))$.
  Given $t\in \T$, we apply (ii) on $(h_1,\rho_1)$ and $p'$, and get a
  state
  $(c_1,\rho'_1)\in R(\lbx{\Pi_\C}{\proj_t(p'),\,t}(\E_{0c},\emptyset,I_\C))$
  with $c_1\in\alpha_t(h_1)$.
  We will note $(l_1,u_1,-)$ the components of $c_1$.
  We construct a state in $\C\times \E$ satisfying (ii) for $p$.
  We first study the case where $t=t'$ and consider several sub-cases
  depending on $s'$:

  \begin{iteMize}{$\bullet$}
  \item
    Case $s'=\sy{yield}$.
    We have $\rho_1=\rho$ and $\alpha_t(h_1)=\alpha_t(h)$.
    We choose $c=(l_1,\emptyset,\weak)$.
    Then $(c,\rho'_1)\in R(\lb{S_\C}{s',\,t}(\set{(c_1,\rho'_1)},\Omega,I_\C))$.
    Moreover, $c\in\alpha_t(h)$.
    We also have, $\fa{c'\in\C}\excl(c_1,c')\Longrightarrow\excl(c,c')$.
    Hence, $\fa{X\in\V}$ either $\rho(X)=\rho'_1(X)$, or $\rho'_1(X)$ 
    comes from some weakly consistent interference not in exclusion with 
    $c_1$, and so, not in exclusion with $c$.
    As a consequence, $(c,\rho'_1)$ satisfies (ii) for $p'$.
    \medskip

  \item
    Case $s'=\sy{lock}(m)$.
    We choose $c=(l_1\cup\sset{m},\emptyset,\weak)$.
    This ensures as before that $c\in\alpha_t(h)$.
    Moreover, $\rho_1=\rho$.
    We now construct $\rho'$ such that
    $$(c,\rho')\in R(\lb{S_\C}{s',\,t}(\set{(c_1,\rho'_1)},\Omega,I_\C))$$
    and
    $$\fa{X\in\V}\rho(X)=\rho'(X)\text{ or }\ex{t'',c'}t'\neq t'',\excl(c,c'),\,(t'',c',X,\rho(X))\in I_\C\enspace.$$
    \begin{iteMize}{$-$}
    \item 
      If $\rho'_1(X)=\rho_1(X)$, then we take 
      $\rho'(X)=\rho'_1(X)=\rho_1(X)=\rho(X)$.
      \smallskip

    \item
      Otherwise, we know that $\rho(X)=\rho_1(X)$ comes from a
      weakly consistent interference compatible with $c_1$:
      $\ex{t'',c'}t'\neq t'',\,\excl(c_1,c'),\,(t'',c',X,\rho_1(X))\in I_\C$.
      If $\excl(c,c')$ as well, then the same weakly consistent interference 
      is compatible with $c$ and can be applied to $\rho'$. 
      We can thus set $\rho'(X)=\rho'_1(X)$.
      \smallskip

    \item 
      Otherwise, as $\excl(c,c')$ does not hold, then either $m\in l'$ or
      $m\in u'$, where $(l',u',-)=c'$.

      Assume that $m\in l'$, i.e., $\rho_1(X)$ was written to $X$ by thread
      $t''$ while holding the mutex $m$.
      Because $R(\lbx{\Pi_\H}{p}(\E_{0h},\emptyset))\neq\emptyset$, the mutex
      $m$ is unlocked before $t'$ executes $s'=\sy{lock}(m)$, so,
      $t''$ executes $\sy{unlock}(m)$ at some point in an environment 
      mapping $X$ to $\rho_1(X)$.
      Note that $\lb{S_\C}{\sy{unlock}(m),\,t''}$ calls
      $\funout$ to convert the weakly consistent interference 
      $(t'',c',X,\rho_1(X))\in I_\C$ to a well synchronized interference
      $(t'',(l'\setminus\sset{m},u',\sync(m)),X,\rho_1(X))\in I_\C$.
      This interference is then imported by $\lb{S_\C}{\sy{lock}(m),\,t'}$
      through $\funin$. Thus:
      $$(c,\rho'_1[X\mapsto \rho_1(X)])\in R(\lbx{\Pi_\C}{\proj_t(p),\,t}(\E_{0c},\emptyset,I_\C))$$
      and we can set $\rho'(X)=\rho_1(X)=\rho(X)$.

      The case where $m\in u'$ is similar, except that the weakly consistent
      interference is converted to a well synchronized one by a statement
      $\lb{S_\C}{\sy{yield},\,t''}$ or $\lb{S_\C}{\sy{unlock}(m'),\,t''}$
      for an arbitrary mutex $m'$, in an environment where $X$ maps to
      $\rho_1(X)$.

      \smallskip

    \end{iteMize}
    In all three cases, $(c,\rho')$ satisfies (ii) for $p'$.

    \medskip

  \item
    Case $s'=\sy{unlock}(m)$.
    We have $\rho_1=\rho$.
    We choose $c=(l_1\setminus\sset{m},u_1,\weak)$, which implies
    $c\in\alpha_t(h)$.
    Moreover, as in the case of $\sy{yield}$,
    $\fa{c'}\excl(c_1,c')\Longrightarrow\excl(c,c')$.
    Similarly, $(c,\rho'_1)$ satisfies (ii) for $p'$.
    \medskip

  \item
    Case $s'=X\leftarrow\sy{islocked}(m)$.
    We have $\fa{Y\neq X}\rho_1(Y)=\rho(Y)$ and $\rho(X)\in\sset{0,1}$.
    When $X\leftarrow\sy{islocked}(m)$ is interpreted as $X\leftarrow[0,1]$, 
    the result is straightforward: 
    we set $c=c_1$ and $\rho'=\rho'_1[X\mapsto\rho(X)]$.
    Otherwise, if $\rho(X)=0$, we set $c=(l_1,u_1\cup\sset{m},\weak)$ and, if
    $\rho(X)=1$, we set $c=(l_1,u_1\setminus\sset{m},\weak)$, so that
    $c\in\alpha_t(h)$.
    Moreover, when $\rho(X)=0$, then $\rho'$ is constructed as in the case
    of $\sy{lock}(m)$, except that $\rho'(X)$ is set to 0.
    Likewise, the case $\rho(X)=1$ is similar to that of $\sy{yield}$ and
    $\sy{unlock}(m)$, except that we also set $\rho'=\rho'_1[X\mapsto 1]$.
    In all cases $(c,\rho')$ satisfies (ii) for $p'$.
    \bigskip

  \end{iteMize}

  \noindent We now study the case $t\neq t'$, which implies
  $\proj_t(p)=\proj_t(p')$.  We prove that, in each case,
  $(c_1,\rho'_1)$ satisfies (ii) for $p$:

  \begin{iteMize}{$\bullet$}

  \item
    Case $s'=\sy{yield}$.
    We have $\rho_1=\rho$ and $\alpha_t(h_1)=\alpha_t(h)$.
    \smallskip

  \item
    Case $s'=\sy{lock}(m)$.
    We have $\rho_1=\rho$.
    In order ensure that $c_1\in\alpha_t(h)$, we need to ensure that
    $m\notin u_1$.
    We note that, by definition of $\lbp{S_\C}$, a configuration with 
    $m\in u_1$ can only be reached if the control path $\proj_t(p')$ 
    executed by the thread $t$ contains some $X\leftarrow\sy{islocked}(m)$ 
    statement not followed by any $\sy{lock}(m)$ nor $\sy{yield}$ statement, 
    and no thread $t''>t$ locks $m$.
    We deduce that $t'<t$.
    Hence, the interleaved control path $p'$ preempts the thread $t$
    after a non-blocking operation to schedule a lower-priority thread $t'$.
    This is forbidden by the $\enabled$ function: we have
    $R(\enabled_{t'}(\lbx{\Pi_\H}{p'}(\E_{0h},\emptyset)))=\emptyset$,
    hence $R(\lbx{\Pi_\H}{p}(\E_{0h},\emptyset))=\emptyset$.
    \smallskip

  \item
    Case $s'=\sy{unlock}(m)$.
    We have $\rho_1=\rho$ and $\alpha_t(h_1)\subseteq\alpha_t(h)$.
    \smallskip

  \item
    Case $s'=X\leftarrow\sy{islocked}(m)$.
    We have $\alpha_t(h_1)=\alpha_t(h)$.
    Moreover, $\fa{Y\neq X}\rho_1(Y)=\rho(Y)$ and $\rho(X)\in\sset{0,1}$.
    To prove that (ii) holds, it is sufficient to prove that
    $\ex{t'',c'}t\neq t'',\,(t'',c',X,\rho(X))\in I_\C$ and 
    $\excl(c_1,c')$, so that the value of $\rho(X)$ can be seen as an 
    interference from some $t''$ in $t$.
    We choose $t''=t'$.
    Moreover, as $c'$, we choose the configuration obtained by
    applying the recurrence hypothesis (ii) to $p'$ on $(h_1,\rho_1)$, but $t'$
    instead of $t$.
    We deduce then by, definition of
    $\lb{S_\C}{X\leftarrow\sy{islocked}(m),\,t'}$ on the configuration $c'$,
    that there exist interferences $(t',c',X,0),(t',c',X,1)\in I_\C$.
    \smallskip

  \end{iteMize}
  This ends the proof that $\lbp{P_\H}\subseteq\lbp{P_\C}$.

  \medskip

  We now prove that $\lbp{P'_\H}\subseteq\lbp{P_\C}$.
  The proof is similar to \ref{proof:weakinterf}.
  Given an original path $p$ and the transformed path $p'$ of a thread $t$, 
  given any $(R,\Omega,I)$ such that $I$ is consistent with
  fresh and local variables, we prove:
  $$\begin{array}{l@{\quad}l}
    \text{(i)}   & \Omega(\lbx{\Pi_\C}{p',t}(R,\Omega,I)) \subseteq \Omega(\lbx{\Pi_\C}{p,t}(R,\Omega,I)) \\[8pt]
    \text{(ii)}  & \fa{(t',c,X,v)\in I(\lbx{\Pi_\C}{p',t}(R,\Omega,I))}\\[2pt]&(t',c,X,v)\in I(\lbx{\Pi_\C}{p,t}(R,\Omega,I))\text{ or }t=t'\wedge X\in\V_l(t)\\[8pt]
    \text{(iii)} & \fa{(c,\rho')\in R(\lbx{\Pi_\C}{p',t}(R,\Omega,I))}\ex{\rho\in\E}(c,\rho)\in R(\lbx{\Pi_\C}{p,t}(R,\Omega,I)),\\[2pt]&\fa{X\in\V}\rho(X)=\rho'(X)\text{ or }X\in \V_f\\
  \end{array}$$
  Note, in particular, that in (ii) and (iii), the configuration $c$ is the same
  for the original path $p$ and the transformed path $p'$.

  We first consider the case of acceptable elementary operations from 
  Def.~\ref{def:transform}.
  As they involve guards and assignments only, not synchronization primitives, 
  and because $\lb{S_\C}{X\leftarrow e,\,t}$ and $\lb{S_\C}{e\bowtie 0?,\,t}$ 
  never change
  the partitioning, the proof of (i)---(iii) for all elementary transformations 
  is identical to that of proof~\ref{proof:weakinterf}.

  We now prove that, if (i)--(iii) hold for a pair $p,p'$, then they also hold
  for a pair $p\cdot s,\,p'\cdot s$ for any statement $s$.
  The case where $s$ is an assignment or guard is also identical to that of
  proof~\ref{proof:weakinterf}, for the same reason.
  We now consider the case of synchronization primitives.
  (i) holds because synchronization primitives do not change the set of 
  errors.
  The proof that (ii) holds is similar to the case of an assignment.
  Indeed, any interference added by $s$ after $p'$ has the form
  $(t,c,X,\rho'(X))$ for some state $(c,\rho')\in R(\lbx{\Pi_\C}{p',t}(R,\Omega,I))$.
  Due to (iii), there is some $(c,\rho)\in R(\lbx{\Pi_\C}{p,t}(R,\Omega,I))$
  where, either $X\in\V_f$ or $\rho(X)=\rho'(X)$.
  When $X\notin\V_f$,  we note that  any $(t,c,X,\rho'(X))$ 
  added by $s$ after $p'$ is also added by $s$ after $p$.
  Thus, the extra interferences in $p'$ do not violate (ii).
  For the proof of (iii), take $(c,\rho')\in R(\lbx{\Pi_\C}{p'\cdot s,\,t}(R,\Omega,I))$.
  There exists some $(c,\rho'_1)\in R(\lbx{\Pi_\C}{p',t}(R,\Omega,I))$
  where, for all $X\in\V$, either $\rho'(X)=\rho'_1(X)$, or there is a
  well synchronized interference $(t',c',X,\rho'(X))$ with $t\neq t'$.
  By applying (ii) to the pair $p,p'$, all these interferences are also in
  $I(\lbx{\Pi_\C}{p,t}(R,\Omega,I))$.
  Thus, by applying (iii) to the pair $p,p'$, we get a state
  $(c,\rho_1)\in R(\lbx{\Pi_\C}{p,t}(R,\Omega,I))$
  that also satisfies (iii) for the pair $p\cdot s,\,p'\cdot s$.

  As in proof~\ref{proof:weakinterf}, the fact that, for any $p,p',p'',q$, 
  if the pairs $p,p'$ 
  and $p',p''$ satisfy (i)--(iii), then so do the pair $p,p''$ and the pair
  $q\cdot p,\,q\cdot p'$ is straightforward. This completes the proof that elementary
  path transformations can be applied in sequence and applied in a context
  containing any primitive statements (even synchronization ones) before and
  after the transformed part.

  The end of the proof is identical to that of proof~\ref{proof:weakinterf}.
  We compare the fixpoints $\lfp F$ and $\lfp F'$ that compute
  respectively the semantics
  of the original program $\lb{S_\C}{\body_t,\,t}=\lbx{\Pi_\C}{\pi(\body_t),\,t}$
  and the transformed program $\lbx{\Pi_\C}{\pi'(t),\,t}$.
  Then, (i) and (ii) imply that $F'(X)\sqsubseteq_\C F(X)$, and so,
  $\lfp F'\sqsubseteq_\C \lfp F$, except for interferences on local or free
  variables.
  In particular, $\Omega(\lfp F')\subseteq\Omega(\lfp F)$.
  The interference semantics of the original program contains all the 
  errors that can occur in any program obtained by acceptable thread
  transformations.
  As a consequence, $\lbp{P'_\H}\subseteq\lbp{P_\C}$.
\qed


\thmproof{soundsched}{$\lbp{P_\C}\subseteq\lbp{P_\C^\s}.$}
\proof 
  We remark that $\lbp{P_\C}$ and $\lbp{P_\I}$, and so,
  $\lbp{P_\C^\s}$ and $\lbp{P_\I^\s}$, are similar.
  In particular, the definitions for non-primitive statements and
  the fixpoint computation of interferences have the exact same structure.
  Hence, the proof \ref{proof:soundproc} applies
  directly to prove the soundness of non-primitive statements as a consequence
  of the soundness of primitive statements.
  Moreover, for assignments and tests, the proof of soundness is identical to
  the proof \ref{proof:soundproc}, but componentwise for each $c\in\C$.
  To prove the soundness of synchronization statements, we first observe
  the soundness of $\funin^\s$ and $\funout^\s$ in that:
  $\fa{t\in\T,\,l,u\subseteq\M,\,m\in \M,\,V^\s\in\E^\s,\,I^\s\in\I^\s}$
  $\fa{\rho\in\gamma_\E(V^\s)}$
  $$\begin{array}{l}
    \funin(t,l,u,m,\rho,\gamma_\I(I^\s))\subseteq\gamma_\E(\funin^\s(t,l,u,m,V^\s,I^\s))\\
    \funout(t,l,u,m,\rho,\gamma_\I(I^\s))\subseteq\gamma_\I(\funout^\s(t,l,u,m,V^\s,I^\s))\enspace.
  \end{array}$$
  Secondly, we observe that $\lbp{S_\C^\s}$ first reorganizes (without loss
  of information) the sets of environment tuples $R\subseteq \C\times \E$ 
  and interference tuples $I\subseteq \T\times\C\times\V\times\R$ 
  appearing in $\lbp{S_\C}$ as functions
  $R'\deq \lbd{c}\setst{\rho}{(c,\rho)\in R}$ and
  $I'\deq \lbd{t,c,X}\setst{v}{(t,c,X,v)\in I}$.
  Then, it applies an abstraction in, respectively, $\E^\s$ and $\RR^\s$
  by replacing the set union $\cup$ by its abstract counterparts
  $\cup^\s_\E$ and $\cup^\s_\I$.
\qed

\end{document}